\documentclass[12pt]{article}
\pdfoutput=1
\usepackage{epsf,amsfonts,amssymb,epsfig,amsmath,graphics,cite}
\usepackage{color}
\usepackage{hyperref,graphicx,subfig}
\usepackage{booktabs}
\usepackage{float}
\usepackage{mathrsfs}
\usepackage{empheq}
\usepackage{enumerate}
 
\makeatletter

\newcommand{\Rmnum}[1]{\expandafter\@slowromancap\romannumeral #1@}
\makeatother

\newcommand{\nn}{\notag }
\def\be{\begin{equation}}
\def\ee{\end{equation}}

\newcommand{\ii}{\mathrm{i}}
\newcommand{\ex}{\mathrm{e}}
\newcommand{\diff}{\mathrm{d}}
\newcommand{\R}{\mathbb{R}}
\newcommand{\Z}{\mathbb{Z}}
\newcommand{\vol}{\mathrm{vol}}
\newcommand{\Vol}{\mathrm{Vol}}
\newcommand{\C}{\mathbb{C}}

\newcommand{\kf}{\mathfrak{p}}

\newcommand{\Y}{Y_{2n+1}}
\newcommand{\X}{X_{2n-1}}
\newcommand{\newnu}{\upsilon}
\newcommand{\StoX}{T}
\newcommand{\Npm}{N^{X_\pm}}
\newcommand{\Np}{N^{X_+}}
\newcommand{\Nm}{N^{X_-}}
\newcommand{\Nas}{M}
\newcommand{\NAS}{M}

\numberwithin{equation}{section}       

\begin{document}

\begin{titlepage}

\begin{flushright}
Imperial/TP/2022/JG/04
\end{flushright}

\vskip 1cm

\begin{center}


{\Large \bf Gravitational Blocks, Spindles\\ 
\vskip 0.2cm
and GK Geometry}

\vskip 1cm
{Andrea Boido$^{\mathrm{a}}$, Jerome P. Gauntlett$^{\mathrm{b}}$, Dario Martelli$^{\mathrm{c,d}}$ and James Sparks$^{\mathrm{a}}$}

\vskip 1cm

 ${}^{\mathrm{a}}$\textit{Mathematical Institute, University of Oxford,\\
Andrew Wiles Building, Radcliffe Observatory Quarter,\\
Woodstock Road, Oxford, OX2 6GG, U.K.\\}

\vskip 0.2cm

${}^{\mathrm{b}}$\textit{Blackett Laboratory, Imperial College, \\
Prince Consort Rd., London, SW7 2AZ, U.K.\\}

\vskip 0.2cm

${}^{\mathrm{c}}$\textit{Dipartimento di Matematica ``Giuseppe Peano'', Universit\`a di Torino,\\
Via Carlo Alberto 10, 10123 Torino, Italy}

\vskip 0.2cm

${}^{\mathrm{d}}$\textit{INFN, Sezione di Torino, \\
 Via Pietro Giuria 1, 10125 Torino, Italy}

\vskip 0.2 cm

\end{center}

\vskip 0.5 cm

\begin{abstract}
\noindent We derive a gravitational block formula for the supersymmetric action for a general class 
of supersymmetric AdS solutions, described by GK geometry. Extremal points of this action 
describe supersymmetric AdS$_3$ solutions of type IIB supergravity, sourced by D3-branes, 
and supersymmetric AdS$_2$ solutions of $D=11$ supergravity, sourced by M2-branes. 
In both cases, the branes are also wrapped over a two-dimensional 
orbifold known as a spindle, or a two-sphere. We develop various geometric methods
for computing the gravitational block contributions, allowing us to recover
previously known results for various explicit supergravity solutions, and to significantly
generalize these results to other compactifications. 
For the AdS$_3$ solutions we give a general proof 
that our off-shell supersymmetric action agrees with an appropriate 
off-shell $c$-function in the dual field theory, establishing a very general exact result in holography. 
For the AdS$_2$ solutions our gravitational block formula allows us to obtain the entropy for
supersymmetric, magnetically charged and accelerating black holes in AdS$_4$.

\end{abstract}

\end{titlepage}

\pagestyle{plain}
\setcounter{page}{1}
\newcounter{bean}
\baselineskip18pt

\tableofcontents

\newpage

\section{Introduction}\label{sec:intro}

The programme of identifying and studying novel geometric structures associated with supersymmetric AdS solutions of string/M-theory has led to enormous progress in our understanding of the AdS/CFT correspondence. A particularly rich arena is provided 
by supersymmetric AdS$_3\times Y_7$ solutions of type IIB supergravity \cite{Kim:2005ez} 
and AdS$_2\times Y_9$ solutions of $D=11$ supergravity \cite{Kim:2006qu}. The AdS$_3\times Y_7$ solutions, which are dual to $d=2$,
$\mathcal{N}=(0,2)$ SCFTs, are supported by five-form flux and are associated with D3-branes wrapping two-dimensional, compact surfaces.
The  AdS$_2\times Y_9$ solutions are supported by electric four-form flux and are associated with M2-branes wrapping such surfaces and, furthermore, arise as the near horizon limit of supersymmetric black hole solutions. The geometry on $Y_7$ and $Y_9$ was further understood in \cite{Gauntlett:2007ts} and, moreover, extended to general odd dimensions $Y_{2n+1}$;  it is referred to as
GK geometry.

Initial progress in studying GK geometry arose from constructing explicit solutions \cite{Gauntlett:2006af,Gauntlett:2006qw,Gauntlett:2006ns,Gauntlett:2007ts}. However, such explicit constructions represent only a small 
fraction of a much larger landscape of solutions, and new techniques are needed in order to study the whole family and to elucidate 
properties of the dual field theories. Inspired by techniques that were developed to study Sasaki-Einstein ($SE$) manifolds some time ago \cite{Martelli:2005tp,Martelli:2006yb}, a geometric extremal problem characterizing GK-geometries was formulated in \cite{Couzens:2018wnk} and further developed in \cite{Gauntlett:2018dpc,Hosseini:2019use,Hosseini:2019ddy,Gauntlett:2019roi,Gauntlett:2019pqg}. In particular, these new techniques enable one to obtain quantities of physical interest for AdS$_3\times Y_7$ and AdS$_2\times Y_9$  backgrounds, including the central charge of the dual to $d=2$, $\mathcal{N}=(0,2)$ SCFTs and the entropy of the supersymmetric black holes, respectively, without needing explicit solutions.

A rich sub-class of such AdS$_3\times Y_7$ and AdS$_2\times Y_9$ solutions are closely associated with
AdS$_5\times SE_5$ and AdS$_4\times SE_7$ solutions, respectively. Recalling that the latter solutions arise from D3-branes and M2-branes 
sitting at the apex of the corresponding Calabi-Yau $CY_3$ and $CY_4$ cone geometries, one can further wrap the D3 and M2-branes on
a compact Riemann surface. Generically this results in a GK geometry which consists of the $SE$ geometry fibred over the Riemann surface.
This setup has been studied in \cite{Gauntlett:2018dpc,Hosseini:2019use,Hosseini:2019ddy,Gauntlett:2019roi,Gauntlett:2019pqg} and, in particular, significant progress has been made in the context of toric $SE$ manifolds, for which additional algebraic tools have been 
developed \cite{Gauntlett:2018dpc,Gauntlett:2019roi,Gauntlett:2019pqg}. In the type IIB context, one can interpret the AdS$_3\times Y_7$ solutions as being dual to the $d=4$, $\mathcal{N}=1$ SCFTs which are associated with AdS$_5\times SE_5$, that are then reduced on the Riemann surface and flow to $d=2$, $\mathcal{N}=(0,2)$ SCFTs in the IR. The geometric extremization of the GK geometry in the toric context has then been precisely identified with $c$-extremization \cite{Benini:2012cz,Benini:2013cda} of the dual field theory in \cite{Gauntlett:2018dpc,Hosseini:2019use}. In the $D=11$ context there is an analogous picture, with the AdS$_2\times Y_9$ solutions also being associated with the near horizon limit of supersymmetric black holes living in AdS$_4\times SE_7$. In this context the geometric extremization of the GK geometry corresponds to $I$-extremization \cite{Benini:2015eyy,Benini:2016rke}
in the field theory, as discussed in \cite{Gauntlett:2019roi,Hosseini:2019ddy,Kim:2019umc}, 
and this has increased the scope of recovering the entropy of supersymmetric AdS black holes from the dual field theory, vastly extending \cite{Benini:2015eyy}.

Until recently, the construction of supersymmetric AdS solutions associated with branes wrapping two-dimensional surfaces has focussed on compact Riemann surfaces, as discussed in the previous paragraph, with supersymmetry preserved via a topological twist \cite{Maldacena:2000mw}. However, it has
recently been shown \cite{Ferrero:2020laf,Ferrero:2020twa} that one can wrap D3 and M2-branes on certain two-dimensional orbifolds known as spindles, and moreover, the supergravity solutions preserve supersymmetry in a novel way, called the anti-twist.
The spindle solutions of \cite{Ferrero:2020laf,Ferrero:2020twa} were constructed in minimal $D=5,4$ gauged supergravity theories and then uplifted on Sasaki-Einstein manifolds $SE_5, SE_7$, respectively, to obtain AdS$_3\times Y_7$ solutions of type IIB and AdS$_2\times Y_9$ solutions  of $D=11$ supergravity. More general solutions have been found in STU-type gauged supergravities which can be uplifted on spheres, $SE_5=S^5$ or $SE_7=S^7$ (or specific orbifolds, thereof) \cite{Hosseini:2021fge,Boido:2021szx,Ferrero:2021ovq,Couzens:2021rlk,Ferrero:2021etw,Couzens:2021cpk}. For the STU class it was shown in \cite{Ferrero:2021etw} that in addition to the anti-twist solutions there are also topological twist solutions, which we will simply refer to
as twist solutions.\footnote{Note that for the usual AdS solutions based on Riemann surfaces with a topological twist, the metric on the Riemann surface has constant curvature and the Killing spinors are constant on the Riemann surface. Neither of these features is present for twist solutions associated with spindles.}  
The fact that supersymmetry is preserved in just one of these two ways is related to the global properties of spin$^c$ spinors on spindles  
with an azimuthal rotation symmetry \cite{Ferrero:2021etw}.  

The GK geometry for the AdS$_3\times Y_7$ and AdS$_2\times Y_9$ solutions obtained by uplifting
these gauged supergravity solutions involving spindles, consists of fibrations of Sasaki-Einstein manifolds over spindles. 
A most striking feature is that the orbifold singularities of the spindles get resolved in the uplift, and the
resulting GK geometries are completely smooth. 
In fact, these smooth solutions had already been constructed some time ago using a quite different perspective \cite{Gauntlett:2006af,Gauntlett:2006ns}. While not relevant for this paper, we note that other constructions of solutions associated with various branes wrapped on spindles as well as higher dimensional orbifolds, have also been made.\footnote{Solutions corresponding to branes wrapped on spindles appear in \cite{Ferrero:2021wvk,Ferrero:2021etw} (M5-branes),     \cite{Faedo:2021nub,Giri:2021xta} (D4-branes),  \cite{Couzens:2022yiv} (D2-branes),
while \cite{Arav:2022lzo} discusses compactifying the Leigh-Strassler theory on a spindle. Solutions for D4 or M5-branes wrapped on four-dimensional orbifolds are contained in 
\cite{Faedo:2021nub,Giri:2021xta,Cheung:2022ilc,Suh:2022olh,Couzens:2022lvg,Faedo:2022rqx}.
A related construction, involving branes wrapped on disks with orbifold singularities, has been presented in \cite{Bah:2021mzw,Bah:2021hei,Couzens:2021tnv,Suh:2021ifj,Suh:2021aik,Suh:2021hef,Karndumri:2022wpu,Couzens:2022yjl}.
We also highlight that some of these solutions still have orbifold singularities in $D=10,11$. }

The principal aim of this paper is to synthesize the general geometric extremization techniques for studying
GK geometries that have been developed with the recent progress in constructing explicit solutions associated to branes wrapped on spindles. The formalism we develop will allow us to study a much broader class of configurations of D3 and M2-branes wrapping spindles, for which supergravity solutions are very unlikely to be ever found in explicit form. In particular, by solving the geometric extremization problem one can extract key properties of the solutions
and of the holographic dual field theories. 

The key object in GK geometric extremization \cite{Couzens:2018wnk} is the supersymmetric action, $S_{\rm SUSY}$, which has to be
suitably extremized in the space of R-symmetry Killing vectors. For AdS$_3\times Y_7$ solutions $S_{\rm SUSY}$ is proportional to the ``trial central charge'', which after extremization becomes the central charge of the dual $d=2$ SCFT. 
For AdS$_2\times Y_9$ solutions $S_{\rm SUSY}$ is proportional to the ``trial entropy''  associated with the dual $d=1$ SCFT;
on-shell it is associated with the entropy of supersymmetric black holes which have the AdS$_2$ solution as a near horizon limit.
One of our main results is to demonstrate that for GK geometry consisting of fibrations of Sasaki-Einstein manifolds over spindles,
we can write $S_{\rm SUSY}$ in a ``gravitational block'' form.
Furthermore, this rewriting is also applicable for 
fibrations of Sasaki-Einstein manifolds over two-spheres, thus providing a new perspective on some of the results
for GK geometry discussed in \cite{Gauntlett:2018dpc,Hosseini:2019use,Hosseini:2019ddy,Gauntlett:2019roi,Gauntlett:2019pqg}.

 The idea of gravitational blocks was first proposed in \cite{Hosseini:2019iad} in the context of 
 supersymmetric black holes and black strings in AdS$_4\times S^7$ and AdS$_5\times S^5$, respectively, which carry electric and 
 magnetic charges as well as non-trivial rotation and with spherical horizons. There  it was shown that the entropy can be obtained by extremizing certain entropy functions that are obtained by summing (``gluing'') basic building blocks. This observation was inspired by the factorization of 
 partition functions of ${\cal N}=2$ field theories in $d=3$, proven in \cite{Beem:2012mb} (for later developments in field theory see also  \cite{Crew:2020jyf,Hosseini:2021mnn,Hosseini:2022vho}). In subsequent work 
\cite{Hosseini:2021fge,Faedo:2021nub} it was shown that for the class of
explicit supergravity solutions associated with M2 and D3-branes\footnote{D4 and M5-branes were also considered in \cite{Faedo:2021nub}.} wrapping spindles, obtained by uplifting solutions in $D=4,5$ STU gauged supergravity theories on 
$S^7$, $S^5$, respectively,
it is possible to write the off-shell trial entropy and trial central charge in the form of gravitational blocks, both for the
twist and the anti-twist classes. Here we will systematically derive the results of 
\cite{Hosseini:2021fge,Faedo:2021nub}, and moreover show that they can be extended
to the whole class of GK geometry consisting of fibrations of arbitrary Sasaki-Einstein manifolds over spindles, and, furthermore, over two-spheres.

We now turn to an outline of the paper, also highlighting some of the key results.
We begin in section \ref{sec:review} by reviewing some key aspects of GK geometry on $Y_{2n+1}$, $n\ge 3$, including summarizing the extremal problem involving the supersymmetric action $S_{\mathrm{SUSY}}$ as a function of the R-symmetry Killing vector $\xi$. In  particular, the extremal problem involves imposing some flux quantization conditions, which in the case of AdS$_3$ and AdS$_2$ solutions
is associated with flux quantization in the type IIB and $D=11$ supergravity solutions, respectively.
In the remainder of the paper we focus on GK geometry of the fibred form
\begin{align}
\label{fibrationformintro}
\X\, \hookrightarrow\, \Y \, \stackrel{\pi}{\longrightarrow}\, \Sigma\, ,
\end{align}
where $\X$ are Sasakian fibers and $\Sigma=\mathbb{WCP}^1_{[m_-,m_+]}$ is a spindle, {\it i.e.} a weighted 
projective space with co-prime weights $m_\pm \in\mathbb{N}$, with an azimuthal symmetry.
While we focus throughout on spindles, our main results are also applicable to replacing the spindle with a smooth two-sphere by
setting $m_\pm =1$.
We assume that the fibers have a $U(1)^s$ isometry (the ``flavour symmetry") so that we can write
\begin{align}
\xi = \sum_{\mu=0}^s b_\mu\, \partial_{\varphi_\mu}\, ,
\end{align}
where $(b_\mu) =(b_0,b_1,b_2,\ldots,b_s)\in\R^{s+1}$. Here  $\partial_{\varphi_0}$ denotes the Killing vector generating azimutal rotations of the spindle 
(uplifted to $\Y$), 
while    $\partial_{\varphi_i}$ with $i=1,\dots, s$ is a basis for the $U(1)^s$ action on the fibers $\X$.
The two fibers located at the north and south poles of the spindle $\Sigma$, denoted by $X_\pm$, respectively,
are orbifolds  $X_\pm\equiv \X/\Z_{m_\pm}$ and they play an important role in our analysis.

We analyse this setup in section \ref{sec:fibred} where we prove our main result
 \begin{align}
 \label{ssusytomaster}
	\boxed{
	S_{\text{SUSY}}	= \frac{2\pi b_1}{b_0} \left(
		\mathcal{V}_{2n-1}^+ - \mathcal{V}_{2n-1}^- \right) 
		}
\end{align}
where $\mathcal{V}_{2n-1}^\pm$
are ``master volumes" \cite{Gauntlett:2018dpc,Gauntlett:2019roi,Gauntlett:2019pqg} of the covering spaces $\X$ of the fibers $X_\pm$, respectively. 
With $\xi_\pm$  the orthogonal projection of the R-symmetry vector 
$\xi$  onto the directions tangent to the fibres $X_\pm$ over the two poles and, similarly
$J_\pm\equiv J|_{X_\pm}$  the transverse K\"ahler class of the GK geometry restricted to the fibres at the poles, we have 
$\mathcal{V}_{2n-1}^\pm  =    \mathcal{V}_{2n-1} ( \xi_\pm ;  [J_\pm])$.
We refer to (\ref{ssusytomaster}) as the 
``gravitational block" decomposition of $S_{\text{SUSY}}$.	

We also define geometric R-charges, $R_a^\pm$, which are associated
with certain supersymmetric submanifolds $S^{\pm}_{a}$ of dimension $(2n-3)$ in $Y_{2n+1}$. More precisely, the latter 
are defined as $U(1)^s$-invariant codimension two submanifolds $S_a\subset\X$, 
whose cones are divisors in the Calabi-Yau cone $\X$.\footnote{Later in section \ref{sec:toricoverspindle} we will take these to be precisely the toric divisors 
when $C(\X)$ is a toric Calabi-Yau cone, with the index $a=1,\ldots,d$ running over the 
number of facets $d$ of the associated polyhedral cone in $\R^n$. But for now we may 
take $a$ to be a general index, labelling any set of $U(1)^s$-invariant divisors in $C(\X)$. 
}
These in turn define codimension four submanifolds $S^{\pm}_a\subset\Y$, as the copies 
of $S_a$ in the fibres $X_\pm=X_{2n-1}/\mathbb{Z}_{m_\pm}$ over the two poles of the spindle. 
For AdS$_3\times Y_7$ solutions when $n=3$, the latter are 
three-dimensional supersymmetric submanifolds in the fibres 
$X_{5}/\mathbb{Z}_{m_\pm}$ and the geometric R-charges are dual to the R-charges of baryonic operators associated with D3-branes wrapping these 
submanifolds. Similarly, for AdS$_2\times Y_9$ solutions when $n=4$, they are five-dimensional supersymmetric submanifolds in the fibres $X_7/\mathbb{Z}_{m_\pm}$ and the geometric R-charges are dual to the R-charges of baryonic operators associated with M5-branes wrapping these submanifolds. 

The remainder of the paper considers further special cases where we can make additional progress. 
In sections \ref{nobar} and \ref{someexamplessec} we consider what we refer\footnote{This was called ``mesonic twist" in \cite{Hosseini:2019ddy}.}
 to as the ``flavour twist", which is defined by imposing a certain restriction on the quantized fluxes so that they are determined by the fibration structure and the choice of $\xi$. A particular case is provided by examples for which the fibre $\X$ has 
 no ``baryonic symmetries" {\it i.e. } $H^2(\X,\mathbb{R})\cong 0$. For the flavour twist we show that for $n=3$ the off-shell trial central 
 charge for AdS$_3\times Y_7$ solutions can be written as
\begin{align}
\mathscr{Z}=\frac{1}{b_0 } \left(\frac{1}{\left.\Vol_S(X_{5})\right |_{b_i^{(+)}}}-\frac{1}{ \left.\Vol_S(X_{5})\right |_{b_i^{(-)}}}\right){3\pi^3 N^{2}}\,,
\end{align}
while for $n=4$ the off-shell entropy for AdS$_2\times Y_9$ solutions has the form
\begin{equation}
	\mathscr{S}=\frac{1}{b_0} \left(\frac{1}{\sqrt{\left.\Vol_S(X_{7})\right |_{b_i^{(+)}}}}-\frac{\sigma}{\sqrt{ \left.\Vol_S(X_{7})\right |_{b_i^{(-)}}}}\right)\frac{8\pi^3 N^{3/2}}{3\sqrt{6 } }\,.
 \end{equation}
 Here $\Vol_S(\X)|_{b_i}$ is the Sasaki-volume of $\X$ as a function of the Reeb vector $b_i$ and $\sigma=\pm1$ is associated with
 the two ways of preserving supersymmetry on the spindle, the twist and the anti-twist \cite{Ferrero:2021etw}.
Also $N=m_+m_-\mathcal{N}_0$ with $\mathcal{N}_0\in\mathbb{N}$, 
 can be understood as the flux through the Sasaki-Einstein space for the associated 
 AdS$_5\times SE_5$ and AdS$_4\times SE_7$ solutions.
   A special sub-case of the flavour twist is what we shall call the universal anti-twist;  for $n=3$ and $4$ this corresponds to the 
explicit AdS$_3\times Y_7$ and AdS$_2\times Y_9$ solutions constructed using minimal gauged supergravity in $D=5$ and $D=4$ in \cite{Ferrero:2020laf}
and \cite{Ferrero:2020twa}, respectively, both of which are in the anti-twist class, and we find exact agreement. Section \ref{someexamplessec} illustrates with some specific examples of the flavour twist.

We then switch gears in sections \ref{sec:fibretoric} - \ref{secftmatch} to study the case when the fibres $\X$, and hence $\Y$, are toric.
Section \ref{sec:fibretoric} reviews some basic features of toric GK geometry on $\Y$ \cite{Gauntlett:2018dpc,Gauntlett:2019roi,Gauntlett:2019pqg}. In section \ref{sec:toricoverspindle} we discuss how
$S_{\text{SUSY}}$ in \eqref{ssusytomaster} can be written algebraically in terms of the toric data of the fibre $\X$.
We also obtain similarly explicit expressions for the geometric R-charges, $R_a^\pm$, in terms of $\mathcal{V}_{2n-1}$ (see 
\eqref{Rapmequiv}). In section \ref{secftmatch} we focus on AdS$_3\times Y_7$ solutions 
and prove the remarkable result
\begin{equation}
		\mathscr{Z}	= \frac{1}{b_0}  \sum_{a < b < c}  (\vec v_{a}, \vec v_{b}, \vec{v}_c) \left(R_a^+ R_b^+ R_c^+ - R_a^- R_b^- R_c^-\right) {3 N^2} \,,
\end{equation}
where $R_a^\pm=R_a^\pm(b_i^\pm)$ are the geometric R-charges and $(\vec v_{a}, \vec v_{b}, \vec{v}_c)$ is a $3\times 3$ determinant,  with $\vec v_{a}$ the toric data for the fibre $\X$.
The AdS$_3\times Y_7$ solutions can be interpreted as being dual to the $\mathcal{N}=1$, $d=4$ SCFT, which is dual to 
AdS$_5\times SE_5$, that is then compactified on the spindle with magnetic fluxes switched on.
We determine the explicit map between the field theory variables involved in $c$-extremization and those appearing in
the extremization of the GK geometry. We also illustrate some of our formalism explicitly by considering some examples.

In section \ref{bhsec9} we discuss some features of the AdS$_2\times Y_9$ solutions when interpreted as the near horizon
limit of supersymmetric and accelerating black holes in AdS$_4\times SE_7$. This complements the recent discussion in
\cite{Boido:2022iye}.

Section \ref{sec:disc} concludes with some discussion. Appendix \ref{app:A} 
and \ref{app:matching} contain some technical material
relevant for sections \ref{sec:toricoverspindle}
and \ref{secftmatch}, respectively. 
Appendix \ref{app:KK} discusses some aspects of Kaluza-Klein reduction of
type IIB and $D=11$ supergravity on $SE_5$ and $SE_7$ spaces, respectively, which illuminates some subtleties
concerning flavour and baryonic symmetries discussed in section \ref{bhsec9}.

\section{AdS solutions from GK geometry}\label{sec:review}

From a physics perspective, we are interested in a class of supersymmetric AdS$_3\times Y_7$ solutions of type IIB string theory 
 supported by D3-brane flux \cite{Kim:2005ez}
and AdS$_2\times Y_9$ solutions of M-theory supported by M2-brane flux \cite{Kim:2006qu}.  
In both cases the internal space $\Y$ is equipped with
a GK geometry \cite{Gauntlett:2007ts}
with $n=3$, $n=4$, respectively. In  \cite{Couzens:2018wnk} solutions 
to the supergravity equations of motion 
were shown to be critical points of a certain finite-dimensional extremal problem for GK geometry. 
In the next two subsections we briefly review these constructions, in general (odd) dimension for $\Y$, 
before then specializing to the above two cases of physical interest.

\subsection{GK geometry}\label{sec:GK}
We begin by discussing some salient features of GK geometry, 
referring to \cite{Gauntlett:2007ts} for more details.

GK geometry \cite{Gauntlett:2007ts} is defined on odd-dimensional Riemannian manifolds $Y=Y_{2n+1}$, of 
dimension $2n+1$, where $n\geq 3$. 
The metric on $\Y$ is equipped with a unit norm Killing vector field $\xi$, called the R-symmetry vector. 
Since $\xi$ is nowhere vanishing, it defines a foliation $\mathcal{F}_\xi$ of $\Y$, and in local coordinates we may write
\begin{align}\label{Reeb}
\xi = \frac{1}{c}\partial_z\, , \qquad \eta = c\, (\diff z+P)\, ,
\end{align}
where we have defined $c\equiv \frac{1}{2}(n-2)$, and $\eta$ is the Killing one-form dual to $\xi$. The metric on $\Y$ then takes the form
\begin{align}\label{metricY}
\diff s^2_{\Y} = \eta^2 + \ex^B\,  \diff s^2_T\, ,
\end{align}
where $\diff s^2_T$ is a K\"ahler metric transverse to $\mathcal{F}_\xi$. This K\"ahler metric has a transverse K\"ahler two-form 
$J$, Ricci two-form $\rho=\diff P$, and Ricci scalar $R$. Moreover, the conformal factor $\ex^B$ in \eqref{metricY} is  given by 
\begin{align}
\ex^B = \frac{c^2}{2}R\, .
\end{align}
In particular this means that we require the K\"ahler metric to have positive scalar curvature, $R>0$.

The metric cone over $\Y$ is by definition $C(\Y)\equiv \mathbb{R}_{>0}\times \Y$, with conical metric 
$\diff s^2_{C(\Y)} = \diff r^2 + r^2\, \diff s^2_{\Y}$. 
There is also an equivalent characterization of GK geometry in terms of the cone geometry \cite{Gauntlett:2007ts}.
For a GK geometry on $\Y$ 
the cone has an integrable complex structure, and moreover we require 
there to exist a nowhere vanishing holomorphic $(n+1,0)$-form $\Psi$, 
which is closed $\diff\Psi=0$. In this sense $C(\Y)$ is then Calabi-Yau, 
having vanishing first Chern class, although the conical metric $\diff s^2_{C(\Y)}$ is 
neither K\"ahler nor Ricci-flat. We also require that $\Psi$ has definite 
charge under $\xi$, satisfying
\begin{align}\label{Psicharge}
\mathcal{L}_\xi \Psi = \frac{\ii}{c}\, \Psi\, .
\end{align}
In particular this condition implies that $\xi$ is a holomorphic vector field on $C(\Y)$. 

A GK geometry becomes ``on-shell'', satisfying 
the supergravity equations of motion in the string/M-theory applications 
described in sections \ref{sec:D3setup} and \ref{sec:M2setup},  
 if the transverse K\"ahler metric satisfies
the non-linear partial differential equation
\begin{align}\label{PDE}
\Box R = \frac{1}{2}R^2 - R_{ab}R^{ab}\, ,
\end{align}
where $R_{ab}$ denotes the Ricci tensor for the K\"ahler metric, and $\Box$ is the Laplacian operator. 

\subsection{The extremal problem}\label{sec:extremal}

We are interested in the following extremal problem in GK geometry, introduced in~\cite{Couzens:2018wnk}. 
We fix a complex cone $C(\Y)=\mathbb{R}_{>0}\times \Y$ with holomorphic 
volume form $\Psi$, together with a holomorphic $U(1)^{s+1}$ action. 
Here $s\geq 0$ necessarily, as $\xi$ generates at least a $U(1)$ action, 
and in this paper we will be interested in the case that $s\geq 1$ and
there is at least a $U(1)^2$ action. 
We take  corresponding generating vector fields $\partial_{\varphi_\mu}$, $\mu=0,1,\ldots,s$, with each 
$\varphi_\mu$ having period $2\pi$. Moreover, we choose this basis so that 
the holomorphic  volume form has unit charge under 
$\partial_{\varphi_1}$, and is uncharged under $\partial_{\varphi_{\hat{\mu}}}$, 
$\hat{\mu}=0,2,\ldots,s$. 
Notice here that we have 
singled out the $\partial_{\varphi_0}$ direction, as well as the $\partial_{\varphi_1}$ direction. The reason for this notation will become clear in
section~\ref{sec:fibred}.
A choice of holomorphic R-symmetry vector $\xi$ may then be 
written as 
\begin{align}\label{Reebbasis}
\xi = \sum_{\mu=0}^s b_\mu\, \partial_{\varphi_\mu}\, ,
\end{align}
where we may regard the coefficients as defining a 
vector $(b_\mu) =(b_0,b_1,b_2,\ldots,b_s)\in\R^{s+1}$.\footnote{As is standard in the physics literature, $b_\mu$  here 
denotes both the vector, and the $\mu$-th component of this vector. When this abuse of notation might lead to potential 
confusion, we write $(b_\mu)\in\R^{s+1}$ for the vector.} Given the condition \eqref{Psicharge}, 
we must then set 
\begin{align}\label{setb1}
b_1 = \frac{1}{c}= \frac{2}{n-2}\, .
\end{align}
A choice of $\xi$ determines the foliation $\mathcal{F}_\xi$, and 
we then further choose a transverse K\"ahler metric with basic 
cohomology class $[J]\in H^{1,1}_B(\mathcal{F}_\xi)$. 
It will often be convenient to note from \eqref{Reeb} that
\begin{align}\label{detarho}
\diff \eta = c \rho\, = \frac{1}{b_1}\rho\, ,
\end{align}
where $[\rho]\in H^{1,1}_B(\mathcal{F}_\xi)$ also defines a 
basic cohomology class.

Given this data, we may define the following \emph{supersymmetric 
action}
\begin{align}\label{SSUSY}
S_{\mathrm{SUSY}} \equiv \int_{\Y}\, \eta \wedge \rho \wedge \frac{J^{n-1}}{(n-1)!}\, .
\end{align}
It is straightforward to show \cite{Couzens:2018wnk} that this is a positive multiple 
of the integral of the scalar curvature of the transverse metric $R>0$ over $\Y$, and thus 
$S_{\mathrm{SUSY}}>0$ is a necessary condition for a regular on-shell solution. 
A {necessary} condition for the transverse K\"ahler metric to solve the PDE \eqref{PDE} is
the constraint equation
\begin{align}\label{constraint}
\int_{\Y}\, \eta \wedge \rho^2 \wedge \frac{J^{n-2}}{(n-2)!} = 0\, ,
\end{align}
which is equivalent to imposing that the integral of \eqref{PDE} over $\Y$ holds. 
We also impose the \emph{flux quantization conditions}
\begin{align}\label{fluxquantize}
\int_{\Sigma_\alpha}\, \eta\wedge \rho \wedge \frac{J^{n-2}}{(n-2)!}= \nu_n\, \mathcal{N}_\alpha\, .
\end{align}
Here $\Sigma_\alpha\subset Y_{2n+1}$ are  codimension two submanifolds, tangent to $\xi$, which form a basis for the free part of $H_{2n-1}(\Y;\Z)$, $\nu_n$ 
are certain real constants that are given in the physical cases of interest of $n=3$, $n=4$ 
in subsections \ref{sec:D3setup}, \ref{sec:M2setup} below, and 
$\mathcal{N}_\alpha\in\Z$ are the quantized fluxes. 
For a given $\xi$, it is important to notice that the quantities 
\eqref{SSUSY}, \eqref{constraint}, \eqref{fluxquantize} 
depend only on the basic cohomology classes $[J], [\rho]\in H^{1,1}_B(\mathcal{F}_\xi)$, 
and not on the choice of K\"ahler metric itself. 

The extremal problem we are interested in is to extremize 
the supersymmetric action \eqref{SSUSY}, subject to imposing 
the constraints \eqref{constraint}, \eqref{fluxquantize}, for fixed 
flux numbers $\mathcal{N}_\alpha$.  Here the parameter space 
is the choice  
of $\xi$, parametrized by $(b_\mu)\in\R^{s+1}$ via \eqref{Reebbasis}, 
and the choice of transverse K\"ahler class. 
The number of constraints \eqref{constraint}, \eqref{fluxquantize} 
is the same as the number of transverse K\"ahler class parameters.
Assuming one can eliminate the latter, the supersymmetric action 
\eqref{SSUSY} then effectively becomes a function only of $b_\mu$ and the flux numbers $\mathcal{N}_\alpha\in\Z$. While we don't have a general theorem 
to this effect, we shall see later that this is the case in various examples.
The main result of \cite{Couzens:2018wnk} is that GK geometries 
that solve the PDE \eqref{PDE} are necessarily solutions to this extremal 
problem. It is an important outstanding problem to determine sufficient conditions for when solutions of
the extremal problem guarantee that one in fact has a GK geometry solving the PDE; for some further discussion 
see \cite{Couzens:2018wnk,Gauntlett:2018dpc}.

\subsection{AdS$_3$ solutions from D3-branes}\label{sec:D3setup}

Setting $n=3$ in the above gives internal spaces $Y_7$ that are associated to supersymmetric 
solutions of type IIB supergravity of the form \cite{Kim:2005ez} 
\begin{align}
\diff s^2_{10} & = L^2\, \ex^{-B/2}\left(\diff s^2_{\mathrm{AdS}_3} + \diff s^2_{Y_7}\right)\, ,\nn\\
F_5 & = - L^4\left(\mathrm{vol}_{\mathrm{AdS}_3}\wedge F + *_{Y_7}\,  F\right)\, .
\end{align}
Here we have introduced the closed two-form
\begin{align}\label{Fflux}
F  \equiv -\frac{1}{c}J + \diff(\ex^{-B}\eta)\, ,
\end{align}
where $c=\tfrac{1}{2}(n-2)=\tfrac{1}{2}$,
$\diff s^2_{\mathrm{AdS}_3}$ denotes the unit radius metric on AdS$_3$, and $L>0$ is a constant.
In fact, if desired one can absorb $L$ in the transverse K\"ahler geometry via the rescaling $J\to L^{-4}J$, which implies $\ex^{-B}\to L^{-4}\ex^{-B}$.

The fact that only the five-form flux $F_5$ in type IIB is non-zero implies that these solutions 
are in some sense supported only by D3-branes. This flux is properly 
quantized via \eqref{fluxquantize}, satisfying
\begin{align}\label{fluxquant3}
\frac{1}{(2\pi\ell_s)^4g_s}\int_{\Sigma_\alpha}\, F_5 = \mathcal{N}_\alpha\, ,
\end{align}
 provided we take the constant $\nu_3$ to be
\begin{align}
\nu_3= \frac{2(2\pi\ell_s)^4 g_s}{L^4}\, ,
\end{align}
where $\ell_s$ is the string length, and $g_s$ denotes the string coupling constant. 
Furthermore, the extremal value of the supersymmetric 
action \eqref{SSUSY} determines the central charge 
$c_{\mathrm{sugra}}$ of the dual field theory. In fact 
\cite{Couzens:2018wnk} introduced a ``trial central charge'' 
$\mathscr{Z}$, defined by
\begin{align}\label{offshellz}
\mathscr{Z}\, \equiv\, \frac{3L^8}{(2\pi)^6g_s^2\ell_s^8}\, S_{\mathrm{SUSY}}= \frac{12(2\pi)^2}{\nu_3^2}\, S_{\mathrm{SUSY}}\, , 
\end{align}
and on-shell, i.e. after extremizing, one has has 
\begin{align}
\mathscr{Z}_\text{os}= c_{\mathrm{sugra}}\, . 
\end{align}

\subsection{AdS$_2$ solutions from M2-branes}\label{sec:M2setup}

Setting instead $n=4$  gives internal spaces $Y_9$ that are associated to supersymmetric 
solutions of $D=11$ supergravity of the form \cite{Kim:2006qu}
\begin{align}
\diff s^2_{11} & = L^2\, \ex^{-2B/3}\left(\diff s^2_{\mathrm{AdS}_2} + \diff s^2_{Y_9}\right)\, ,\nn\\
G & = - L^3\, \mathrm{vol}_{\mathrm{AdS}_2}\wedge F \, ,
\end{align}
where $F$ is again given by \eqref{Fflux}, but with now $c=\tfrac{1}{2}(n-2)=1$. 
Again $L>0$ is a constant which can be absorbed into the transverse K\"ahler geometry via $J\to L^{-3}J$ (this is
what is done in \cite{Boido:2022iye}).

We may interpret the fact that the flux $G$ is zero when restricted to $Y_9$,
as meaning there is only M2-brane flux, and no M5-brane flux, sourcing 
these solutions.
The Hodge dual seven-form $*_{11}G$ is properly 
quantized via \eqref{fluxquantize}, satisfying
\begin{align}\label{fluxquant4}
\frac{1}{(2\pi \ell_p)^6}\int_{\Sigma_\alpha}\, *_{11}G= \mathcal{N}_\alpha\, ,
\end{align}
 provided we take the constant $\nu_4$ to be
\begin{align}
\nu_4= \frac{(2\pi \ell_p)^6}{L^6}\, ,
\end{align}
where $\ell_p$ is the eleven-dimensional Planck length. 
In this case we define a ``trial entropy'' $\mathscr{S}$ via
\begin{align}\label{ads2entess}
\mathscr{S}\, \equiv\, \frac{4\pi L^9}{(2\pi)^8 \ell_p^9}\, S_{\mathrm{SUSY}}
=\frac{2(2\pi)^2}{\nu_4^{3/2}}\, S_{\mathrm{SUSY}}\, \, ,
\end{align}
where the supersymmetric action is given by \eqref{SSUSY} with $n=4$. 
When the $D=11$ solution arises as the near horizon limit of a 
supersymmetric black hole, it was argued in \cite{Couzens:2018wnk} 
that the on-shell value, $\mathscr{S}_\text{os}$, is the entropy 
of the black hole. 
More generally, one expects this quantity to 
be the logarithm of a supersymmetric partition function 
of the dual superconformal quantum mechanics. 

\section{Spindly gravitational blocks}\label{sec:fibred}

In the remainder of the paper we will be interested in studying the extremal problem 
described in section \ref{sec:review}, in the special case that the internal space 
$\Y$ takes the fibred form
\begin{align}\label{XoverY}
\X\, \hookrightarrow\, \Y \, \stackrel{\pi}{\longrightarrow}\, \Sigma\, .
\end{align}
Here $\Y$ projects under the projection map $\pi$ to a two-dimensional surface $\Sigma$, 
with Sasakian fibre $\X$.

Physically this corresponds to the following set-up. When the fibres $X=X_{2n-1}$ 
of $Y=Y_{2n+1}$ are Sasaki-Einstein manifolds, taking $n=3$, $n=4$ leads to associated 
AdS$_5\times X_5$ and AdS$_4\times X_7$ solutions of type IIB 
supergravity and $D=11$ supergravity, respectively. These are the near horizon limits 
of $N$ D3-branes or $N$ M2-branes placed at the Calabi-Yau cone 
singularities of $C(\X)$, respectively. We may then interpret 
the fibration \eqref{XoverY} as wrapping the D3-branes or M2-branes 
over the two-dimensional surface $\Sigma$, with a general 
partial topological twist/fibration. The resulting low-energy 
effective theories on these wrapped branes, in dimensions $d=2$ and $d=1$ respectively, then 
flow to superconformal fixed points, with near horizon 
holographic duals given by AdS$_3\times Y_7$ and AdS$_2\times Y_9$, 
respectively. 

The case where $\Sigma=\Sigma_g$ is a smooth Riemann 
surface of genus $g$, and where the R-symmetry vector $\xi$ is tangent 
to toric fibres $X$, was studied in \cite{Gauntlett:2018dpc, Gauntlett:2019roi}. 
Here we are interested in generalizing this set-up by taking
$\Sigma=\mathbb{WCP}^1_{[m_-,m_+]}$ to be a spindle, or equivalently a weighted projective space with co-prime weights 
$m_\pm \in\mathbb{N}$. Moreover, we take general fibres (not just toric),
where crucially $\xi$ now has a component that 
is also tangent to $\Sigma$.\footnote{More precisely, 
$\pi_*\xi$ is a non-zero vector field on $\Sigma$.} 
The action of $\xi$ on $\Sigma$ is simply
rotation about the poles, which are orbifold points modelled 
locally by $\C/\Z_{m_\pm}$.  This includes the special case that $\Sigma=S^2$ 
is a two-sphere\footnote{In practice, for the $S^2$ case the extremal $\xi$ does not have a component
tangent to $S^2$, as in \cite{Gauntlett:2018dpc, Gauntlett:2019roi}. From a physics perspective, for $n=3$ one can understand
this as being associated with the lack of mixing of the non-abelian isometries of the $S^2$ with the R-symmetry vector in $c$-extremization.}, when $m_\pm =1$.

The main result of this section will be that the supersymmetric 
action \eqref{SSUSY}, together with the associated 
constraint \eqref{constraint} and flux quantization conditions 
\eqref{fluxquantize}, \emph{localize} to integrals 
over the fibres $X_\pm$ over the poles of the spindle.\footnote{In general 
these fibres will be topologically $X_\pm \equiv \X/\Z_{m_\pm}$, rather than
copies of the generic fibre $\X$, as we will explain in section \ref{sec:fibre} below. 
Strictly speaking, \eqref{XoverY} is then not a fibration.} 
We refer to these contributions as \emph{gravitational blocks}, 
since they generalize similar (conjectured) formulas that have already 
appeared in the literature -- see   \cite{Hosseini:2019iad, Hosseini:2021fge, Faedo:2021nub, Hosseini:2021mnn, Hristov:2021qsw, Cassani:2021dwa, Ferrero:2021ovq}.  As well as deriving these 
gravitational block formulas, we  will also relate physical quantities 
in the parent $d=4$ and $d=3$ field theories on the 
D3-branes and M2-branes, respectively, to physical quantities 
in the compactified $d=2$ and $d=1$ theories in the IR. 
In turn, this will allow us to prove
certain relations in holography.

\subsection{Fibrations over a spindle}\label{sec:fibre}

We may construct a spindle $\Sigma$ straightforwardly 
by gluing copies of $\C/\Z_{m_+}$ and
$\C/\Z_{m_-}$, along their common $S^1$ boundary. 
The fibred geometries $\Y$ that we are interested in may then 
be realized via a modification of this gluing construction. 

We begin with the product $\C\times \X$, where $X_{2n-1}$ will be the base 
of a Calabi-Yau cone $C(\X)$ with $\dim_\C C(\X) = n$. The conical metric is
$\diff s^2_{C(\X)}=\diff r^2 + r^2\, \diff s^2_{\X}$, with $r>0$, and we denote by 
$\Omega$ the closed holomorphic $(n,0)$-form on $C(\X)$. 
We also suppose that $C(\X)$  is equipped with a holomorphic 
$U(1)^s$ action, with the Reeb vector field $\mathcal{J}(r\partial_r)$ 
generating a subgroup of this action, where $\mathcal{J}$ is the complex structure tensor of 
$C(\X)$. We take a basis $\partial_{\psi_i}$, $i=1,\ldots,s$, of holomorphic vector 
fields that generates the $U(1)^s$ action, such that the holomorphic 
volume form $\Omega$ has charge $1$ under $\partial_{\psi_1}$, and is 
uncharged under $\partial_{\psi_i}$, $i=2,\ldots,s$. The corresponding 
coordinates $\psi_i$ on $\X=\{r=1\}\subset C(\X)$ here have period $2\pi$.\footnote{Compare 
to the discussion at the start of section \ref{sec:extremal}. The 
reason for using the notation $\psi_i$ here, rather than $\varphi_i$,  
will become clear momentarily.}

Given this set-up, the space $\Y$ may be constructed by 
taking $(\C\times \X)/\Z_{m_+}$ and gluing it to $(\C\times \X)/\Z_{m_-}$,
where the local orbifold groups $\Z_{m_\pm}$ act on both $\C$ and the ``fibres'' $\X$. 
It then remains to specify this action, and also make precise how we glue. 
We may accomplish both together by first choosing two
homomorphisms $h_\pm:U(1)\rightarrow U(1)^s$. This is equivalent 
to specifying integers $\alpha^\pm_i\in\Z$, $i=1,\ldots,s$, so that explicitly
\begin{align}
h_\pm(\omega)= (\omega^{\alpha^\pm_1},\ldots,\omega^{\alpha^\pm_s}) \, \in \, U(1)^s\, ,
\end{align}
with $\omega\in U(1)$ a unit norm complex number. Given a point $(z,x)\in \C\times \X$, 
where $z\in\C$ and $x$ is a point in $\X$, the $\Z_{m_\pm}$ orbifold actions are then defined by taking the generators $\Gamma_\pm$ to be
\begin{align}\label{orbiaction}
\Gamma_\pm (z,x)  \equiv (\omega_\pm\cdot z, h_\pm(\omega_\pm)\cdot x)\, \in \, \C\times \X\, ,
\end{align}
respectively. Here $\omega_\pm \equiv \ex^{2\pi \ii/m_\pm}$ are primitive 
$m_\pm$-th roots of unity, and the action of $h_\pm \in U(1)^s$ on the 
point $x\in X$ is via the assumed holomorphic $U(1)^s$ action on $\X$. 
Quotienting $\C\times \X$ by \eqref{orbiaction} then defines $(\C\times \X)/\Z_{m_\pm}$. 
Notice here that any two choices of $\alpha^\pm_i$ that agree modulo 
$m_\pm$ (respectively for upper and lower signs) give the same $\Z_{m_\pm}$ quotient. 
That is, $(\C\times \X)/\Z_{m_\pm}$ depends only on $\alpha^\pm_i\in \Z_{m_\pm}$, where the  latter then
specify homomorphisms from $\Z_{m_\pm}\rightarrow U(1)^s$. However, in the gluing construction 
of the two local models we describe next it will be important to regard $\alpha^\pm_i\in \Z$, 
and not just defined mod $m_\pm$. 

The homomorphisms $h_\pm$ also specify diffeomorphisms $\Phi_\pm$ (which may be 
thought of as large $U(1)^s$ gauge transformations) of $S^1\times \X\subset \C\times \X$, where 
$S^1 =\{z\in \C\mid |z|=1\}\cong U(1)$ are the unit  norm complex numbers. Specifically,
\begin{align}\label{diffeo}
\Phi_\pm(z,x)\, \equiv  \, (z,h_\pm^{-1}(z)\cdot x)\, \in \, S^1\times \X\, .
\end{align}
Notice that the composition $\Phi_\pm \circ \Gamma_\pm\circ \Phi^{-1}_\pm$ 
maps $(z,x)\in S^1\times \X$ to $(\omega_\pm \cdot z,x)$.  Thus, after composing with the diffeomorphism
$\Phi_\pm$, the quotient acts only on the $z$ coordinate, 
showing 
that   $(S^1\times \X)/\Z_{m_\pm}\cong 
S^1/\Z_{m_\pm}\times \X\cong 
S^1\times \X$, which is the boundary of $(\C\times \X)/\Z_{m_\pm}$. 
Both models then have the same boundary $S^1\times \X$, and
we glue these boundaries with the identity diffeomorphism, after reversing the orientation 
of $(\C\times \X)/\Z_{m_-}$. 
This constructs the total space $\Y$.

We may make the above discussion more explicit by introducing coordinates. 
We write $z_\pm = |z_\pm|\ex^{\ii \hat\phi^\pm}$ as complex coordinates 
on each copy of $\C$, where $\hat\phi^\pm$ have period $2\pi$ before quotienting. 
The diffeomorphism/large gauge transformation \eqref{diffeo} is then implemented by the
coordinate transformation
\begin{align}\label{diffeocoords}
\phi^\pm  \equiv \hat\phi^\pm\,  , \qquad \varphi^\pm_i  \equiv \psi^\pm_i - \alpha^\pm_i \hat\phi_\pm\, .
\end{align}
Here $\psi^\pm_i$ are $2\pi$-period angular coordinates on each copy of $\X$, 
and $(\phi^\pm,\varphi^\pm_i)$ are the new angular coordinates on each copy of $S^1 \times\X$. 
It follows that on the quotients $(\C\times \X)/\Z_{m_\pm}$ the
$\phi^\pm$ have periods $2\pi/m_\pm$, respectively. The gluing 
is then accomplished by identifying (the minus sign due to the orientation 
change of the second copy)
\begin{align}\label{spindleglue}
\varphi\ \equiv\ m_+ \phi^+ = -m_-\phi^-\, ,
\end{align}
so that $\varphi$ is a $2\pi$-period coordinate for the azimuthal direction 
of the spindle. The fibres $\X$ are  glued with the identity 
diffeomorphism, meaning we identify\footnote{The fibres $\X$ can be glued with the diffeomorphism
$\varphi^+_i = \varphi^-_i - t_i\varphi$
where the parameters $t_i\in\Z$ give a further diffeomorphism/large gauge transformation when we glue, 
as described in  \cite{Ferrero:2021etw}. However, in terms 
of the original coordinates this leads to the identification
$\psi_i^+ - \alpha_i^+\hat\phi_+ = \psi_i^- - \alpha_i^-\hat\phi_- + t_im_-\hat\phi_-$,
from which we see that we can simply absorb $t_i$ into a redefinition of 
$\alpha_i^-$ via $\alpha_i^-\rightarrow \alpha^-_i - t_i m_-$ (notice this 
leaves $\alpha_i^-$ mod $m_-$ invariant, and so preserves the local model 
$(\C\times \X)/\Z_{m_-}$).  
Thus we can set $t_i=0$ without loss of generality. We could also remove the
remaining redundancy in this construction/description by 
requiring $\alpha_i^+\in \{1,\ldots,m_+\}$, and then take $\alpha_i^-\in \Z$.
}
\begin{align}\label{phipmp}
\varphi^+_i = \varphi^-_i \,.
\end{align}
We then define $\varphi_i\equiv \varphi_i^+=\varphi_i^-$ as the angular coordinates on the fibres $\X$.

\subsection{Twist and anti-twist}\label{sec:volform}

The previous subsection gives a self-contained description for how 
to construct the fibration 
\eqref{XoverY} over a spindle $\Sigma$. However, we 
may also make contact with the discussion in \cite{Ferrero:2021etw}, 
which focused on $U(1)$ fibrations over a spindle, by 
focusing on the $i$th factor of $U(1)\subset U(1)^s$. 
The corresponding  pair of integers $(\alpha^+_i,\alpha^-_i)$ 
specify a $U(1)$ orbibundle over the spindle base $\Sigma$.\footnote{This 
pair of integers was denoted $(m_N,-m_S)$ in  \cite{Ferrero:2021etw}, 
where recall we have without loss of generality set the variable $p=0$ in \cite{Ferrero:2021etw}, absorbing this into $m_N$ or $m_S$ ({\it cf.} equation \eqref{phipmp} 
and the discussion after where $p$ in \cite{Ferrero:2021etw} should be identified with one of the $t_i$). We also note that the analogue of the $\psi_i$ coordinates here were denoted by $\chi$ (for a single $U(1)$)
in \cite{Ferrero:2021etw}.} 
One can introduce a corresponding connection one-form $A_i$, and the discussion in \cite{Ferrero:2021etw}
computes the total flux/Chern number
\begin{align}\label{pis}
\frac{1}{2\pi }\int_\Sigma \, F_i = \frac{\alpha_i^-}{m_-}+\frac{\alpha_i^+}{m_+}
= \frac{p_i}{m_+m_-}\, ,
\end{align}
where $F_i\equiv \diff A_i$, and we have defined 
\begin{align}\label{pidef}
p_i\, \equiv\, \alpha_i^-m_++\alpha_i^+m_- \, .
\end{align}
The corresponding complex line bundle over $\Sigma$ is denoted 
$\mathcal{O}(p_i)$. In particular, different choices of
$\alpha_i^\pm$ with the same $p_i$ in \eqref{pidef} give
isomorphic line bundles, as we shall see.

Conversely, given a choice of $(p_1,\ldots,p_s)\in\Z^s$, we may 
specify the local model data above by first picking coprime 
integers $a_\pm$ satisfying
\begin{align}\label{apm}
a_-m_++a_+ m_- = 1\, ,
\end{align}
which exist by Bezout's lemma for coprime $m_+$, $m_-$, 
and then defining
\begin{align}\label{alphasa}
\alpha^+_i  \equiv a_+ p_i\, , \qquad \alpha^-_i \equiv a_-p_i\, ,\qquad i= 2,\ldots,s\, .
\end{align}
These satisfy \eqref{pidef}, by virtue of \eqref{apm}. Notice that here
we treat the $i=1$ direction differently, as discussed below. 
One can check that different choices in the above construction result 
 in equivalent spaces $\Y$. Explicitly, given a solution $(a_+,a_-)\in \Z^2$ to 
\eqref{apm}, another solution is given by taking
\begin{align}
a_+ \, \mapsto\, a_+ -\kappa m_+\, , \qquad a_-\, \mapsto\, a_- + \kappa m_-\, ,
\end{align} 
where $\kappa\in\Z$ is arbitrary. Via \eqref{alphasa} this in turn shifts $\alpha_i^\pm \rightarrow 
\alpha^\pm_i \mp \kappa p_i  m_\pm \equiv \alpha_i^\pm$  mod $m_\pm$. It follows that different choices of $\kappa$ lead to 
local models at each pole with equivalent quotient spaces $(\C\times \X)/\Z_{m_\pm}$. Moreover, the fact that 
$\alpha_i^\pm/m_\pm$ change by $\mp \kappa p_i$, with opposite 
signs, then implies this change to each local model simply cancels in the gluing construction, to obtain $\Y$. 

The $i=1$ copy of $U(1)$ is special, as we chose the basis so that the 
holomorphic $(n,0)$-form $\Omega$ on $\X$ has charge 1 under this isometry. 
Equivalently, the Killing spinor associated to the GK geometry is charged 
under this direction. This was analysed in some detail in  \cite{Ferrero:2021etw}, 
with the conclusion being that there are precisely two possibilities, 
called the \emph{twist} and \emph{anti-twist}:
\begin{align}\label{twistalpha}
\mbox{twist}: \qquad & \alpha_1^+ =-1\, , \  \alpha_1^- = -1\, , \nn\\
\mbox{anti-twist}: \qquad & \alpha_1^+ = -1\, , \ \alpha_1^- = +1\, .
\end{align} 
Via \eqref{pidef} we then have\footnote{Fixing the \emph{overall} sign of 
$p_1$ here involves a choice of convention, which we have fixed in writing both 
\eqref{twistalpha} and \eqref{twistp}, and also in the discussion of the complex structure 
and holomorphic volume form in the remainder of this subsection. The result of \cite{Ferrero:2021etw}
more generally fixes $|\alpha_1^\pm|=1$, rather than \eqref{twistalpha}, and it is the \emph{relative} sign of 
$\alpha_1^+$ and $\alpha_1^-$ that distinguishes the twist and anti-twist cases.}
\begin{align}\label{twistp}
\mbox{twist}: \qquad & p_1= -m_+-m_-\, ,\nn \\
\mbox{anti-twist}: \qquad & p_1= +m_+-m_-\, .
\end{align}
In particular for the twist case this gives $\mathcal{O}(p_1) = \mathcal{O}(-m_+-m_-)=K_\Sigma$ as the
canonical line orbibundle of the spindle $\Sigma$. 
It will be convenient for the remainder of the paper to 
write
\begin{align}\label{twisty}
\alpha_1^-  =  -\sigma\, , \qquad \boxed{p_1 = -\sigma m_+ - m_-}\, ,
\end{align}
where we have introduced
\begin{align}\label{sigmadef}
\sigma \, \equiv\, \begin{cases} \ +1 & \mbox{twist}\, , \\ \ -1 & \mbox{anti-twist}\, .\end{cases}
\end{align}

The result for the twist case $\sigma=+1$ may also be obtained via the 
following construction. 
 Recall that $\Y$ is a fibration of $\X$ over $\Sigma$, 
and that $\Psi$ denotes the holomorphic $(n+1,0)$-form on $C(\Y)$. The twist case arises 
precisely when $C(\X)$ fibred over $\Sigma$ is also 
Calabi-Yau, in the sense that it admits a holomorphic $(n+1,0)$-form. 
This space is in general only a partial resolution 
of the cone singularity $C(\Y)$, and we denote its holomorphic volume 
form also by~$\Psi$. 
This may then be constructed  locally as the 
wedge product of a $(1,0)$-form on the spindle $\Sigma$ with 
the holomorphic $(n,0)$-form $\Omega$ on $\X\subset C(\X)$, being 
careful to check this glues together appropriately. 

We begin by constructing $\Psi$ in each of the two patches. 
Recall that $z_\pm = |z_\pm|\ex^{\ii \hat\phi_\pm}$ 
are the complex coordinates on each copy of $\C$. We denote by 
$\Omega^\pm$ the $(n,0)$-forms 
on the corresponding copies of $\X$, before quotienting. 
These may further be written as $\Omega^\pm = \ex^{\ii\psi_1^\pm}\hat{\Omega}^\pm$,
where  $\hat\Omega^\pm$ is uncharged under all $\partial_{\psi_i^\pm}$, $i=1,\ldots,s$. 
We may then define
\begin{align}
\Psi^\pm\, \equiv\, \diff z_\pm \wedge \Omega^\pm\, ,
\end{align}
which are holomorphic $(n+1,0)$-forms, before quotienting 
by $\Z_{m_\pm}$. Combining the angular changes of coordinates in section \ref{sec:fibre}, 
we find
\begin{align}\label{Psipm}
\Psi^\pm= \diff \left(|z_\pm|\ex^{\ii \phi^\pm }\right) \wedge \ex^{ \ii\alpha_1^\pm \phi^\pm }\ex^{\ii\varphi_1}\hat\Omega^\pm \,= \, \diff \left(|z_\pm|\ex^{\pm \ii \varphi/m_\pm }\right)\wedge \ex^{\pm \ii\alpha_1^\pm \varphi/m_\pm }\ex^{\ii\varphi_1}\hat\Omega^\pm\, .
\end{align}
In order for the first expression to be invariant under the $\Z_{m_\pm}$ quotient, which recall acts only on the $\phi^\pm$ coordinates, 
we see that we immediately require $\alpha_1^\pm \equiv -1$ mod $m_\pm$. Furthermore, in the second expression both $\varphi$ and $\varphi_1$ 
are globally defined angular coordinates (on the complement of fixed points of the torus action), where recall that \eqref{spindleglue}
implements the gluing of the spindle, while $\varphi_i=\varphi_i^+=\varphi_i^-$ identifies angular coordinates 
on the two copies of $\X$. Requiring $\Psi\equiv \Psi^+=\Psi^-$ to agree on the overlap in \eqref{Psipm}, and moreover be charged 
only under $\partial_{\varphi_1}$, with unit charge, then imposes
precisely the twist condition $\alpha_1^+=\alpha_1^-=-1$ in \eqref{twistalpha}. 
Thus, the twist condition is equivalent to 
the fibration of $C(\X)$ over $\Sigma$ admitting this 
global holomorphic $(n+1,0)$-form. In the 
literature this is often then called a (\emph{partial}) \emph{ 
topological twist}. 
At present there is no similar geometric interpretation of the anti-twist in 
\eqref{twistalpha}. 

Finally, for both the twist and the anti-twist case, notice that defining
 $\partial_{\varphi_0}\equiv \partial_{\varphi}$, together with $\partial_{\varphi_i}$, $i=1,\ldots,s$, these vector fields 
generate a $U(1)^{s+1}$ action on the total space $\Y$, where this basis then satisfies the conditions imposed in the general set-up described in section \ref{sec:extremal}. 
In particular, the holomorphic $(n+1,0)$-form $\Psi$ is \emph{uncharged} 
under $\partial_{\varphi_0}$. In terms of the coordinates
introduced in  the 
construction in section \ref{sec:fibre}, it is helpful to note here that
\begin{align}\label{vectordiffeo}
\partial_{\varphi_0} \, \equiv \, \partial_\varphi= \pm \frac{1}{m_\pm} \partial_{\hat\phi^\pm} \pm  \sum_{i=1}^s\frac{\alpha^\pm_i}{m_\pm}\partial_{\varphi_i}\, ,\qquad 
\partial_{\varphi_i}= \partial_{\varphi_i^\pm}=  \partial_{\psi_i^\pm}\, .
\end{align}
Later in our discussion we will be particularly interested in the vector fields $\zeta_\pm$, that 
by definition rotate the normal directions to the fibres over the poles of the spindle; 
that is, the copies of $X_\pm\equiv \X/\Z_{m_\pm}$ located at $z_\pm=0$, respectively, in the
construction of section \ref{sec:fibre}. These are given by $\zeta_\pm = \partial_{\hat\phi^\pm}$, and 
\eqref{vectordiffeo} immediately gives
\begin{align}\label{zetapm}
\zeta_\pm \, \equiv \, \partial_{\hat\phi^\pm}=  
\begin{cases} \ \ m_+ \partial_{\varphi_0} +\ \ \partial_{\varphi_1} -\sum_{i=2}^s \alpha_i^+ \partial_{\varphi_i} \, \equiv\, \ \ \sum_{\mu=0}^s v_{+\mu} \partial_{\varphi_\mu} \\   - m_- \partial_{\varphi_0} +\sigma \partial_{\varphi_1} -\sum_{i=2}^s \alpha_i^- \partial_{\varphi_i}\, \equiv\, \sigma \sum_{\mu=0}^s v_{-\mu} \partial_{\varphi_\mu}\, ,
\end{cases}
\end{align}
where we have used \eqref{twistalpha} and \eqref{twisty}. We 
may then also read off the ``toric data'' vectors 
\begin{align}\label{vpmmu}
(v_{+\mu})  = ( m_+, 1,-\vec{\alpha}^+)\, , \qquad (v_{-\mu})  = (-\sigma m_-, 1,- \sigma\vec{\alpha}^-)\, ,
\end{align}
where $\vec\alpha^\pm \equiv (\alpha_2^\pm,\ldots,\alpha_s^\pm)$. 
We will make use of this result in section \ref{sec:toricoverspindle}.

For later use, we define $\xi_\pm$ to be
the orthogonal projection of the R-symmetry vector 
$\xi$ given in \eqref{Reebbasis} onto directions tangent to the fibres over the two poles. 
These may then be viewed as R-symmetry vectors for the fibres $X_\pm$.
From \eqref{zetapm} we have
\begin{align}\label{orthreebprojs}
\xi_+\equiv \, \xi-\frac{b_0\zeta_+}{m_+}=\sum_{i=1}^s {b}^{(+)}_i\partial_{\varphi_i}\,,\qquad
\xi_-\equiv\,  \xi+\frac{b_0\zeta_-}{m_-}=\sum_{i=1}^s {b}^{(-)}_i\partial_{\varphi_i}\,,
\end{align}
where we have defined the shifted vectors
\begin{align}\label{twistedblamfirst}
{b}^{(+)}_i&\equiv\,  {b}_i-\frac{b_0}{m_+}{v}_{+i}\,,\qquad\quad
{b}^{(-)}_i\equiv \, {b}_i+\frac{b_0}{\sigma m_-}{v}_{-i}\,.
\end{align}
For later use note that
\begin{align}\label{diffbpbm}
{b}^{(+)}_i-{b}^{(-)}_i=\frac{b_0p_i}{m_+m_-}\,,
\end{align}
where we used \eqref{pidef}.

\subsection{Gravitational block lemma}\label{sec:fixedpoints}

In this section we prove the following general formula
\begin{align}\label{generalblock}
\int_{\Y} \, \eta \wedge \rho \wedge  \Gamma= \frac{2\pi b_1}{b_0} \left(m_+ \int_{X_+}\eta\wedge\Gamma - m_- \int_{X_-}\eta\wedge\Gamma\right)\, .
\end{align}
Here $\Y$ is a GK geometry fibering over a spindle $\Sigma$, as in \eqref{XoverY}, where the fibres over 
the two poles of the spindle are respectively $X_\pm \cong \X/\Z_{m_\pm}$, with the quotient given by \eqref{orbiaction}. 
The orientations of $X_\pm$ here are those naturally induced via the Stokes' theorem argument we introduce shortly, 
but we note that these may not agree with the natural orientation induced by the complex structure on $C(\Y)$ (we shall return to discuss this later).  
Recall that we write 
the R-symmetry vector on $\Y$ as
\begin{align}\label{Reebexpandagain}
\xi = \sum_{\mu=0}^s b_\mu \partial_{\varphi_\mu}\, ,
\end{align}
where the vector field $\partial_{\varphi_0}$ was defined in the previous subsection, and rotates the spindle $\Sigma$, 
while $\partial_{\varphi_i}$ for $i=1,\ldots, s$ are tangent to the fibres $\X$ of \eqref{XoverY}.
The differential form $\Gamma$ in \eqref{generalblock} is then any closed form on $\Y$ that is basic with 
respect to the foliation defined by $\xi$; that is, $\mathcal{L}_\xi \Gamma=0$, 
$\xi\lrcorner \, \Gamma=0$. 

Both the supersymmetric action \eqref{SSUSY} and constraint equation \eqref{constraint} 
take the form of the left hand side of \eqref{generalblock}, but so too does the flux quantization 
condition \eqref{fluxquantize}, for submanifolds  $\Sigma_\alpha\subset \Y$ that themselves fibre
over the spindle direction $\Sigma$. Equation \eqref{generalblock} will form the basis for much 
of the rest of the paper. Here we provide a general differential-geometric proof for arbitrary fibre $\X$.
In later sections we will obtain an alternative proof for the case of toric $\X$.

We begin by noting from \eqref{detarho} that $\rho=b_1 \diff\eta$, so that \eqref{generalblock} is equivalent to
\begin{align}\label{generalblocketas}
\int_{\Y} \, \eta \wedge \diff\eta \wedge  \Gamma= \frac{2\pi }{b_0} \left(m_+ \int_{X_+}\eta\wedge\Gamma - m_- \int_{X_-}\eta\wedge\Gamma\right)\, .
\end{align}
In section \ref{sec:fibre} we introduced on $\Y$ a set of  $2\pi$-period coordinates 
$\varphi_\mu$, $\mu=0,\ldots,s$, which are globally well-defined away from the fixed points 
of the $U(1)^{s+1}$ action on $\Y$. 
In particular, $\varphi_0$ is a $2\pi$-period coordinate for the azimuthal direction 
of the spindle $\Sigma$, while $\varphi_i$, $i=1,\ldots,s$, are coordinates on the fibres $\X$.
$\varphi_0$ is then well-defined on $\Sigma$, except at the poles. 
Each pole of the spindle is locally modelled as a quotient $\C/\Z_{m_\pm}$, where 
we introduced complex coordinates $z_\pm = |z_\pm| \ex^{\ii\hat\phi_\pm}$ 
on the covering space copies of $\C$. Via \eqref{diffeocoords}, \eqref{spindleglue}, we then 
have
\begin{align}\label{anglesagain}
\hat\phi_+ =  \frac{1}{m_+}\varphi_0\, , \qquad \hat\phi_- =  -\frac{1}{m_-}\varphi_0\, .
\end{align}

Given the above notation, we next define the one-form
\begin{align}
\newnu_0 \, \equiv\, \frac{\diff\varphi_0}{2\pi}\, .
\end{align}
{\it A priori} this is a 
one-form on $\Sigma$ that is well-defined except at the poles. We may  
then pull this back to a one-form $\pi^*\newnu_0$ on $\Y$, which in an abuse 
of notation we simply refer to also as $\newnu_0$. The latter is correspondingly then well-defined on $\Y$,
except at the fibres $X_\pm$ over the poles, {\it i.e.} 
precisely where we integrate on the right hand side of \eqref{generalblocketas}.

The form $\newnu_0$ is manifestly closed, but we must be careful at the poles. 
Indeed, if we take the standard complex coordinate $z=|z|\ex^{\ii\phi}$ on 
$\C$, where $\phi$ has period $2\pi$, then $\newnu\equiv\diff\phi/2\pi $ is a smooth one-form
on $\C\setminus\{0\}$. Its exterior derivative is a two-form that is zero 
on $\C\setminus\{0\}$, but we also have
\begin{align}\label{nuintegral}
\int_{S^1_\epsilon} \, \newnu = 1\, ,
\end{align}
where $S^1_\epsilon \, \equiv\, \{|z|=\epsilon\}$ is the circle of radius $\epsilon>0$. 
We may then identify
\begin{align}\label{Diracs}
\diff \newnu = \delta\, \equiv\, \delta(x,y)\, \diff x\wedge\diff y\, ,
\end{align}
as a distribution-valued two-form, 
where $z=x+\ii y$, and $\delta(x,y)$ is the usual Dirac delta function. 
More abstractly, $\diff\newnu=\delta$ is a delta function 
representative of the Poincar\'e dual to the origin $0\in\C$. 
We then have
\begin{align}
1= \int_{D_\epsilon} \, \delta = \int_{D_\epsilon}\, \diff\newnu = \int_{S^1_\epsilon} \, \newnu \, ,
\end{align}
as in \eqref{nuintegral}, where $D_\epsilon \, \equiv\, \{|z|\leq \epsilon\}$ is the disc of radius $\epsilon$. 

We may apply the above analysis to $\newnu_0$ on the spindle, where $\diff\newnu_0$ is zero 
except at the poles of $\Sigma$. These are both orbifold singularities, where 
recall that we introduced a local model 
at each pole, starting with $\C\times \X$ and then quotienting by the $\Z_{m_\pm}$ 
action generated by \eqref{orbiaction}. 
We may then identify
\begin{align}\label{dnu0}
\diff\newnu_0= m_+ \delta_+ - m_- \delta_-\, .
\end{align}
Here the factors of $\pm m_\pm$ come from the corresponding factors in \eqref{anglesagain}, 
and $\delta_\pm$ are defined precisely as in \eqref{Diracs}, in local coordinates $z_\pm$ 
for each covering space $\C$. 
Notice that the origin of $\C$ is fixed under the $\Z_{m_\pm}$ action 
generated by \eqref{orbiaction}, so this group acts purely on the fibres $\X$ 
over the origin in $\C\times \X$, meaning the fibres over the poles are $X_\pm = \X/\Z_{m_\pm}$. 
Using \eqref{dnu0} we may then immediately write
\begin{align}\label{deltaXpm}
\int_{\Y}\, \eta\wedge \diff\newnu_0 \wedge \Gamma = m_+\int_{X_+} \eta \wedge \Gamma - m_-\int_{X_-} \eta \wedge \Gamma \, .
\end{align}
Here $\diff\newnu_0$ is zero except near to  $X_\pm$, which 
are at the origins of each local model $(\C\times \X)/\Z_{m_\pm}$. 
On the other hand, the delta functions in \eqref{dnu0} precisely 
restrict the integral to these copies of $X_\pm$. 
Starting with the left hand side of \eqref{deltaXpm}, we may then integrate by parts, 
using the fact that $\diff \Gamma=0$, to obtain\footnote{One can circumvent the use 
of delta functions by instead cutting out small neighbourhoods 
of $X_\pm$ in $\Y$, to obtain a manifold with two boundary components,  
$(S^1\times \X)/\Z_{m_+}\sqcup\,  (-S^1\times \X)/\Z_{m_-}$, and then
integrating $\newnu_0\wedge\eta\wedge \Gamma$ over this boundary. The integral of 
$\newnu_0$ over the circle factors gives $m_\pm$, resulting in the right hand side of 
\eqref{deltaXpm}. Instead using Stokes' theorem, we have $\diff\newnu_0=0$ 
on the interior of the manifold with boundary, and the only contribution 
is the right hand side of \eqref{byparts}. }
\begin{align}\label{byparts}
\int_{\Y}\, \eta\wedge \diff\newnu_0 \wedge \Gamma= \int_{\Y}\, \newnu_0\wedge \diff\eta\wedge \Gamma 
= \int_{\Y}\, (\xi\lrcorner\, \newnu_0)\, \eta\wedge \diff\eta \wedge \Gamma\, .
\end{align} 
Here in the last step we have used the fact that both $\diff\eta$ and $\Gamma$ are basic, 
so $\xi\lrcorner\, \diff\eta=0=\xi\lrcorner\, \Gamma$, 
while $\xi\lrcorner\, \eta=1$. On the other hand, from \eqref{Reebexpandagain} 
we immediately have $\xi\lrcorner \, \newnu_0=b_0/2\pi$. 
Combining \eqref{byparts} with \eqref{deltaXpm}, we have thus proven 
\eqref{generalblocketas}. 

\subsection{Gravitational block formula}\label{gravblockformrchgetoo}
We can now use these results to refine the extremal problem discussed in section
\ref{sec:extremal} for GK geometries of the fibred form \eqref{XoverY}. Specifically we want to reconsider the
supersymmetric action \eqref{SSUSY}, the constraint equation \eqref{constraint} and the expression for the fluxes \eqref{fluxquantize}.

We first observe that the supersymmetric action \eqref{SSUSY} takes the form 
of \eqref{generalblock}, with $\Gamma=J^{n-1}/(n-1)!$ and hence we
may write
\begin{align}\label{actiongb0}
S_{\mathrm{SUSY}}  = \frac{2\pi b_1}{b_0} \left(m_+ \int_{X_+}\eta\wedge \frac{J^{n-1}}{(n-1)!} - m_- \int_{X_-}\eta\wedge\frac{J^{n-1}}{(n-1)!}\right)\, ,
\end{align}
or
\begin{align}\label{actiongb}
\boxed{
S_{\mathrm{SUSY}}  =  \frac{2\pi b_1}{b_0}\left[m_+ \mathrm{Vol}(X_+) - m_- \mathrm{Vol}(X_-)\right]
}
\end{align}
where $\mathrm{Vol}(X_\pm)$ are the induced volumes 
of the fibres over the poles. Indeed, since 
$X_\pm \cong \X/\Z_{m_\pm}$, we have in both 
cases $m_\pm \mathrm{Vol}(X_\pm) = \mathrm{Vol}(\X)$, 
although the induced volume forms for each copy of $\X$ 
are in general different, and the volumes are different. We shall obtain an explicit 
formula for this for toric $\X$ in section \ref{sec:toricoverspindle}. 
As remarked after equation \eqref{generalblock}, the orientations of $X_\pm$ 
in \eqref{actiongb} may not agree with those induced from the complex structure 
on $C(\Y)$ (or equivalently the transverse complex structure on $\Y$), and correspondingly
the ``volumes'' $\mathrm{Vol}(X_\pm)$ will then be \emph{negative}. 
In fact we shall find that we can take
$\Vol(X_+)>0$ and $\sigma\, \Vol(X_-)>0$.

We next note that there are two preferred fluxes, 
$\Npm$, associated with the fibres $X_\pm \cong \X/\Z_{m_\pm}$
at the north and south poles, respectively. Following \eqref{fluxquantize} we define
 \begin{equation}\label{fluxquantfibre}
	\int_{X_\pm} \eta \wedge \rho \wedge \frac{J^{n-2}}{(n-2)!}\,  \equiv\,  \nu_n \, \Npm \:.
    \end{equation}
We can then apply \eqref{generalblock} to the constraint equation \eqref{constraint}. We set $\Gamma = \rho \wedge \frac{J^{n-2}}{(n-2)!}$, so that the constraint becomes
\begin{equation}
	m_+ \int_{X_+} \eta \wedge \rho \wedge \frac{J^{n-2}}{(n-2)!} - m_- \int_{X_-} \eta \wedge \rho \wedge \frac{J^{n-2}}{(n-2)!} = 0 \:.
\end{equation}
Thus, the constraint equation simply relates the two fluxes $\Npm$ in \eqref{fluxquantfibre}. It is convenient to then define $N\in \mathbb{Z}$ via
\begin{equation}\label{defN}
\boxed{
	N \, \equiv \, m_+ \Np = m_- \Nm
	}
\end{equation}
Notice that since $m_+$ and $m_-$ are assumed co-prime, this means that $N=m_+m_- {\mathcal N}_0$, where ${\mathcal N}_0\in\mathbb{N}$.
Indeed, without loss of generality we will take 
\begin{align}
N>0\,,
\end{align}
and hence with $m_\pm>0$, we also have $N^{X_\pm}>0$.

Before continuing, we now define the geometric R-charges for GK geometries of the fibred form \eqref{XoverY}. The R-charges, $R_a^\pm$, are associated
with certain supersymmetric submanifolds $S^{\pm}_{a}$ of dimension $(2n-3)$ in $Y_{2n+1}$. More precisely, these 
may be defined as follows. We begin with a set of $U(1)^s$-invariant codimension two submanifolds $S_a\subset\X$, 
whose cones are divisors in the Calabi-Yau cone $\X$.\footnote{Later in section \ref{sec:toricoverspindle} we will take these to be precisely the toric divisors 
when $C(\X)$ is a toric Calabi-Yau cone, with the index $a=1,\ldots,d$ running over the 
number of facets $d$ of the associated polyhedral cone in $\R^n$. But for now we may 
take $a$ to be a general index, labelling any set of $U(1)^s$ invariant divisors in $C(\X)$. 
}
These in turn define codimension four submanifolds $S^{\pm}_a\subset\Y$, as the copies 
of $S_a$ in the fibres $X_\pm=X_{2n-1}/\mathbb{Z}_{m_\pm}$ over the two poles of the spindle. 
We then define
\begin{equation}\label{Rpmdef0}
	R_a^\pm \equiv \frac{4 \pi }{\nu_n \Npm} \, \int_{S_a^\pm} \eta \wedge \frac{J^{n-2}}{(n-2)!} \:.
\end{equation}
For AdS$_3\times Y_7$ solutions when $n=3$, the latter are 
three-dimensional supersymmetric sub-manifolds in the fibres 
$X_{5}/\mathbb{Z}_{m_\pm}$ and the geometric R-charges are dual to the R-charges of baryonic operators associated with D3-branes wrapping these 
submanifolds. Similarly, for AdS$_2\times Y_9$ solutions when $n=4$, they are five-dimensional supersymmetric submanifolds in the fibres $X_7/\mathbb{Z}_{m_\pm}$ and the geometric R-charges are dual to the R-charges of baryonic operators associated with M5-branes wrapping these submanifolds. 
We emphasize that these supersymmetric submanifolds exist even in cases with
 $H^2(\X,\mathbb{R})\cong 0$, which we discuss further in section 
\ref{nobar}.
 We also note that there is again an issue of orientation on $S_a$ and we will see later in the toric case 
that we have $R_a^+>0$ and $\sigma R_a^->0$.

We now compute the fluxes through another preferred class of $(2n-1)$-dimensional submanifolds $\Sigma_a \subset Y_{2n+1}$. Specifically, we consider $\Sigma_a$ which are the total spaces of the supersymmetric submanifolds $S_a$ of the fibres $X_{2n-1}$ defined above, that are then fibred over the spindle $\Sigma$:
 \begin{equation}\label{Sigmaa}
 	S_a \, \hookrightarrow \, \Sigma_a \, \rightarrow \, \Sigma \: .
 \end{equation}
The associated flux through this class of cycles is\footnote{We emphasize that the fluxes $\Nas_a$ 
are, in general, distinct from the fluxes $\mathcal{N}_\alpha$ in
\eqref{fluxquantize} -- see \eqref{refeqn}. We also note that even when $H_{2n-3}(\X,\R)=0$, 
so that the homology classes of $S_a$ in the fibres 
$\X$ are necessarily trivial, it does not follow that the $M_a$ are zero. 
Indeed, these are integrals of the flux over $\Sigma_a$, whose homology classes are generically non-trivial due to the fibration.}
\begin{equation}\label{fluxes0}
	  \int_{\Sigma_a} \eta \wedge \rho \wedge \frac{J^{n-2}}{(n-2)!} \equiv \nu_n \Nas_a\:,
\end{equation}
where $\Nas_a\in \mathbb{Z}$. 
In a completely analogous fashion to how we derived the block formula \eqref{generalblock}, we can split the integral in \eqref{fluxes0} into two pieces as
\begin{equation}
	\int_{\Sigma_a} \eta \wedge \rho \wedge \frac{J^{n-2}}{(n-2)!} = \frac{2\pi b_1}{b_0} \left(m_+ \int_{S_a^+} \eta \wedge \frac{J^{n-2}}{(n-2)!} - m_- \int_{S_a^-} \eta \wedge \frac{J^{n-2}}{(n-2)!}\right) ,
\end{equation}
where recall that $S_a^\pm$ are the copies of the $S_a$ in the fibres over the north and south poles of the spindle, respectively. 
We thus conclude that these preferred fluxes are related to the R-charges associated to these submanifolds via
\begin{equation}\label{fluxes}
\boxed{
	\Nas_a = \frac{b_1}{2b_0} \left(R_a^+ - R_a^-\right)N 
	}
\end{equation}

 \section{Block formula for some simpler sub-classes}\label{nobar}
 
We can make further progress by making some additional assumptions on the GK geometry.
In this section we first consider what we call a ``flavour twist" (called a ``mesonic twist" in the toric setting in \cite{Hosseini:2019ddy}) followed by the ``universal anti-twist", which involves an additional assumption.
 
  \subsection{Block formula for $X_{2n-1}$ with a flavour twist}\label{nobarsec1}

First consider the special class of geometries where the fibre $\X$ 
 has no baryonic symmetries, {\it  i.e.}
 $H^2(\X,\mathbb{R})\cong 0$. 
The nomenclature comes from the fact that for $n=3,4$ the field theories dual to the associated AdS$_5\times X_5$ and AdS$_4\times X_7$ geometries then have no baryonic $U(1)$ flavour symmetries.
For this case
 the transverse K\"ahler class of the GK geometry restricted to the fibres at the poles, $[J|_{X_\pm}]$, must necessarily be proportional to $[\rho]$ and we can write
 \begin{equation}\label{kahlerclass}
 	[\left.J\right|_{X_{\pm}}] = \Lambda_{\pm} [\rho] \in H^2_B(\mathcal{F}_{\xi_\pm})\:,
 \end{equation}
 with $\Lambda_\pm \in \mathbb{R}$, 
 and recall that $\xi_\pm$ are the 
 R-symmetry vectors of the fibres $X_\pm$, introduced in \eqref{orthreebprojs}. 
 
 We can also consider a more general class of
 geometries where we don't assume $H^2(\X,\mathbb{R})\cong 0$, but nevertheless 
 \eqref{kahlerclass} is still satisfied. We call this the ``flavour twist". We will see below that this corresponds to fixing the 
 GK geometry $Y_{2n+1}$, including certain constraints on the fluxes $\Nas_a$, in terms of
$\X$, $m_\pm$ and $p_i$.
 
Substituting \eqref{kahlerclass} into the supersymmetric action \eqref{actiongb} we obtain
 \begin{align}\label{Snobar0}
 	S_{\mathrm{SUSY}}  &\, = \,  \frac{2\pi b_1}{ b_0}\left(m_+ \Lambda_+^{n-1}\int_{X_+}\eta\wedge \frac{\rho^{n-1}}{(n-1)!} - m_- \Lambda_-^{n-1}\int_{X_-}\eta\wedge \frac{\rho^{n-1}}{(n-1)!} \right)\nn \\
 		&\, = \,  \frac{\pi (2b_1)^n}{ b_0}\left[m_+\Lambda_+^{n-1}\, \Vol_S(X_+) -  m_-\Lambda_-^{n-1}\, \Vol_S(X_-)\right]\, .
 \end{align}
 Here $\Vol_S(X_\pm)$ is the Sasakian volume, obtained from $\Vol(X_\pm)$ by setting $[J] = \frac{1}{2b_1}[\rho]$, namely
 \begin{equation}\label{volsas}
 	\Vol_S(X_\pm) \, = \,  \frac{1}{(2 b_1)^{n-1}}  \int_{X_\pm} \eta \wedge \frac{\rho^{n-1}}{(n-1)!} \:.
 \end{equation}
These Sasaki volumes can be considered to be functions of trial Reeb vectors $\xi_\pm$ tangent to $X_\pm$. 
In practice, $\Vol_S(X_\pm)$ can be computed by taking the expression for the Sasakian volume of the fibre manifold $\X$ and taking the trial Reeb vector to be the orthogonal projection of the R-symmetry vector $\xi$ on $Y_{2n-1}$ onto the fibres over the poles. Recall $\xi_\pm$ were defined in  \eqref{orthreebprojs}, where we have 
\begin{align}\label{orthreebprojs2}
\xi_+\, \equiv \, \xi-\frac{b_0\zeta_+}{m_+}=\sum_{i=1}^s {b}^{(+)}_i\partial_{\varphi_i}\,,\qquad
\xi_-\, \equiv \,  \xi+\frac{b_0\zeta_-}{m_-}=\sum_{i=1}^s {b}^{(-)}_i\partial_{\varphi_i}\,.
\end{align}
Finally, we should divide by $m_\pm$ in order to take into account the quotient/orbifold singularities, and hence we can write
\begin{equation}\label{Vsas}
	\Vol_S(X_\pm) = \frac{1}{m_\pm} \left.\Vol_S(X_{2n-1}) \right |_{\xi=\xi_\pm} 
	 = \frac{1}{m_\pm} \left.\Vol_S(X_{2n-1}) \right |_{b_i^{(\pm)}} \:.
\end{equation}
Here $\Vol_S(X_{2n-1}) |_{b_i}$ is the standard positive\footnote{The sign is consistent with
the toric formalism in section \ref{sec:toricoverspindle} as well as with the known explicit supergravity solutions.}
 Sasaki volume on $X_{2n-1}$ 
as a function of $b_i$ (leading to a positive 
Sasaki-Einstein volume after extremization).
We thus have
 \begin{equation}\label{Snobar}
 	S_{\mathrm{SUSY}}  
	 		\, = \,  \frac{\pi (2b_1)^n}{ b_0}\left[\Lambda_+^{n-1}\, \left.\Vol_S(\X)\right |_{b_i^{(+)}}  -  \Lambda_-^{n-1}\, \left.\Vol_S(\X)\right |_{b_i^{(-)}} \right]\,.
 \end{equation}

To make further progress, we now consider the expression for the flux quantization \eqref{fluxquantfibre} through the fibres over the poles.
Using the same type of argument we find 
 \begin{align}\label{pmflux}
 		\Npm  &=   \frac{1}{\nu_n} \int_{X_\pm} \eta \wedge \rho \wedge \frac{J^{n-2}}{(n-2)!} \, = \frac{(n-1)}{m_\pm}\,  \frac{(2b_1)^{n-1} }{\nu_n}\, \Lambda_{\pm}^{n-2} \left.\Vol_S(X_{2n-1}) \right |_{b_i^{(\pm)}} \:,
 \end{align}
and we also recall the constraint equation \eqref{defN} given by $N = m_+ \Np = m_- \Nm$.
We can therefore solve for $\Lambda_{\pm}$ in terms of $N$. After substituting into \eqref{Snobar} we obtain
an expression for the off-shell action as a function of the R-symmetry vector $\xi$ given in \eqref{Reebbasis}
and the integer $N$. 
At this point, we can set $b_1= \frac{2}{n-2}$ and the extremization of the GK geometry with respect to the remaining
components of $b_\mu$ can be carried out. 
There is some difference in signs depending on whether $n$ is even or odd, which we will discuss further below.

Before doing that we derive an expression for the geometric R-charges $R_a^\pm$, defined in \eqref{Rpmdef0}. Recall
that $S_a^\pm$ are the copies of the supersymmetric $(2n-3)$-dimensional submanifolds $S_a$ in the fibres over the north and south poles of the spindle, respectively. We can express $R_a^\pm$
in terms of the Sasakian volume of the submanifolds $S_a^\pm$, defined as
 \begin{equation}
 	\Vol_S (S_a^\pm) \equiv \frac{1}{(2b_1)^{n-2}} \, \int_{S_a^\pm} \eta \wedge \frac{\rho^{n-2}}{(n-2)!}
	= \frac{1}{m_\pm} \left.\Vol_S(S_a) \right|_{b_i^{(\pm)}}  \:.
 \end{equation}
Here $\Vol_S(S_a) |_{b_i}$ is the standard positive Sasaki volume of the submanifold $S_a\subset X_{2n-1}$ 
as a function of $b_i$ (leading to a positive result on the Sasaki-Einstein manifold after extremization).
Hence, after eliminating $\Lambda_\pm$ using \eqref{pmflux}, we can conclude that 
 \begin{equation}\label{Rpmnobar}
	\boxed{
	R_a^\pm = \frac{2 \pi }{(n-1)b_1} \left. \frac{\Vol_S(S_a) }{\Vol_S(X_{2n-1}) }\right|_{b_i^{(\pm)}}
	}
	\end{equation}
Notice that using \eqref{fluxes} we can therefore write the fluxes in the form 
 \begin{equation}\label{Naformula}
\boxed{
	\Nas_a = \frac{\pi N}{(n-1) b_0} \left[\frac{\Vol_S(S_a)}{\Vol_S(\X)}\bigg|_{{b}_i^{(+)}} - \frac{\Vol_S(S_a)}{\Vol_S(\X)}\bigg|_{{b}_i^{(-)}}\right]
	}
\end{equation}
This equation must be interpreted carefully. The fluxes $\Nas_a$ on the left hand side 
are integers, while on the right hand side for fixed $\X$ this is a function of the spindle 
data $m_\pm$, the flavour twisting variables $p_i$, and the R-symmetry vector $(b_\mu)=(b_0,b_1,\ldots,b_s)\in\R^{s+1}$. 
Fixing $\X$, $m_\pm$ and $p_i$ fixes the internal space $\Y$, while 
the extremization involves varying over the R-symmetry vector, which of course is not compatible with holding 
the $\Nas_a$ fixed in \eqref{Naformula}. However, the correct interpretation is to extremize the supersymmetric action, 
detailed below in \eqref{cc_nobar}, \eqref{entropy}, to obtain the critical R-symmetry $b_\mu^*$, and then use this to compute 
the associated fluxes for this flavour twist solution using \eqref{Naformula};
in this sense the GK geometry $Y_{2n+1}$ and the fluxes $\Nas_a$ are fixed by
$\X$, $m_\pm$ and $p_i$. A necessary and sufficient 
condition for this to make sense is that the right hand side of \eqref{Naformula} is rational (hence integer for appropriate choice of $N$). 
On the other hand, this is guaranteed to be true if 
  $(b_\mu^*)\in \mathbb{Q}^{s+1}$ is itself rational, as the Sasakian volumes are (up to overall powers of $\pi$ which cancel) rational functions 
  of $b_i$ with rational coefficients \cite{Martelli:2006yb}. 
The fluxes are thus determined in this way for the flavour twist. 
We shall also expand upon this point in section~\ref{sec53}, in the case that $\X$ is toric.

We now return to the issue of signs. 
Recall that $N = m_+ \Np = m_- \Nm>0$ and hence from \eqref{pmflux} we can conclude that for 
odd $n$ we have $\Lambda_\pm\left.\Vol_S(X_{2n-1}) \right |_{b_i^{(\pm)}}>0$
while for even $n$ we have
$\left.\Vol_S(X_{2n-1}) \right |_{b_i^{(\pm)}}>0$. Next consider
$\Vol(X_\pm)$:
\begin{equation}\label{volrelationfirst}
	\Vol(X_\pm)\equiv \int_{X_\pm}\eta\wedge \frac{J^{n-1}}{(n-1)!}=\frac{(2b_1)^{n-1}}{m_\pm}\, \Lambda_\pm^{n-1} \,\Vol_S(\X)|_{b_i^{(\pm)}}  \:.
\end{equation}
Recalling our discussion in section \ref{gravblockformrchgetoo},
by considerations of the toric case (see \eqref{TtoX}), the orientation on $X_\pm$ that we need to take are such that
\begin{equation}\label{signsofvols}
	\Vol(X_+)>0 \:, \qquad \sigma\, \Vol(X_-)>0 \:.
\end{equation}
Thus, for odd $n$ we have: 
\begin{align}\label{oddsigns}
\Lambda_+>0, \quad \sigma\Lambda_->0,\qquad \left.\Vol_S(X_{2n-1}) \right |_{b_i^{(+)}}>0,\quad 
\sigma \left.\Vol_S(X_{2n-1}) \right |_{b_i^{(-)}}>0\,,
\end{align}
while for even $n$ we have:
\begin{align}\label{evensigns}
\Lambda_+>0, \quad \sigma\Lambda_->0,\qquad \left.\Vol_S(X_{2n-1}) \right |_{b_i^{(+)}}>0,\quad 
\left.\Vol_S(X_{2n-1}) \right |_{b_i^{(-)}}>0\,,
\end{align}
which fixes the sign ambiguity in solving \eqref{pmflux}. 

We now further illustrate using the two cases of physical interest when $n=3$ and $n=4$.
The case $n=3$ is associated with AdS$_3\times Y_7$ solutions. From
\eqref{oddsigns} we have
\begin{align}\label{sgnsvolsx5}
\Vol_S(X_5)|_{b_i^{(+)}} >0\,,\qquad \sigma \,\Vol_S(X_5)|_{b_i^{(-)}}>0 \, ,
\end{align}
and 
\begin{align}
\Lambda_\pm=\frac{\nu_3 N}{8 b_1^2\Vol_S(X_5)|_{b_i^{(\pm)}}}
\,.
\end{align}
Using \eqref{Snobar} we therefore find that the off-shell central charge \eqref{offshellz}
can be written
\begin{align}\label{cc_nobar}
\boxed{
\mathscr{Z}=\frac{6\pi^3 N^{2}}{b_1 b_0 } \left(\frac{1}{\left.\Vol_S(X_{5})\right |_{b_i^{(+)}}}-\frac{1}{ \left.\Vol_S(X_{5})\right |_{b_i^{(-)}}}\right)
}
\end{align}
We should set $b_1=2$ when carrying out the extremization.

We now consider the case of $n=4$, associated with AdS$_2\times Y_9$ solutions.
From \eqref{evensigns} we now have
\begin{align}
\Vol_S(X_7)|_{b_i^{(\pm )}} >0\,,
\end{align}
and
\begin{align}
\Lambda_+=\left[\frac{\nu_4 N}{24b_1^3\Vol_S(X_7)|_{b_i^{(+)}}}\right]^{1/2}\,,\qquad 
\Lambda_-=\sigma \left[\frac{\nu_4 N}{24b_1^3\Vol_S(X_7)|_{b_i^{(-)}}}\right]^{1/2}
\end{align}
The off-shell entropy \eqref{ads2entess} now reads
 \begin{equation}\label{entropy}
 	\boxed{
	\mathscr{S}=\frac{8\pi^3 N^{3/2}}{3b_0\sqrt{6b_1 } } \left(\frac{1}{\sqrt{\left.\Vol_S(X_{7})\right |_{b_i^{(+)}}}}-\frac{\sigma}{\sqrt{ \left.\Vol_S(X_{7})\right |_{b_i^{(-)}}}}\right)
	}
 \end{equation}
We should set $b_1=1$ when carrying out the extremization.

It is interesting to make a comparison with fibrations with no baryonic symmetries 
over round two spheres.
To do this we want to consider the twist case, {\it i.e.} $\sigma=1$, set $m_\pm=1$ and take the limit $b_0\to 0$.
First observe that with $\sigma=1$ we have
 \begin{equation}
 	\lim\limits_{b_0 \to 0} \frac{f(b^{(+)}_i) - f(b^{(-)}_i)}{b_0} = -\sum_{i=1}^s \left(\frac{v_{+i}}{m_+} + \frac{v_{-i}}{m_-}\right) \, \frac{\partial f}{\partial b_i} = \sum_{i=1}^s \frac{p_{i}}{m_+ m_-} \, \frac{\partial f}{\partial b_i} \:,
 \end{equation}
 where $f({b}_i)$ is a generic function of the Reeb vector. Defining the operator $\nabla = -\sum_{i=1}^s \frac{p_{i}}{m_+ m_-} \, \frac{\partial}{\partial b_i}$ we thus have
 \begin{equation}
 	\lim\limits_{b_0 \to 0} \mathscr{Z}= - \frac{6 \pi^3N^{2}}{b_1} \, \nabla \frac{1}{\Vol_S(X_{2n-1})} \:,
 \end{equation}
 and
 \begin{equation}\label{limitads2nobaryon}
 	\lim\limits_{b_0 \to 0} \mathscr{S} = - \frac{4 N^{3/2}}{\sqrt{b_1}} \, \nabla \sqrt{\frac{2\pi^6}{27 \,\Vol_S(X_{2n-1})}} \:.
 \end{equation}
In these expressions one should take derivatives with respect to $b_1$ before setting $b_1=2$ or $b_1=1$, respectively.
We also observe that after setting $m_\pm=1$ \eqref{limitads2nobaryon} is the same as equation 
 (3.10) of \cite{Hosseini:2019ddy}, who considered the case of $X_7$ with no baryonic symmetries fibred over a round two-sphere  

 \subsection{Universal anti-twist}\label{unitwist}

The universal anti-twist case is associated with Sasaki-Einstein $\X$ fibred over the spindle and,
moreover, with the twisting in the fibration 
only along the Reeb Killing vector of the Sasaki-Einstein space. We therefore impose the condition 
 \begin{equation}\label{kahlerclass2}
 	[\left.J\right|_{X_{\pm}}] = \Lambda_{\pm} [\rho] \in H^2_B(\mathcal{F}_{\xi_\pm})\:,
 \end{equation}
 where $ \Lambda_{\pm}$ are constants,
to ensure that $\X$ is Sasaki-Einstein. Notice that this is the same condition 
as the flavour twist \eqref{kahlerclass} imposed in the previous subsection. However, in addition 
 we demand that
the fluxes, $p_i$, determining the fibration satisfy
 \begin{align}\label{unitwistcond}
 p_i=\frac{p_1}{b_1^{(+)}}b_i^{(+)}=\frac{p_1}{b_1^{(-)}}b_i^{(-)}\,.
 \end{align}
 We need to check {\it a posteriori} that after extremization this is consistent with the $p_i$ being integer, much 
 as with the discussion of fluxes after equation \eqref{Naformula}. We comment on this further below. 
Note that \eqref{unitwistcond} is consistent with \eqref{diffbpbm} and, in particular, implies that the twisting parameters can be written as
 \begin{align}\label{utexplicbival}
 	p_i = \frac{p_1}{\left[b_1+b_0(-a_-+\sigma a_+)\right]} \, b_i\,,\qquad i=2,\dots,s\,.
 \end{align}
As we will see we shall need to set $\sigma=-1$, and thus just have the anti-twist case, as well as take 
the Sasaki-Einstein manifold $\X$ to be quasi-regular.

Since the universal anti-twist is a special case of the previous subsection, we can carry over all of the previous results of 
section \ref{nobarsec1} and write, for general $n$,
\begin{align}\label{entropy2}
	S_{\mathrm{SUSY}}  &=\frac{\pi (\nu_nN)^{\frac{n-1}{n-2}}}{[ 2 b_1(n-1)^{n-1}]^\frac{1}{n-2}b_0}
	\left[\frac{1}{[\left.\Vol_S(X_{2n-1})\right |_{b_{i}^{(+)}}]^\frac{1}{n-2}}-\frac{k_n}{[\left.\Vol_S(X_{2n-1})\right |_{b_{i}^{(-)}}]^\frac{1}{n-2}}\right]\,,
\end{align}
where $k_n=+1$ if $n$ is odd and $k_n=\sigma$ if is $n$ even. 
As usual, to extremize we should set $b_1=2/(n-2)$. To proceed we extremize over $b_i$ and then $b_0$. For the former, it is useful to use \eqref{unitwistcond} to introduce a new variable
\begin{equation}\label{newvariableri}
	r_i \equiv \frac{n}{b_1^{(+)}} \, b_i^{(+)} = \frac{n}{b_1^{(-)}} \, b_i^{(-)} \,,
\end{equation}
and we note that $r_1 = n$.
We next recall that $\Vol_S(\X)(r_i)$, as a function of the Reeb vector $r_i$, is homogeneous of degree $-n$ in $r_i$.
 Hence, we can write
 \begin{align}
 \left.\Vol_S(X_{2n-1})\right |_{b_{i}^{(\pm)}}=\bigg(\frac{b_1^{(\pm)}}{n}\bigg)^{-n} \left.\Vol_S(X_{2n-1})\right |_{r_i}
 \:,
 \label{genunitwist}
 \end{align}
and after substituting into \eqref{entropy2} we obtain
\begin{equation}
	S_{\mathrm{SUSY}}  =\frac{\pi (\nu_nN)^{\frac{n-1}{n-2}}}{[ 2 b_1 n^n(n-1)^{n-1}]^\frac{1}{n-2}b_0} \, \frac{1}{[\left.\Vol_S(X_{2n-1})\right |_{r_i}]^\frac{1}{n-2}}
	\left[(b_1^{(+)})^\frac{n}{n-2}-k_n\,(b_1^{(-)})^\frac{n}{n-2}\right].
\end{equation}
At this point we can set $b_1=2/(n-2)$ and proceed with the extremization. It is clear that extremizing over $b_i$, $i=2,3,\dots$, is the same as extremizing over $r_i$, $i=2,3,\dots$. Thus, we immediately deduce that $S_{\mathrm{SUSY}}$ is extremized over $b_i$ when $\left.\Vol_S(X_{2n-1})\right |_{r_i}$ is extremized over $r_i$ and, recalling that $r_1=n$, this results in nothing but the Sasaki-Einstein volume {\it i.e.} $\Vol_{SE}(X_{2n-1})$.
This leads to the off-shell action
   \begin{align}\label{osutwist1}
S_{\mathrm{SUSY}} &= \left(\frac{n-2}{4 n^n (n-1)^{n-1}}\right)^\frac{1}{n-2}\frac{\pi (\nu_nN)^{\frac{n-1}{n-2}}}{\Vol_{SE}(X_{2n-1})^{\frac{1}{n-2}}}\frac{1}{b_0} \left[(b_1^{(+)})^\frac{n}{n-2}-k_n (b_1^{(-)})^\frac{n}{n-2}\right]\,,
 \end{align}
with the extremization over $b_0$ still to be carried out. We will explicitly extremize over $b_0$ for
the two cases of physical interest below and favourably compare with known results for explicit supergravity solutions.
In particular we find that only the anti-twist case with $\sigma=-1$ gives rise to a positive on-shell value for $S_{\mathrm{SUSY}}$.
 
For the geometric R-charges, from \eqref{Rpmnobar} and \eqref{newvariableri} we have
  \begin{equation}\label{Rpmnobarut}
	R_a^\pm = \frac{2 \pi }{(n-1)b_1} \left. \frac{\Vol_S(S_a) }{\Vol_S(X_{2n-1}) }\right|_{\frac{b_{1}^{(\pm)}r_i}{n} }\,,
	\end{equation}
	where $\Vol_S(S_a)$ is the volume of $S_a$ with respect to the Sasakian metric.
Now $\Vol_S(X_{2n-1})(r_i)$ and $\Vol_S(S_a)(r_i)$ are homogeneous of degree $-n$ and $-(n-1)$ in $r_i$, respectively. Thus,
by a similar scaling argument we deduce that after extremizing over $b_i$ the off-shell R-charges, as a function of $b_0$, are given by
   \begin{equation}\label{Rpmnobarut2}
	R_a^\pm = \frac{2 \pi b_1^{(\pm)}  }{n(n-1)b_1} \frac{\Vol_{SE}(S_a) }{\Vol_{SE}(X_{2n-1})} = \frac{b_1^{(\pm)}}{b_1}\, {R}_a\,,
	\end{equation}
where we introduced 
\begin{equation}
	{R}_a \equiv  \frac {2\pi }{n(n-1)} \frac{\Vol_{SE}(S_a) }{\Vol_{SE}(X_{2n-1}) }\,.
\end{equation}
For $n=3$ and $n=4$ we have ${R}_a\equiv {R}_a^{\mathrm{4d}}$ and ${R}_a\equiv {R}_a^{\mathrm{3d}}$, respectively, where
${R}_a^{\mathrm{4d}}$ and ${R}_a^{\mathrm{3d}}$ are the R-charges associated with the AdS$_5\times SE_5$ and 
AdS$_4\times SE_7$ solutions, respectively.
Note that combining \eqref{fluxes} with \eqref{diffbpbm}, the equations above for $R_a^\pm$ allows us to find the fluxes (with $\sigma=-1$)
\begin{equation}\label{Nauniversaltwist}
	\Nas_a = \frac{b_1}{2b_0} (R_a^+ - R_a^-)N = \frac{1}{2b_0}\,(b_1^{(+)}-b_1^{(-)}) {R}_a  N= \frac{m_+ - m_-}{2m_+ m_-}{R}_a N \:,
\end{equation}
and we also have (with $\sigma=-1$)
\begin{align}\label{Nauniversaltwist2}
R^+_a+R^-_a=\left(2-\frac{b_0}{b_1}\chi\right)R_a \:,
\end{align}
where $\chi=(m_-+m_+)/(m_-m_+)$ is the orbifold Euler character of the spindle. In both 
\eqref{Nauniversaltwist}, \eqref{Nauniversaltwist2} we have set $\sigma=-1$, as it is needed for the universal anti-twist case.
Note that in the special case that the $SE$ space is toric, we have $\sum_a R_a=2$ and hence we then also have (with $\sigma=-1$)
\begin{align}\label{Nauniversaltwist3}
\frac{1}{2}\sum_a(R^+_a+R^-_a)=2-\frac{b_0}{b_1}\chi\,.
\end{align}

Finally, we should return to the issue of everything being properly quantized. Combining \eqref{utexplicbival} with 
\eqref{newvariableri} and evaluating on-shell, we have
\begin{align}\label{piri}
p_i = \frac{p_1}{n} r_i^*\, ,
\end{align}
where $r_i^*$ is the critical Reeb vector for the Sasaki-Einstein metric on $\X$. 
The latter is thus necessarily quasi-regular, meaning that $(r_i^*)\in\mathbb{Q}^s$ is rational. 
One may then ensure that $p_i\in\Z$ in \eqref{piri} by appropriately choosing 
$p_1=m_+-m_-\in\Z$, which is a constraint on the choice of spindle $\Sigma$.
Note that a similar conclusion was reached for the universal twist solutions in \cite{Gauntlett:2018dpc}, 
where there $b_0=0$ and $\Sigma=\Sigma_g$ is a smooth Riemann surface. In that case the 
universal twist requires an appropriate divisibility property for the genus $g$ of the Riemann surface. 
Such issues were also effectively discussed in 
\cite{Ferrero:2020laf}
for the explicit AdS$_3\times Y_7$  universal anti-twist  solutions described in 
the next subsection, where $\Z_k$ quotients along the 
Reeb $U(1)$ isometry of the Sasaki-Einstein metric on $\X$  were also considered\footnote{For such quotients the effective
$U(1)^s$ action is not the same as on the unquotiented space.}. 

Finally, since the R-charges $R_a$ are rational for rational Reeb vector $r_i^*$, 
one may ensure the fluxes $\Nas_a$ in \eqref{Nauniversaltwist} are integer by appropriately choosing 
$N$. Again, {\it cf.} the discussion of the universal twist in  \cite{Gauntlett:2018dpc}. 

 \subsubsection{AdS$_3\times Y_7$ case}
 For the $n=3$ case, from \eqref{offshellz} and \eqref{osutwist1}, we therefore have 
  \begin{align}\label{offshellz2}
\mathscr{Z}&= \frac{\pi^3 N^2}{9\Vol(SE_{5})}\frac{1}{b_0} \left[(b_1^{(+)})^3- (b_1^{(-)})^3\right]\, .
  \end{align}
Notice that this expression is quadratic in $b_0$. If we extremize over $b_0$ we find that it is only in the anti-twist case, 
$\sigma=-1$, that we get a positive on-shell central charge. So, setting $\sigma=-1$ we find the extremal value
\begin{align}\label{b0extd3ucase}
b_{0*}=\frac{3 m_- m_+(m_-+m_+)}{m_-^2+m_- m_+ +m_+^2}\,,
\end{align} 
and one can check that \eqref{sgnsvolsx5} is satisfied, provided that we take
$m_+>m_-$ (which is associated with taking $N>0$ and \eqref{signsofvols}).
We can express the on-shell central charge
in terms of, $a_{\mathrm{4d}}$, the central charge of the $d=4$ SCFT, which is given by
\begin{align}\label{4dcc}
a_{\mathrm{4d}}=\frac{\pi^3 N^2}{4\Vol(SE_{5})}\,,
\end{align}
and we find
  \begin{align}\label{offshellz3}
\mathscr{Z}_{\text{os}}&=\frac{4(m_+-m_-)^3}{3 m_- m_+(m_-^2+m_- m_+ +m_+^2)}a_{\mathrm{4d}}\,,
  \end{align}
which is positive for $m_+>m_-$.
The results \eqref{offshellz3} and \eqref{b0extd3ucase} are in precise alignment with the results (21), (22) for the explicit 
supergravity solutions constructed in minimal gauged supergravity in \cite{Ferrero:2020laf} (after identifying $m_\pm$ with $n_\mp$).
 
From \eqref{Rpmnobarut2}, the (off-shell) geometric R-charges can be written in the form    \begin{equation}\label{Rpmnobarut3}
	R_a^\pm 
	=\frac{1}{2} \,b_1^{(\pm)} R_a^{\mathrm{4d}}\,,
	\end{equation}
	and hence  
	 \begin{align}\label{rpmads3caseos}
	R^+_a=\frac{(m_+-m_-)(m_-+2 m_+)}{2(m_-^2+m_- m_+ +m_+^2)}R_a^{\mathrm{4d}}\,,\quad
	R^-_a=-\frac{(m_+-m_-)(2m_-+ m_+)}{2(m_-^2+m_- m_+ +m_+^2)}R_a^{\mathrm{4d}}\,,
\end{align}
where $R_a^{\mathrm{4d}}$ are the four-dimensional R-charges associated with the AdS$_5\times SE_5$ solution.
Notice that with $R_a^{\mathrm{4d}}>0$ and $m_+>m_-$ we have $R^+_a>0$ and $\sigma R^-_a>0$ (with $\sigma=-1$ for this case as we have seen).
We will argue more generally in the toric setting that we always have $R^+_a>0$ and $\sigma R^-_a>0$.

  \subsubsection{AdS$_2\times Y_9$ case}\label{utwsitads2y9}
 For $n=4$ using \eqref{ads2entess} and \eqref{osutwist1} we get the off-shell entropy
  \begin{align}\label{osutwist0}
 \mathscr{S} &= \frac{N^{3/2}\pi^3}{6^{3/2}\Vol(SE_7)^{1/2}}\frac{1}{b_0} \left[(b_1^{(+)})^2-\sigma (b_1^{(-)})^2\right]\,,
 \end{align}
 which is to be extremized over $b_0$.
 If $\sigma=+1$ there are no extrema, so we again conclude that we again must have the anti-twist case with $\sigma=-1$.
   
 We can express the off-shell result in terms of the four-dimensional Newton constant $G_{(4)}$ (in the conventions of \cite{Ferrero:2020twa}), defined by
 \begin{align}
 \frac{1}{G_{(4)}}=\frac{2^{3/2}\pi^2}{3^{3/2}\Vol(SE_7)^{1/2}}N^{3/2}\,.
 \end{align}
 Recall that the free energy on the three-sphere, $ \mathcal{F}_{S^3}$, 
 of the $d=3$ SCFT dual to the AdS$_4\times SE_7$ solution is given by
 \begin{align}
 \mathcal{F}_{S^3}=\frac{\pi}{2G_{(4)}}\,.
 \end{align}
 We then can express the off-shell entropy (with $\sigma=-1$) as
    \begin{align}\label{osutwist}
 \mathscr{S} 
  &= \frac{\pi}{8G_{(4)}}\frac{1}{b_0} \left[(b_1^{(+)})^2+(b_1^{(-)})^2\right]\,.
 \end{align}
Extremizing over $b_0$ (and setting $b_1=1$) we find two solutions but only one gives a positive entropy. Focusing on this extremum we find
 \begin{align}
b_{0*}= \frac{\sqrt{2}m_- m_+}{\sqrt{m_-^2+m_+^2}}\,,
 \end{align}
 with the associated on-shell entropy given by
  \begin{align}\label{exuatads2}
 \mathscr{S}_{\text{os}} = \frac{\pi} {4G_{(4)} } \frac{ \sqrt{2}\sqrt{m_-^2+m_+^2}- m_+-m_-}{m_-m_+}\,.
 \end{align}
This agrees with the entropy for the supersymmetric, accelerating and magnetically charged black holes that were constructed in minimal gauged supergravity in (1.1) of \cite{Ferrero:2020twa}.

From \eqref{Rpmnobarut2}, the (off-shell) geometric R-charges can be written in the form 
    \begin{equation}\label{Rpmnobarut311}
	R_a^\pm 
	=b_1^{(\pm)} R_a^{\mathrm{3d}}\,,
	\end{equation}
and hence 
	 \begin{align}
	R^+_a=\frac{\sqrt{m_-^2+m_+^2}-\sqrt{2}m_-}{\sqrt{m_-^2+m_+^2}}R_a^{\mathrm{3d}}\,,\quad
	R^-_a=\frac{\sqrt{m_-^2+m_+^2}-\sqrt{2}m_+}{\sqrt{m_-^2+m_+^2}}R_a^{\mathrm{3d}}\,,
\end{align}
where $R_a^{\mathrm{3d}}$ are the four-dimensional R-charges associated with the AdS$_4\times SE_7$ solution.
Notice that with $m_+>m_-$ and $R_a^{\mathrm{3d}}>0$ we again have
$R^+_a>0$ and $\sigma R^-_a>0$ (with $\sigma=-1$).

Finally,  as in \cite{Boido:2022iye} 
we can also make a comparison of the off-shell entropy \eqref{osutwist} with an off-shell entropy function that
 was constructed in \cite{Ferrero:2020twa}. More specifically, \cite{Ferrero:2020twa} considered a complex locus of 
 supersymmetric, accelerating and magnetically\footnote{In fact the calculation in \cite{Ferrero:2020twa} was carried out for a larger family 
 of black hole solutions with additional electric charge and rotation.}  charged black holes that were non-extremal and then holographically calculated 
 the on-shell action $I$ at the AdS$_4$ boundary. After suitable extremization $I$ gives rise to the on-shell entropy.
 To make the connection, note that we can also write the off-shell action \eqref{osutwist} with $b_1=1$ as
  \begin{align}
 \mathscr{S} =\mp  \frac{1}{2 \ii G_{(4)}} \left( \frac{\varphi^2}{\omega}+(G_{(4)}Q_m)^2\omega   \right)\, ,
 \end{align}
 with 
 \begin{align}
(G_{(4)} Q_m)=\frac{m_--m_+}{4m_- m_+}\, ,\qquad
 \varphi=\frac{\chi}{4}\omega\pm \pi \ii\,,\qquad \omega=\mp 2\pi \ii \, b_0\, ,
 \end{align}
 and we see this is precisely the same as (4.41) of \cite{Cassani:2021dwa} with $ \mathscr{S} =-I$.
We also highlight that in the special case that the $SE_7$ is toric,
 using \eqref{Nauniversaltwist3} we can also write
 \begin{align}
 \varphi=\pm\frac{\pi \ii}{4} \sum_a(R_a^++R_a^-)\,.
 \end{align}
The off-shell action \eqref{osutwist} also agrees with (5.20) of \cite{Cassani:2021dwa}, with $ \mathscr{S} =-I$,
after setting $a_1=1$ and $a_2=-b_0$.

\section{Some examples}\label{someexamplessec}

In this section we use the general results of the previous section to calculate some physical quantities for particular examples
 of AdS$_3\times Y_7$ and AdS$_2\times Y_9$ solutions. For the AdS$_3\times Y_7$ solutions we consider
$X_5=S^5$ and compare with known results associated with explicit supergravity solutions. For the AdS$_2\times Y_9$ solutions we consider 
$X_7=S^7$, for which we can also compare with known supergravity results, as well as $X_7=V^{5,2}$ for which explicit supergravity solutions are not known.
Note that the cases $X_5=S^5$ and $X_7=S^7$ can also be treated with the toric methods that we develop in sections 
\ref{sec:fibretoric} and \ref{sec:toricoverspindle}.

\subsection{AdS$_3\times Y_7$ solutions with $X_5=S^5$}\label{sec3s5ex}

Since $H^2(S^5,\mathbb{R})\cong 0$ the case of $X_5=S^5$ is an example with no baryonic symmetries, as in section 
\ref{nobar}. In a suitable basis, compatible with the toric data we will use for this example later, the Sasakian volume of $S^5$ 
as a function of the Reeb vector can be written as
\begin{equation}\label{s5volsass5}
	\Vol_S(S^5) = \frac{\pi^3}{b_2\, b_3\, (b_1 - b_2 - b_3)}\:.
\end{equation}
As is well known there are three supersymmetric three-submanifolds $S_a$ which are $U(1)^3$ invariant and associated with baryonic operators 
in $\mathcal{N}=4$ SYM theory dual to AdS$_5\times S^5$. The Sasakian volume of these submanifolds are given by
\begin{equation}
	\Vol_S(S_1) = \frac{2\pi^2}{b_2 \, b_3}\:, \qquad\qquad  \Vol_S(S_a) = \frac{2\pi^2}{b_a\, (b_1 - b_2 - b_3)}\:, \quad a=2,3 \:.
\end{equation}
It will be convenient below to briefly recall that we obtain the Sasaki-Einstein volume by setting $b_1=3$ in \eqref{s5volsass5} 
and then extremizing over $b_2,b_3$. The extremal values are $b_2=b_3=1$ with $\Vol_{SE}(S^5)=\pi^3$ and 
$\Vol_{SE}(S_a)=2\pi^2$.

We now consider the GK geometry associated with the AdS$_3\times Y_7$ solutions.
The off-shell central charge can be obtained from \eqref{cc_nobar} and reads
\begin{equation}\label{ZS5}
	\mathscr{Z}=\frac{6 N^{2}}{b_1 b_0 } \left[b_2^{(+)}b_3^{(+)}  \left(b_1^{(+)} - b_2^{(+)} - b_3^{(+)}\right) -b_2^{(-)}b_3^{(-)}  \left(b_1^{(-)} - b_2^{(-)} - b_3^{(-)}\right)\right]\,.
\end{equation}
Similarly, the off-shell geometric R-charges are obtained from \eqref{Rpmnobar}  and are given by
\begin{equation}\label{rchgess5examp1}
	R_1^\pm = \frac{2}{b_1} \left(b_1^{(\pm)} - b_2^{(\pm)} - b_3^{(\pm)}\right) , \qquad\qquad R_a^\pm = \frac{2 b_a^{(\pm)}}{b_1} \:,\quad a=2,3 \:.
\end{equation}
Observe that for this example, which is also toric, we have
\begin{align}\label{s5rchgefacts}
\sum_{a=1}^d R_a^\pm&=\frac{2b_1^{(\pm)}}{b_1}\,,\nn\\
\frac{1}{2}\sum_{a=1}^d \left(R_a^++ R_a^-\right) &= 2 - \frac{m_--\sigma m_+}{m_+m_-}\frac{b_0}{b_1}\, ,
\end{align}
in agreement with the general toric results 
\eqref{Rpmidentities12}, \eqref{Rpmidentities123}.

Recall that there  are two preferred fluxes, $\Npm$, associated with the fibres $X_\pm \cong S^5 /\Z_{m_\pm}$
at the north and south poles, respectively, and from \eqref{defN} we have 
$N \equiv  m_+ \Np = m_- \Nm$. There are no further cohomologically non-trivial five-form fluxes. However, we 
also have three five-form fluxes $\Nas_a$ defined as the integral of the five-form flux through the three submanifolds $\Sigma_a$, obtained
as the three supersymmetric submanifolds $S_a$ of $S^5$ fibred over the spindle, as in \eqref{fluxes0}.
Using \eqref{fluxes} we obtain the following expression for the fluxes $\Nas_a$:
\begin{equation}
	\kf_a\equiv \frac{\Nas_a}{N} = \frac{1}{  m_+ m_-}\big\{p_1 - ( p_2 + p_3),  p_2,p_3\big\}\,,
\end{equation}
where as usual, $p_1 =-\sigma m_+ -  m_-$. 
For this example, 
we have $\sum_{a=1}^d \Nas_a=\frac{p_1}{ m_+  m_-}N$, which one can check is in agreement with the general result \eqref{homfluxes}
for toric $X_7$.
We also note that the variables $\kf_a$ we have introduced can be identified 
with the background magnetic fluxes of the dual $d=4$ SCFT, in this case $\mathcal{N}=4$ SYM theory, as we discuss in more detail in the toric setting in section \ref{secftmatch}.

To obtain the on-shell values for the central charge and R-charges we should set $b_1=2$ and then extremize $\mathscr{Z}$
over $b_0,b_2$ and $b_3$. 
The expression for $\mathscr{Z}$ is quadratic in $b_0,b_2,b_3$ and we find that 
the extremal values for $b_0,b_2$ and $b_3$ depend on the spindle data $m_\pm$, $\sigma$ as well as $a_\pm$ (subject to \eqref{alphasa}). However, the on-shell values of $\mathscr{Z}$ and $R_a^\pm$ are independent of $a_\pm$.

There is a $\mathbb{Z}_3$ symmetry permuting the $\kf_a$ but not $(p_1,p_2,p_3)$: the latter is a consequence of the fact that we singled out the $p_1$ direction in constructing the fibration. It is therefore illuminating to express the on-shell values 
of $\mathscr{Z}$ and $R_a^\pm$  in terms of $\kf_a$. For the on-shell central charge we find 
  \begin{align}\label{cC3first}
c_{\mathrm{sugra}}\equiv \mathscr{Z}_\text{os} =
\frac{6m_-^2m_+^2 \kf_1 \kf_2 \kf_3}{m_-^2+m_+^2-m_-^2m_+^2(\kf_1^2+\kf_2^2+\kf_3^2) }N^2\,.
   \end{align}
In the case of the twist $\sigma=+1$ we have exact agreement with the supergravity solutions of \cite{Ferrero:2021etw}.\footnote{We should identify $\kf_a$ with $p_a/(m_-m_+)$ there, as well as $m_-,m_+$ with $n_1,n_2$, respectively, and notice that the constraint 
\eqref{constonpss5} on our $\kf_a$ is twist case A of \cite{Ferrero:2021etw} when $\sigma=+1$ and
anti-twist case B when $\sigma=-1$.} For the anti-twist case $\sigma=-1$ we likewise have agreement with the supergravity solutions found in \cite{Boido:2021szx, Ferrero:2021etw}, 
and for the  special anti-twist case when 
$\kf_1=\kf_2=\kf_3=(\frac{m_+-m_-}{3m_-m_+})$ we recover the result that we derived in 
\eqref{offshellz3} for the universal anti-twist after using $a_{\mathrm{4d}}=N^2/4$; this is also the result found for the supergravity solutions constructed in \cite{Ferrero:2020laf}. 

Starting with \eqref{Rpmnobar}, the on-shell expressions for the R-charges can be written 
 \begin{align}
   R^+_a=&\ 
-Cm_+\left(\kf_1[\sigma +m_- \kf_1],\kf_2[\sigma +m_- \kf_2],\kf_3[\sigma +m_- \kf_3]\right)\,,\nn\\
 R^-_a=&\ -Cm_- \left(\kf_1[1+m_+\kf_1],\kf_2[1+ m_+\kf_2],\kf_3[1+m_+\kf_3]\right)\,,
  \end{align}
 where
 \begin{align}
 C=\frac{2m_-m_+}{m_-^2+m_+^2-m_-^2m_+^2(\kf_1^2+\kf_2^2+\kf_3^2)}\,.
 \end{align}
 This is a new result which could be checked using the explicit supergravity solutions.
It is also interesting to observe that if we demand, on-shell, $c_{\mathrm{sugra}}>0$ as well as 
 $R^+_a>0$ and $\sigma R^-_a>0$ then, for the anti-twist case ($\sigma=-1$) we find we must have $\kf_1>0$, $\kf_2>0$ and $\kf_3=-\kf_1-\kf_2+(m_+-m_-)/(m_+ m_-)>0$ (which, in particular, implies $m_+>m_-$); this is in alignment with the result found for the explicit supergravity solutions as in (3.30) of \cite{Ferrero:2021etw}. Similarly for the twist case ($\sigma=+1$) we find that we must have any two of
 $(\kf_1,\kf_2,\kf_3)$ to be positive, now with $\kf_3=-\kf_1-\kf_2-(m_++m_-)/(m_+ m_-)>0$ which is also consistent\footnote{Here we do not find a condition $m_+>m_-$ as stated in (3.31) of \cite{Ferrero:2021etw}; the resolution is that twist solutions can also be found using the analysis of \cite{Ferrero:2021etw} with $m_+<m_-$ after relabelling.} with the result found for the explicit supergravity solutions, see (3.31) of \cite{Ferrero:2021etw}. 
  
  For the universal anti-twist case with $\kf_1=\kf_2=\kf_3=(\frac{m_+-m_-}{3m_-m_+})$ and $\sigma=-1$, the geometric R-charges simplify to give
     \begin{align}
   R^+_a=&\frac{(m_+-m_-)(m_-+2 m_+)}{3(m_-^2+m_- m_+ +m_+^2)}\,,\nn\\
      R^-_a=&-\frac{(m_+-m_-)(m_++2 m_-)}{3(m_-^2+m_- m_+ +m_+^2)}\,.
  \end{align}
This is in exact agreement with the general result of \eqref{Rpmnobarut3} after using the fact that $R_a^{\mathrm{4d}}=2/3$.

Finally, we can prove a conjecture for the central charge stated in \cite{Hosseini:2021fge,Faedo:2021nub}.
To see this, observe that we can use \eqref{ZS5}, \eqref{rchgess5examp1} to write
the off-shell central charge as
\begin{equation}\label{zexpphi52}
 		\mathscr{Z}=  \frac{3 N^{2}}{b_0 }\frac{2}{b_1}
		 \Bigg[\prod_{a=1}^3\bigg(\varphi_a+\mathfrak{p}_a\frac{b_0}{2}\bigg)-\prod_{a=1}^3\bigg(\varphi_a-\mathfrak{p}_a\frac{b_0}{2}\bigg) \Bigg]\, ,
		  \end{equation}
where 
\begin{align}
\varphi_a\equiv \frac{b_1}{4} \left(R_a^+ + R_a^-\right)\,,
\end{align}
and $\sum_a\varphi_a=\frac{1}{2}(b_1^{(+)}+b_1^{(-)})=b_1-\frac{b_0}{2}\frac{m_--\sigma m_+}{m_-m_+}$. We also recall 
$\sum\mathfrak{p}_a=-\frac{\sigma m_++m_-}{m_+ m_-}$ and that we should set $b_1=2$ before extremizing. 
This is then in agreement with e.g. \cite{Faedo:2021nub} provided
that we identify our $(m_\pm,\varphi_a,b_0,\mathfrak{p}_a)$ with their $(n_\mp,\varphi_a,-2\epsilon,-\mathfrak{n}_a)$.

 \subsection{Examples of AdS$_2 \times Y_9$ solutions }\label{ads2solsexamples}
 
We now consider some examples of AdS$_2 \times Y_9$ solutions of $D=11$ supergravity with
 $Y_9$ a fibration of $X_7$ over a spindle, specifically, $X_7=S^7$ and also $X_7=V^{5,2}$.
 Both of these cases are examples with no baryonic symmetries with $H^2(X_7,\mathbb{R})\cong 0$ and  
can be analysed using the results of section \ref{nobar}. The $S^7$ case is toric and can also be treated using the toric formalism, similar to the analysis for $S^5$ later in section~\ref{torics5example}.
 
 \subsubsection{$X_7=S^7$ example}
 
In a suitable basis, the Sasakian volume of $S^7$, as a function of the Reeb vector, is given by
 \begin{equation}\label{volsass7}
 	\Vol_S(S^7) \, = \,  \frac{\pi^4}{3\, b_2 b_3 b_4 (b_1 - b_2 - b_3 - b_4)} \:.
 \end{equation}
 There are four supersymmetric five-submanifolds $S_a$, which are $U(1)^4$ invariant and associated with 
 baryonic operators in the SCFT dual to AdS$_4\times S^7$. The Sasakian volumes of these
supersymmetric submanifolds are given by
  \begin{equation}
 	\Vol_S(S_1) \, = \,  \frac{\pi^3}{ b_2 b_3 b_4} \,,\qquad
	 	\Vol_S(S_a) \, = \,  \frac{\pi^3b_a}{b_2 b_3 b_4(b_1 - b_2 - b_3 - b_4)}\:, \quad a=2,3,4  \,.
 \end{equation}
 The Sasaki-Einstein volume can be obtained by setting $b_1=4$ and then extremizing \eqref{volsass7} over $b_2,b_3,b_4$.
 The extremal point has $b_2=b_3=b_4=1$ with $\Vol_{SE}(S^7) =\pi^4/3$ and $\Vol_{SE}(S_a)=\pi^3$.
 
We now consider the GK geometry associated with AdS$_2\times Y_9$ solutions.
The off-shell entropy function can be obtained from \eqref{entropy} and takes the form
\begin{align}\label{entropyS7}
 		\mathscr{S} &\,= \, \frac{8\pi N^{3/2}}{3b_0\sqrt{2b_1} } \Bigg[\sqrt{b_2^{(+)} b_3^{(+)} b_4^{(+)} \left(b_1^{(+)} - b_2^{(+)} - b_3^{(+)} - b_4^{(+)}\right)} \nn\\
 		&\qquad\qquad\qquad\qquad - \sigma\,\sqrt{b_2^{(-)} b_3^{(-)} b_4^{(-)} \left(b_1^{(-)} - b_2^{(-)} - b_3^{(-)} - b_4^{(-)}\right)}\Bigg] \: .
 \end{align}
The off-shell geometric R-charges are obtained from \eqref{Rpmnobar} and are given by
\begin{equation}\label{RchargesS7}
	R_1^\pm = \frac{2}{b_1}\left(b_1^{(\pm)} - b_2^{(\pm)} - b_3^{(\pm)} - b_4^{(\pm)}\right) \:, \qquad\qquad R_a^\pm = \frac{2b_a^{(\pm)}}{b_1} \:, \quad a=2,3,4 \:.
\end{equation}
Observe that for this example, which is also toric, we have
\begin{align}\label{s7rchgefacts}
\sum_{a=1}^d R_a^\pm&=\frac{2b_1^{(\pm)}}{b_1}\,,\nn\\
\frac{1}{2}\sum_{a=1}^d \left(R_a^++ R_a^-\right) &= 2 - \frac{m_--\sigma m_+}{m_+m_-}\frac{b_0}{b_1}\,,
\end{align}
in agreement with the general toric results 
\eqref{Rpmidentities12}, \eqref{Rpmidentities123}.
Also, using \eqref{fluxes} we can find the following expression for the preferred five-form fluxes $\Nas_a$ defined in \eqref{fluxes0}.
\begin{equation}\label{expfluxess7case}
	\kf_a\equiv \frac{\Nas_a}{N} = \frac{1}{  m_+ m_-}\big\{p_1 - ( p_2 + p_3 + p_4), \, p_2,\,p_3,\,p_4\big\}\,,
\end{equation}
where, as usual, $p_1 =-\sigma m_+ -  m_-$. For this example,
we have $\sum_{a=1}^d \Nas_a=\frac{p_1}{ m_+  m_-}N$, which one can check is in agreement with the general result \eqref{homfluxes}
for toric $X_7$.

 To obtain the on-shell values for $\mathscr{S}$ and $R_a^\pm$, we should set  $b_1 = 1$, 
 as appropriate for an AdS$_2\times Y_9$ solution, and then extremize \eqref{entropyS7} as a function of $b_2$, $b_3$, $b_4$, and $b_0$.
 Performing such a computation analytically with general fluxes is difficult due to the presence of the square roots. We will look at a simpler case when the fluxes are pairwise equal, {\it i.e.} $\Nas_1 = \Nas_2$ and $\Nas_3=\Nas_4$, which amounts to setting
 \begin{equation}\label{pairwise}
 	p_2 = \frac{1}{2}\,( p_1 - 2p_3 ) \:, \qquad p_4 = p_3 \:.
 \end{equation}
This choice corresponds to considering fibrations of the $S^7$ that maintain $U(2)\times U(2)\subset SO(8)$ symmetry
and will allow us to compare with some explicit supergravity solutions.
Notice in passing that this implies that $p_1 =-\sigma m_+ -  m_-$ must be even. Consistent with this $U(2)\times U(2)$ symmetry
we now make the further assumption that we can carry out the extremization over the locus
\begin{align}
b_2=\frac{1}{2}(b_1-2b_3)+\frac{b_0}{2}(\sigma a_+-a_-)\,,\qquad
b_4=b_3\,,
\end{align}
which implies 
 \begin{align}
		b_2^{(\pm)} &= \frac{1}{2}\, (b_1^{(\pm)}-2b_3^{(\pm)})\,,\qquad
			b_4^{(\pm)} = b_3^{(\pm)}\,.
 \end{align}
 
 With these restrictions, the arguments of the square roots in \eqref{entropyS7} become perfect squares, so that
\begin{align}\label{entropyPairwise}
 		\mathscr{S} &\,= \, \frac{4\pi N^{3/2}}{3b_0\sqrt{2} } \bigg[ {\delta_+}b_3^{(+)}(b_1^{(+)}-2b_3^{(+)})  - {\delta_-}b_3^{(-)}(b_1^{(-)}-2b_3^{(-)})\bigg] \: ,
 \end{align}
 where $\delta_{\pm}$ are signs defined by
 \begin{equation}\label{bpmcondss7}
\delta_+ \equiv \text{sign}\left[b_3^{(+)}(b_1^{(+)}-2b_3^{(+)})\right], \qquad \delta_- \equiv \sigma\,\text{sign}\left[b_3^{(-)}(b_1^{(-)}-2b_3^{(-)})\right] .
 \end{equation}
 We now need to extremize over $b_0,b_3$ with $b_1=2$.
By an explicit calculation we find that when $\delta_+ =  \delta_-$, \eqref{entropyPairwise} is linear in $b_3$ and $b_0$ and so it does not have any extremum. Hence we set 
\begin{align}\label{deltasS7}
\delta_+ = -\delta_- \equiv \delta\,.
\end{align}
We then find two different set of values for $(b_0,\,b_3)$ that extremize the off-shell entropy \eqref{entropyPairwise}
 \begin{equation}
 {b}_0^* = \mp \frac{2\sigma m_+ m_-}{\mathcal{D}} \:, \qquad 
{b}_3^* = \frac{1}{4}\pm\frac{\sigma m_--m_+ +  
	4p_3\sigma (1-2a_- m_+)}{4\mathcal{D}} \:,  
 \end{equation} 
 where
 \begin{equation}
 	\mathcal{D} = \sqrt{-16 p_3^2 -8p_3(\sigma m_++ m_-)+(\sigma m_--m_+)^2} \:,
 \end{equation}
with the corresponding entropy given by
\begin{align}
\mathscr{S}_{\text{os}} &= -\delta \frac{\pi N^{3/2}}{3\sqrt{2} } \left(\frac{m_--\sigma m_+\pm \sigma \mathcal{D}}{m_- m_+}\right)\,.
\end{align}

 Clearly there are restrictions on $m_\pm > 0$, $p_3$ and $\sigma$ to ensure that $\mathscr{S}_{\text{os}}$ is real and positive.
The conditions \eqref{bpmcondss7}, \eqref{deltasS7}, and the positivity of
 the entropy imply
\begin{align}
	\delta = 1 \: \quad \Rightarrow  \quad 
	\begin{cases}
		b_3^{(+)}(b_1^{(+)}-2b_3^{(+)})|_{{b}_0^*,{b}_3^*} > 0 \\
		\sigma\, b_3^{(-)}(b_1^{(-)}-2b_3^{(-)})|_{{b}_0^*,{b}_3^*} < 0 \\
		m_-- \sigma m_+\pm \sigma\mathcal{D} < 0\,,
	\end{cases} \label{ineq1} \\[6pt]
	\delta = -1 \: \quad \Rightarrow \quad 
	\begin{cases}
		b_3^{(+)}(b_1^{(+)}-2b_3^{(+)})|_{{b}_0^*,{b}_3^*} < 0 \\
		\sigma\, b_3^{(-)}(b_1^{(-)}-2b_3^{(-)})|_{{b}_0^*,{b}_3^*} > 0 \\
		m_-- \sigma m_+\pm \sigma\mathcal{D} > 0\,.
	\end{cases} \label{ineq2}
\end{align}
For the twist case, $\sigma=+1$, we find that it is not possible to satisfy these conditions.
Hence, there are no twist solutions with $X_7 = S^7$ and pairwise equal fluxes. This conclusion was also reached within the context
of a specific class of local supergravity solutions in \cite{Ferrero:2021etw,Couzens:2021cpk}.

We now consider the anti-twist case,  $\sigma=-1$. In this case there are no possibilities for the lower sign extremal solution 
in \eqref{ineq1}, \eqref{ineq2}, but there are possibilities for the upper sign. In particular we find
\begin{equation}\label{b0b3exts7}
		{b}_0^* =  \frac{2 m_+ m_-}{\mathcal{D}} \:, \qquad 
		{b}_3^* = \frac{1}{4} - \frac{ m_-+m_+ +  
			4p_3 (1-2a_- m_+)}{4\mathcal{D}} \:,  
	\end{equation} 
	and
	\begin{equation}\label{entrS7anti}
	\mathscr{S}_{\text{os}} = -\delta \frac{\pi N^{3/2}}{3\sqrt{2} } \left(\frac{m_- +  m_+ - \mathcal{D}}{m_- m_+}\right) ,
	\end{equation}
	with
	\begin{align}
			\delta &= 1 \qquad \text{for:} \qquad \frac{1}{2}\,(m_+ - m_-) < p_3 < 0 \qquad\text{or}\qquad
			0 < p_3 < \frac{1}{2}\,(m_+ - m_-)\,,\nn\\
			\delta &= -1 \quad\, \text{for:} \qquad\qquad -\frac{m_-}{2} < p_3 < \min \left[0,\,\frac{1}{2}\,(m_+ - m_-)\right]\nn\\ &\quad\qquad \qquad\qquad  \text{or} 
			 \qquad\max \left[0,\,\frac{1}{2}\,(m_+ - m_-)\right] < p_3 < \frac{m_+}{2} \:.
	\end{align}
	Notice that in both cases $m_+ - m_-$ is even.
	For these classes of anti-twist solutions we can also compute the on-shell R-charges and find
	\begin{align}\label{rpmexs7ads2}
			R_1^+ = R_2^+ &= \frac{1}{2} + \frac{1}{2\mathcal{D}} \left(m_+ - 3 m_- - 4 p_3\right) , \nn\\
			R_3^+ = R_4^+ &= \frac{1}{2} - \frac{1}{2\mathcal{D}} \left(m_++ m_-  - 4 p_3\right) , \nn\\
			R_1^- = R_2^- &= \frac{1}{2} - \frac{1}{2\mathcal{D}} \left(3 m_+ - m_- - 4 p_3\right) , \nn\\
			R_3^- = R_4^- &= \frac{1}{2} - \frac{1}{2\mathcal{D}} \left( m_+ + m_- + 4 p_3\right) . 
	\end{align}
	Note that \eqref{entrS7anti} is consistent with the analysis of explicit supergravity solutions
	studied in \cite{Ferrero:2021ovq}.\footnote{The fluxes there are 2 times the fluxes here.}
	The expressions for the geometric R-charges \eqref{rpmexs7ads2} are a new result, which could, in principle, be compared with
	a calculation using the supergravity solutions.
	If we now impose the conditions $R^+_a>0$ and $\sigma R^-_a>0$ (here with $\sigma=-1$) we find that we are only left with the case
\begin{align}\label{posrchges7}
			\delta &= 1 \qquad \text{for:} \qquad 0 < p_3 < \frac{1}{2}\,(m_+ - m_-)\,.
				\end{align}

	We can also make a comparison with the analysis of explicit supergravity solution in \cite{Couzens:2021cpk}. An expression for
	the on-shell entropy for general fluxes was given in  equation (6.4) of \cite{Couzens:2021cpk} and reads, in their notation,
 \begin{equation}\label{entrchris}
 	\mathscr{S}^{\text{os}}_{\text{\cite{Couzens:2021cpk}}} = \frac{2 \pi}{3 m_{+} m_{-}} N^{3 / 2} \sqrt{ -\sigma m_{+} m_{-}+\hat{P}^{(2)}+\sigma \sqrt{\left( -\sigma m_{+} m_{-}+\hat{P}^{(2)}\right)^{2}-4 \hat{P}^{(4)}}} \:,
 \end{equation}
 where
 \begin{equation}
 	\hat{P}^{(2)} = (m_+ m_-)^2 \sum_{a < b} \mathfrak{p}_a \mathfrak{p}_b \:, \qquad \hat{P}^{(4)} = (m_+ m_-)^4 \prod_{a=1}^4 \mathfrak{p}_a \:.
 \end{equation}
 We expect that this agrees with our general result; we will explicitly 
check it agrees in the special case of pairwise equal fluxes.
Specifically, from \eqref{expfluxess7case}, \eqref{pairwise} we substitute
 \begin{equation}\label{pairwise2}
 	\kf_1=\kf_2 = \frac{m_+-m_--2p_3}{2m_+m_-}\,,\qquad
	 	\kf_3=\kf_4 = \frac{p_3}{m_+m_-}\,,
	 \end{equation}
	 to find that \eqref{entrchris} can be simplified to give
	 \begin{equation}
		\mathscr{S}^{\text{os}}_{\text{\cite{Couzens:2021cpk}}} = \frac{2 \pi}{3 m_{+} m_{-}} N^{3 / 2} \sqrt{\frac{(m_- +  m_+ - \mathcal{D})^2}{8}} \:.
	\end{equation}
Taking the square root we get agreement with the two distinct branches, $\delta=\pm1$, above. Here we ruled
out the $\delta=-1$ branch and one can do the same using the analysis of 
\cite{Couzens:2021cpk}. In addition we believe that our result
\eqref{posrchges7} arising from the positivity conditions R-charges is in accord with the existence 
of supergravity solutions of \cite{Couzens:2021cpk} (up to a relabelling of the parameters in the solutions and using (2.17) of \cite{Couzens:2021cpk}).

We can now consider the further special sub-case of the universal anti-twist  which has, with $\sigma=-1$,
\begin{align}
p_2=p_3=p_4=\frac{1}{4}p_1=\frac{1}{4}(m_+-m_-)\,.
\end{align}
In this case the fluxes in \eqref{expfluxess7case} read
 \begin{equation}
 	\kf_a\equiv \frac{\Nas_a}{N}= \frac{ p_1}{4 m_+ m_-}\big\{1,1,1,1\big\} \,.
 \end{equation}
We can also
check that the resulting extremal values for $ b_2,  b_3, b_4$ and $b_0$ 
are
consistent with \eqref{utexplicbival}. The on-shell entropy \eqref{entrS7anti} and  R-charges \eqref{rpmexs7ads2} also agree with those given in section \ref{utwsitads2y9}.

Finally, we can prove a conjecture for the entropy stated in \cite{Faedo:2021nub} (see also \cite{Hosseini:2021fge}).
To see this observe that we can use \eqref{entropyS7}, \eqref{RchargesS7} to write
the off-shell entropy function, not assuming the $U(2)^2$ symmetry condition \eqref{pairwise}, as
\begin{equation}\label{entropyS72}
 		\mathscr{S}=  \frac{2\pi N^{3/2}}{3b_0\sqrt{2 b_1} }
		 \Bigg[\sqrt{\prod_{a=1}^4(\varphi_a+\mathfrak{p}_ab_0)}-\sigma\,\sqrt{\prod_{a=1}^4(\varphi_a-\mathfrak{p}_ab_0)} \Bigg]\, ,
		  \end{equation}
where 
\begin{align}
\varphi_a\equiv \frac{b_1}{2} \left(R_a^+ + R_a^-\right)\,,
\end{align}
and $\sum_a\varphi_a=b_1^{(+)}+b_1^{(-)}=2b_1-b_0\frac{m_--\sigma m_+}{m_-m_+}$. We also recall 
$\sum\mathfrak{p}_a=-\frac{\sigma m_++m_-}{m_+ m_-}$ and that we should set $b_1=1$ before extremizing. 
This is then in agreement with \cite{Faedo:2021nub} provided
that we identify our $(m_\pm,\varphi_a,b_0,\mathfrak{p}_a)$ with their $(n_\mp, \varphi_a,-\epsilon,-\mathfrak{n}_a)$. 

 \subsubsection{$X_7=V^{5,2}$ example}%
We now consider an example of $X_7$ where there is currently no known explicit supergravity solution. 
Specifically, we consider the (non-toric) Sasaki manifold $V^{5,2} = SO(5)/SO(3)$. This manifold has 
$SO(5)\times U(1)$ isometry, with a maximal torus $U(1)^3$, and hence this is an example with $s=3$. 
The Sasakian volume can be computed using the Hilbert series method and reads \cite{Hosseini:2019ddy}
 \begin{equation}\label{v52sasvol}
 	\Vol_S(V^{5,2}) = \frac{54\pi^4}{(b_1 - b_2) (b_1 + b_2) (b_1 - b_3) (b_1 + b_3)} \:.
 \end{equation}
 There are five supersymmetric five-submanifolds $S_a$, which are $U(1)^3$ invariant, and are associated with 
 baryonic operators in the SCFT dual to AdS$_4\times V^{5,2}$. The Sasakian volumes of these
supersymmetric submanifolds are given by
  \begin{align}
 	\Vol_S(S_1) & =   \frac{b_1\pi^3}{(b_1 - b_2) (b_1 + b_2) (b_1 - b_3) (b_1 + b_3)}  \,,\nn\\
	 	\Vol_S(S_2) & =   \frac{\pi^3}{(b_1 - b_2) (b_1 + b_2) (b_1 + b_3)}  \,,\quad
		 	\Vol_S(S_3)  =   \frac{\pi^3}{ (b_1 + b_2) (b_1 - b_3) (b_1 + b_3)}  \,,\nn\\
			 	\Vol_S(S_4) & =   \frac{\pi^3}{(b_1 - b_2) (b_1 - b_3) (b_1 + b_3)}  \,,\quad
				 	\Vol_S(S_5)  =   \frac{\pi^3}{(b_1 - b_2) (b_1 + b_2) (b_1 - b_3) }  \,.
 \end{align}
 The Sasaki-Einstein volume can be obtained by setting $b_1=4$ and then extremizing \eqref{volsass7} over $b_2,b_3$.
 The extremal point has $b_2=b_3=0$ with $\Vol_{SE}(V^{5,2}) =(27/128)\pi^4$ and $\Vol_{SE}(S_a)=\pi^3/64$.

We now consider the GK geometry associated with AdS$_2\times Y_9$ solutions.
The off-shell entropy function can be obtained from \eqref{entropy} and takes the form
 \begin{align}\label{entropyV52}
 		\mathscr{S} &\,= \, \frac{4\pi N^{3/2}}{27b_0\sqrt{b_1} } \Bigg[\sqrt{(b_1^{(+)} - b_2^{(+)}) (b_1^{(+)} + b_2^{(+)}) (b_1^{(+)} - b_3^{(+)}) (b_1^{(+)} + b_3^{(+)})} \nn\\
 		&\qquad\qquad\qquad -\sigma\, \sqrt{(b_1^{(-)} - b_2^{(-)}) (b_1^{(-)} + b_2^{(-)}) (b_1^{(-)} - b_3^{(-)}) (b_1^{(-)} + b_3^{(+)})}\Bigg] \: .
 \end{align}
 The off-shell geometric R-charges are obtained from \eqref{Rpmnobar} and are given by
 \begin{align}
 		R_1^\pm = \frac{2}{3} \,\frac{b_1^{(\pm)}}{b_1} \:, \qquad R_2^\pm &= \frac{2}{3}\,\frac{b_1^{(\pm)} - b_3^{(\pm)}}{b_1} \:, \qquad R_3^\pm = \frac{2}{3}\,\frac{b_1^{(\pm)} - b_2^{(\pm)}}{b_1} \:, \nn\\
 		R_4^\pm &= \frac{2}{3}\,\frac{b_1^{(\pm)} + b_2^{(\pm)}}{b_1} \:, \qquad R_5^\pm = \frac{2}{3}\,\frac{b_1^{(\pm)} + b_3^{(\pm)}}{b_1} \: .
 \end{align}
From \eqref{fluxes} we obtain the fluxes
 \begin{equation}\label{fluxesv52}
 	\kf_a\equiv \frac{\Nas_a}{N}= \frac{ 1}{3 m_+ m_-}\big\{p_1, \,p_1 - p_3,\,p_1 - p_2,\,p_1 + p_2,\,p_1 + p_3\big\} \,,
 \end{equation}
with $p_1 =-\sigma m_+ -  m_-$. 
Note that
\begin{align}\label{V52fluxsum}
\Nas_1 + \Nas_2 + \Nas_5 = \Nas_1 + \Nas_3 + \Nas_4= \frac{p_1}{m_+m_-}N\, .
\end{align}
These equations play an analogous role to the general formula \eqref{homfluxes} valid in the toric case, that we discuss later. 
Indeed, both \eqref{homfluxes} and \eqref{V52fluxsum} may be interpreted in field theory 
as the condition that the superpotential is twisted via the line bundle $\mathcal{O}(p_1)$, where 
the flux terms on the left hand side correspond to the twistings of individual fields 
that appear in the superpotential. For $V^{5,2}$ the particular combination of terms in \eqref{V52fluxsum} may be determined from the 
field theory dual worked out in \cite{Martelli:2009ga}, with particular fields being associated with particular divisors $C(S_a)\subset C(V^{5,2})$.

To obtain the on-shell values for $\mathscr{S}$ and $R_a^\pm$, we should set  $b_1 = 1$ and then extremize \eqref{entropyV52} as a function of $b_2$, $b_3$ and $b_0$. As for the $S^7$ case, the case of general fluxes is involved, hence we will consider the case when four of the fluxes are pairwise equal, namely $\Nas_2 = \Nas_3$ and $\Nas_4 = \Nas_5$, which holds when
\begin{align}\label{v52p2p3}
p_2 = p_3\,.
\end{align}
This choice corresponds to considering fibrations of the $V^{5,2}$ that maintain $SU(2)\times SU(2)\times U(1)\subset SO(5)\times U(1)$ isometry.
In this case the entropy \eqref{entropyV52} becomes symmetric under the exchange $b_2 \leftrightarrow b_3$, 
and we may assume that at the extremal point
\begin{align}\label{v52b2b3}
b_2 = b_3\,.
\end{align}
With \eqref{v52p2p3} and restricting to the locus \eqref{v52b2b3} we find that
the \eqref{entropyV52} simplifies to
 \begin{equation}\label{entropyV52simp}
 	\mathscr{S} = \frac{4\pi N^{3/2}}{27b_0} \left[\delta_+ (b_1^{(+)} - b_2^{(+)})(b_1^{(+)} + b_2^{(+)}) - \delta_- (b_1^{(-)} - b_2^{(-)})(b_1^{(-)} + b_2^{(-)}) \right] ,
 \end{equation}
 where the signs are defined as
 \begin{equation}
 	\delta_+ \equiv \text{sign}\left[(b_1^{(+)} - b_2^{(+)})(b_1^{(+)} + b_2^{(+)})\right], \quad \delta_- \equiv \sigma \,\text{sign}\left[(b_1^{(-)} - b_2^{(-)})(b_1^{(-)} + b_2^{(-)})\right].
 \end{equation}

  By an explicit calculation we find that when $\delta_+ =  \delta_-$, \eqref{entropyV52simp} is linear in $b_2$ and $b_0$ and so it does not have any extremum. Hence we take
  \begin{align}\label{deltaV52}
  \delta_+ =  -\delta_- \equiv \delta\,.
  \end{align}
We then find that \eqref{entropyV52simp} is extremized for
 \begin{equation}\label{extbsv52}
 	{b}_0^* = \pm\frac{2m_+ m_-}{\sqrt{2(m_+^2+m_-^2)-p_2^2}} \:, \qquad {b}_2^* = \mp\frac{p_2(a_+ m_- - a_- m_+)}{\sqrt{2(m_+^2+m_-^2)-p_2^2}} \:.  
 \end{equation} 
Interestingly, ${b}_0^*$ and ${b}_2^*$ do not depend on whether we are in the twist or in the anti-twist case. The on-shell entropy reads
\begin{equation}\label{entropyp2equalp3}
 	\mathscr{S}_{\text{os}}= \delta\, \frac{8 \pi  N^{3/2} }{27 m_+ m_-} \left[\sigma m_+ - m_- \pm\sqrt{2 \left(m_-^2+m_+^2\right)-p_2^2}\right] .
 \end{equation}
 
   Clearly we must have $p_2^2\le 2 \left(m_-^2+m_+^2\right)$ and there are additional restrictions that are needed in order to have a positive entropy. Mimicking what we did for the case $X_7 = S^7$, we need to analyse the cases of $\delta=\pm1$ and the two signs  
\eqref{extbsv52}, for each case of the twist, $\sigma=+1$, and the anti-twist, $\sigma=-1$. We find that there are no possibilities for
choosing $m_\pm>0$, $ p_2$ for
the twist case and hence there are no AdS$_2\times Y_9$ solutions in the twist class for $X_7 = V^{5,2}$ and $p_2 = p_3$.
   
For the anti-twist case with $\sigma=-1$ we find there are no possibilities associated with the lower sign in the extremal values in
\eqref{extbsv52}, \eqref{entropyp2equalp3}. However, there are possibilities associated with the upper sign. In particular we find
 	\begin{equation}\label{v52bext}
 		{b}_0^* = \frac{2m_+ m_-}{\sqrt{2(m_+^2+m_-^2)-p_2^2}} \:, \qquad {b}_2^* = -\frac{p_2(a_+ m_- - a_- m_+)}{\sqrt{2(m_+^2+m_-^2)-p_2^2}} \: ,  
 	\end{equation} 
	and
 	\begin{equation}\label{entrV52anti}
 		\mathscr{S}_{\text{os}}= -\delta\, \frac{8 \pi  N^{3/2} }{27 m_+ m_-} \left[ m_+ + m_- -\sqrt{2 \left(m_-^2+m_+^2\right)-p_2^2}\right] ,
 	\end{equation}
 	with
 	\begin{align}
 			\delta &= 1 \qquad \,\,\,\,\text{for} \qquad -|m_+ - m_-| < p_2 < |m_+ - m_-| ,\nn\\[6pt]
 			\delta &= -1 \qquad \text{for} \qquad  -(m_+ + m_-) < p_2 < -|m_+ - m_-| \nn\\
 			& \qquad\qquad\qquad\quad \text{or} \qquad |m_+ - m_-| < p_2 < (m_+ + m_-) \:.
 	\end{align}
 	Notice that the condition $p_2^2\le 2 \left(m_-^2+m_+^2\right)$ is automatically satisfied.
 	For this class, with the upper sign, the on-shell R-charges are given by
 	\begin{align}\label{osrchv52}
 		R^\pm_1&=\frac{2}{3}-\frac{4m_\mp}{3\sqrt{2(m_+^2+m_-^2)-p_2^2}}\,,\nn\\
 		R^+_2=R^+_3&=\frac{2}{3}-\frac{2(2m_-+p_2)}{3\sqrt{2(m_+^2+m_-^2)-p_2^2}}\,,\nn\\
 		R^+_4=R^+_5&=\frac{2}{3}-\frac{2(2m_--p_2)}{3\sqrt{2(m_+^2+m_-^2)-p_2^2}}\,,\nn\\
		R^-_2=R^-_3&=\frac{2}{3}-\frac{2(2m_+-p_2)}{3\sqrt{2(m_+^2+m_-^2)-p_2^2}}\,,\nn\\
 		R^-_4=R^-_5&=\frac{2}{3}-\frac{2(2m_++p_2)}{3\sqrt{2(m_+^2+m_-^2)-p_2^2}}\,.
 	\end{align}
	If we now impose the conditions $R^+_a>0$ and $\sigma R^-_a>0$ (here with $\sigma=-1$) we find that we are only left with the case of the upper sign with
\begin{align}\label{posrchgesv52}
			\delta &= 1 \qquad \text{for:} \qquad 0 < p_3 < \frac{1}{2}\,(m_+ - m_-)\,.
				\end{align}
It is natural to conjecture that these are sufficient constraints on the parameters in
order for the AdS$_2\times Y_9$ solutions to actually exist.

We can now consider the further special sub-case of the universal anti-twist  which has
\begin{align}
p_2=p_3=0\,.
\end{align}
In this case the fluxes in \eqref{fluxesv52} read
 \begin{equation}
 	\kf_a\equiv \frac{\Nas_a}{N}= \frac{ p_1}{3 m_+ m_-}\big\{1,1,1,1,1\big\} \,.
 \end{equation}
From \eqref{v52bext} we see that 
the extremal value has $ b_2^*=b_3^*=0$, consistent with \eqref{utexplicbival}. The extremal value of $b_0$, the on-shell entropy \eqref{entrV52anti} and  R-charges \eqref{osrchv52} agree with those given in section \ref{utwsitads2y9}.
 

\section{Toric GK geometry}\label{sec:fibretoric}

In this section we recall some general properties of the GK geometry associated with 
toric $Y_{2n+1}$, and in particular discuss the master volume $\mathcal{V}_{2n+1}$ of $Y_{2n+1}$ which allows one to carry out the geometric extremization algebraically. 
We follow \cite{Gauntlett:2018dpc,Gauntlett:2019roi,Gauntlett:2019pqg}, where further details can
be found.

\subsection{Toric K\"ahler cones}\label{sec:toric}

We begin with a toric K\"ahler cone ${C}(Y_{2n+1})$
in real dimension $2(n+1)$. Thus we have a K\"ahler metric of the form
\begin{align}\label{kcone}
\diff s^2_{C(Y_{2n+1})} &= \diff r^2 + r^2\diff s^2_{2n+1}~,
\end{align}
with a $U(1)^{n+1}$ action generated by holomorphic Killing vectors
$\partial_{\varphi_\mu}$, $\mu=0,1,\dots, n$,  with each $\varphi_\mu$ having period $2\pi$. 
We assume that $C(Y_{2n+1})$ is Gorenstein meaning that it admits a global 
holomorphic $(n+1,0)$-form $\Psi_{(n+1,0)}$. As before we choose a basis so that this holomorphic volume form has unit charge
under $\partial_{\varphi_1}$ and is uncharged under $\partial_{\varphi_{\hat{\mu}}}$, $\hat{\mu} = 0,2,\ldots, n$.

The manifold $Y_{2n+1}$ is embedded at $r=1$. 
The complex structure of the cone pairs the radial vector $r\partial_r$ with the Killing vector field
$\xi$ tangent to $Y_{2n+1}$, \emph{i.e.} $\xi=\mathcal{J}(r\partial_r)$.
We can write 
\begin{align}\label{Reebbasis2}
\xi &= \sum_{\mu=0}^{n} b_\mu\partial_{\varphi_\mu}~,
\end{align}
and the vector $(b_\mu)=(b_0, b_1,\ldots,b_{n})\in \mathbb{R}^{n+1}$ then parametrizes the choice of R-symmetry vector $\xi$. 
A given vector field $\xi$ defines a foliation of $Y_{2n+1}$ which we denote by $\mathcal{F}_\xi$.
We also have 
\begin{align}\label{b1charge}
\mathcal{L}_\xi \Psi_{(n+1,0)} = \ii b_1 \Psi_{(n+1,0)}\, .
\end{align}

Similarly, the complex structure pairs the one-form $\eta$, dual to the Killing vector $\xi$,
with $\diff r/r$. For K\"ahler cones we have
\begin{align}\label{detaSasakian}
\diff\eta &= 2J_S~,
\end{align}
where $J_S$ is the transverse K\"ahler form. In this case
$\eta$ is a contact one-form 
on $Y_{2n+1}$ and $\xi$, satisfying $\xi\lrcorner \, \eta=1$, is then also called the Reeb vector field. 
The associated Sasakian metric on $Y_{2n+1}$ is given by
\begin{align}\label{splitmetric}
\diff s^2_{\Y} &= \eta^2 + \diff s^2_{2n}({J_S})~.
\end{align}
We also note that that \eqref{detarho} immediately gives the cohomology relation
\begin{align}\label{bcoh}
[\diff\eta] = \frac{1}{b_1}[\rho]\in H^2_B(\mathcal{F}_\xi)\, , 
\end{align}
where $\rho$ denotes the Ricci two-form of the transverse K\"ahler metric $\diff s^2_{2n}({J_S})$.

We next define the moment map coordinates 
\begin{align}\label{yi}
y^\mu & \equiv  \tfrac{1}{2}r^2\partial_{\varphi_\mu}\lrcorner\,  \eta~, \qquad \mu=0,1,\ldots,n~.
\end{align}
These span the moment map polyhedral cone $\mathcal{C}\subset \R^{n+1}$, where the 
$y^\mu$ are Euclidean coordinates on $\R^{n+1}$. 
The polyhedral cone $\mathcal{C}$, which is convex, is determined by $D$ vectors 
$v_{A\mu}$ with $A=1,\dots D$ and for each $A$ we have $(v_{A\mu})\in\Z^{n+1}$ . Specifically, these are inward pointing and primitive
normals to the $D$ facets of the cone and we have
\begin{align}
\mathcal{C} &= \left\{(y^\mu)\in \R^{n+1}\ \mid \ 
\sum_{\mu=0}^n y^\mu v_{A\mu} \geq 0~, \quad A=1,\ldots, D\right\}~.
\end{align}
If we change basis for the $U(1)^{n+1}$ action with an $SL(n+1,\Z)$ transformation, then there is a corresponding $SL(n+1,\Z)$
transformation on the moment map coordinates and hence on 
the toric data, \emph{i.e.} the vectors $v_{A\mu}$.
In the basis above, satisfying \eqref{b1charge}, the Gorenstein condition implies that $(v_{A\mu=1}) =1$ for all $A$.

Geometrically, $C(Y_{2n+1})$ fibres over the polyhedral cone $\mathcal{C}$. 
There is a trivial $U(1)^{n+1}$ fibration over the interior $\mathcal{C}_{\mathrm{int}}$ 
of  $\mathcal{C}$, with the $D$ normal vectors $v_{A\mu}$ to each 
bounding facet $\{\sum_{\mu=0}^n y^\mu{v}_{A\mu}=0\}\subset \partial\mathcal{C}$ 
specifying which $U(1)\subset U(1)^{n+1}$ collapses along that facet. Each facet is also the image under the moment map of a 
toric divisor $\mathcal{D}_A$ in $C(Y_{2n+1})$, where $\mathcal{D}_A$ is a complex codimension one submanifold that is invariant under 
the action of the $U(1)^{n+1}$ torus. 
We define  $T_A$ to be the $U(1)^{n+1}$ invariant $(2n-1)$-cycle in $Y_{2n+1}$ associated with the toric divisor 
$\mathcal{D}_A$ on the cone. Since $\dim H_{2n-1}(Y_{2n+1},\mathbb{R})=D-(n+1)$, the toric
$(2n-1)$-cycles $[T_A]\in H_{2n-1}(Y_{2n+1},\mathbb{Z})$ are not independent in
$H_{2n-1}(Y_{2n+1},\mathbb{Z})$,  satisfying the $(n+1)$ relations
\begin{equation}\label{constraint0}
	\sum_{A=1}^D {v}_{A\mu} [T_{A}]= 0~, \qquad \mu =0,1,\dots,n \:.
\end{equation}

We next define the Reeb cone, $\mathcal{C}^*$, to be the dual cone 
to $\mathcal{C}$. Then for a K\"ahler cone metric on $C(Y_{2n+1})$ 
we necessarily have $({b}_\mu)\in \mathcal{C}_{\mathrm{int}}^*$, where 
$\mathcal{C}^*_{\mathrm{int}}$ is the open interior of the Reeb cone \cite{Martelli:2005tp}.
The image of $Y_{2n+1}=\{r=1\}$ under the moment map is then the compact, convex $n$-dimensional polytope
\begin{align}\label{Psas}
{P} = {P}(b_\mu) = \mathcal{C}\cap H(b_\mu)\, ,
\end{align}
where the 
Reeb hyperplane defined by
\begin{align}\label{ReebH}
H=H(b_\mu) \equiv \left\{(y^\mu)\in \R^{n+1}\mid \sum_{\mu=0}^n y^\mu b_\mu = \tfrac{1}{2}\right\}\, .
\end{align}

\subsection{The master volume}\label{sec:master}

We first fix a choice of toric K\"ahler cone metric on the complex cone $C(Y_{2n+1})$. As described in the previous subsection, this allows us to introduce coordinates $(y^\mu, \varphi_\mu)$ on $C(Y_{2n+1})$.  For a fixed choice of such complex cone, with Reeb vector $\xi$ given by (\ref{Reebbasis}),
we would then like to study a more general class of transversely K\"ahler metrics of the form (\ref{splitmetric}) with $J$, in general, no longer equal to $J_S$.

Of central interest is the ``master volume'' defined by
\begin{align}
\label{defVmaaster}
\mathcal{V}_{2n+1} \equiv   \int_{Y_{2n+1}}\eta\wedge \frac{J^n}{n!}~,
\end{align}
which is considered to be a function both of the vector $\xi$, specified by $b_\mu$, and the transverse K\"ahler class $[J]\in H^2_B(\mathcal{F}_\xi)$. 
We can introduce $c_A\in H^2_B(\mathcal{F}_\xi)$ to be basic representatives 
of integral classes in $H^2(Y_{2n+1},\mathbb{Z})$, which are Poincar\'e dual to 
the $D$ toric divisors $\mathcal{D}_A$ on $C(Y_{2n+1})$. This allows us to write
\begin{align}\label{omegalambda}
[J] = -2\pi\sum_{A=1}^D \lambda_Ac_A \in  H^2_B(\mathcal{F}_\xi) \, ,
\end{align}
with the real parameters $\lambda_A$ determining the transverse K\"ahler class.
The $c_A$ are not all independent and $[J]$ in fact only depends on $D-n$  of the $D$  parameters $\{ \lambda_A\}$, as 
we shall see shortly. 
It is also useful to note that the first Chern class of the foliation can be written in terms of the $c_A$ as 
\begin{align}\label{rhoca}
[\rho] &= 2\pi\sum_{A=1}^D c_A  \in H^2_B(\mathcal{F}_\xi)~.
\end{align}
We then see that in the special case in which 
\be
\label{lambdaSas}
\lambda_A  =  -\frac{1}{2b_1}\,, \qquad A=1,\dots D\,,
\ee
we recover the Sasakian  K\"ahler class $[\rho]= 2 b_1 [J_S]$ and the master volume (\ref{defVmaaster})
reduces to the Sasakian volume.

The master volume (\ref{defVmaaster}) may be written as 
\begin{align}\label{VEuc}
\mathcal{V}_{2n+1} = \frac{(2\pi)^{n+1}}{|{b}|}\Vol(\mathcal{P})~,
\end{align}
where $|{b}|\equiv ({\sum_{\mu=0}^n b_\mu b_\mu})^{1/2}$.
The factor of $(2\pi)^{n+1}$ arises by integrating over the torus $U(1)^{n+1}$, while $\Vol (\mathcal{P})$ is 
the Euclidean volume of the compact, convex $n$-dimensional polytope
\begin{align}\label{generalP0}
\mathcal{P} & = \mathcal{P}(b_\mu;\lambda_a ) \nn\\
\, & \equiv \, \left\{(y^\mu)\in H(b_\mu) \ \mid \ 
\sum_{\mu=0}^n({y}^\mu-{y}^\mu_*)v_{A\mu} \geq \lambda_A~, \  A=1,\ldots,D\right\}~,
\end{align}
with $y^\mu_*$ an arbitrary point in the Reeb hyperplane. Different choices for $y^\mu_*$ can be reabsorbed into a re-definition of the $\lambda_A$.
It is convenient to take
\begin{align}\label{originP0}
(y^\mu_*) \equiv \left(0,\frac{1}{2b_1},0,\ldots ,0\right) \in H~.
\end{align}
Notice then that with $\lambda_A  =  -\frac{1}{2b_1}$ the polytope $\mathcal{P}$
is the same as the Sasaki polytope $P$ given in \eqref{Psas}.

The master volume $ \mathcal{V}_{2n+1} $ is homogeneous of degree $n$ in the $\lambda_A$, and
we have
 \begin{align}\label{mvapphere}
\mathcal{V}_{2n+1} \, \equiv  \, \int_{Y_{2n+1}}\eta\wedge \tfrac{1}{n!}  J^n= (-2\pi)^n\sum_{A_1,\dots, A_n=1}^D \tfrac{1}{n!}I_{A_1\dots A_n} \lambda_{A_1}\dots \lambda_{A_n}~,
 \end{align}
 where the ``intersection'' numbers $I_{A_1\dots A_n}$ are defined as
\begin{align}\label{intersection}
 I_{A_1\dots A_n} \, \equiv \,  \int_{Y_{2n+1}}\eta\wedge c_{A_1}\wedge\dots  \wedge c_{A_n} &=  \frac{1}{(-2\pi)^n}\frac{\partial^n\mathcal{V}_{2n+1}}{\partial\lambda_{A_1}\dots  \partial\lambda_{A_n}}~.
\end{align}
We may then calculate
\begin{align}\label{lamderv}
 \int_{Y_{2n+1}}\eta\wedge \rho^k \wedge \tfrac{1}{(n-k)!}{J^{n-k}}
 =(-1)^k\sum_{A_1,\dots, A_k=1}^D \frac{\partial^k\mathcal{V}_{2n+1}}{\partial\lambda_{A_1}\dots \partial\lambda_{A_k}}
\,.
 \end{align}
We are also interested in integrating over $T_A$, 
the $(2n-1)$-dimensional and $U(1)^{n+1}$ invariant cycle in $Y_{2n+1}$ associated with a toric divisor $\mathcal{D}_A$ on the cone and Poincar\'e dual to $c_A$.
We have
\begin{align}\label{SaVnew}
\int_{T_A} \eta\wedge \rho^k\wedge\tfrac{1}{(n-k-1)!}J^{n-k-1} &= \int_{Y_{2n+1}}\eta\wedge \rho^k\wedge\tfrac{1}{(n-k-1)!}J^{n-k-1} \wedge c_A \nn\\
&=\frac{(-1)^{k+1}}{2\pi}\sum_{B_1,\dots,B_k=1}^D \frac{\partial^{k+1} \mathcal{V}_{2n+1}}{\partial \lambda_A\partial \lambda_{B_1}\dots\partial\lambda_{B_k}}\,.
\end{align}
We can obtain similar formulas for integrals over higher-dimensional cycles that are Poincar\'e
dual to products of the $c_A$. For example, consider the codimension four cycle $T_{B_1B_2}$ that is Poincar\'e dual to $c_{B_1}\wedge c_{B_2}$. We then find, for example,
\begin{align}\label{UBBnew}
\int_{T_{A_1A_2}} \eta\wedge\rho^k\wedge \tfrac{1}{(n-k-2)!}J^{n-2} &= \int_{Y_{2n+1}}\eta\wedge \tfrac{1}{(n-2)!}J^{n-2} \wedge c_{A_1}\wedge c_{A_2} \nn\\
&=\frac{(-1)^k}{(2\pi)^2} \sum_{B_1,\dots,B_k=1}^D \frac{\partial^{s+2} \mathcal{V}_{2n+1}}{\partial \lambda_{A_1}\partial \lambda_{A_2}\partial \lambda_{B_1}\dots\partial\lambda_{B_k}}
\end{align}
It can also be shown that the
master volume $ \mathcal{V}_{2n+1} $ is homogeneous of degree $-1$ in~$b_\mu$.

The master volume satisfies the important identity
\begin{align}\label{id7dvol}
\sum_{A=1}^D\bigg(v_{A\mu}-\frac{b_\mu}{b_1}\bigg)
\frac{\partial\mathcal{V}_{2n+1}}{\partial\lambda_A}=0\,,
\qquad \mu =0,1,\dots,n\,.
\end{align}
 It follows from this, together with homogeneity, that the
master volume is invariant under the ``gauge" transformations 
\begin{align}\label{lamgt}
\lambda_A\ \to\ \lambda_A+\sum_{\mu =0}^{n}\gamma^\mu(v_{A\mu} b_1-b_\mu)\,,
\end{align}
for arbitrary constants $\gamma^\mu$. Since the transformation
parametrized by $\gamma^1$ is trivial, we see that
the master volume only depends on $D-n$ of the $D$ parameters $\{\lambda_A\}$, as noted above.\footnote{Note that since the $\lambda_A$ specify a K\"ahler class, they satisfy some positivity constraints, and some care is required in utilizing these gauge transformations.}

It is possible to obtain very explicit formulas for the master volume in low dimensions in terms of
$v_{A\mu}$, $b_\mu$ and $\lambda_A$. Specifically,
in dimensions $n=1,2$ and $3$ the relevant formulae for $Y_3,Y_5$ and $Y_7$ were 
derived in \cite{Gauntlett:2019pqg},\cite{Gauntlett:2018dpc} and \cite{Gauntlett:2019roi}, respectively.\footnote{Note that the minus sign in (2.23) of \cite{Gauntlett:2019roi} should be a plus sign.}

Finally, we note that the  formulae in this section assume that the polyhedral
cone $\mathcal{C}$ is convex, since we started the section with a cone that 
 admits a toric K\"ahler cone metric. However, this convexity
condition is, in general, too restrictive. Indeed, for applications to AdS$_2$ and AdS$_3$ solutions of interest many explicit supergravity solutions are known that are associated
with ``non-convex toric cones'', as defined in \cite{Couzens:2018wnk}, which in particular have toric data
which do not define a convex polyhedral cone. Based on the consistent picture that has emerged, including recovering results for the known explicit supergravity solutions,
it is expected that the key formulae in this section are also applicable to non-convex toric cones, and we will assume this is the case in the sequel. 


\subsection{Geometric extremization for toric $Y_{2n+1}$}\label{sec:sumgeomext}
When $Y_{2n+1}$ is toric, the geometric extremization of section \ref{sec:extremal} for the associated GK geometry on 
$Y_{2n+1}$ can be carried out using the master volume $\mathcal{V}_{2n+1}(b_\mu; \lambda_A)$. 
In order to carry this out we need expressions for
the off-shell supersymmetric action, the constraint equation and flux quantization conditions. These can all be 
expressed in terms of derivatives of the master volume as follows:
\begin{align}\label{tent_cc}
	S_\text{SUSY}&=-\sum_{A=1}^{D} \frac{\partial \mathcal{V}_{2n+1}}{\partial \lambda_{A}}\,,\nn\\
	0&= \sum_{A,B=1}^{D} \frac{\partial^{2} \mathcal{V}_{2n+1}}{\partial \lambda_{A} \partial \lambda_{B}}\,, \nn\\
	\nu_n \NAS_A&=
	 -\frac{1}{(2 \pi)} \frac{\partial S_\text{SUSY} }{\partial \lambda_{A} }=
	 \frac{1}{2 \pi} \sum_{B=1}^{D} \frac{\partial^{2} \mathcal{V}_{2n+1}}{\partial \lambda_{A} \partial \lambda_{B}} \:,
\end{align}
with $\NAS_A\in\mathbb{Z}$. The $\NAS_A$ are not all independent\footnote{Note that the $\NAS_A$ are a linear combination
of the (independent) quantized fluxes $\mathcal{N}_\alpha$ in \eqref{fluxquantize}, which are associated with 
the flux through a basis for the free part of $H_{2n-1}(\Y;\Z)$.}
 and, as a consequence of
\eqref{constraint0}, satisfy $(n+1)$ linear relations given by
\begin{equation}\label{constraintfluxes}
	\sum_{A=1}^{D} {v}_{A\mu}\NAS_{A}= 0~, \qquad \mu=0,1,\dots,n \:.
\end{equation}
Notice that the $\mu=1$ component in \eqref{constraintfluxes}, $\sum_A \NAS_A=0$, is actually the constraint equation given in \eqref{tent_cc}. 
Since the master volume is homogeneous of degree $n$ in the $\lambda_A$, it is also possible to write
\begin{align}
	S_\text{SUSY}&=-\frac{2\pi}{n-1}\nu_n\sum_{A=1}^D\lambda_A \NAS_A\,.
	\end{align}
	   
The geometric extremization principle that is to be implemented for the
GK geometry on $Y_{2n+1}$ is now simple to state.
To obtain the on-shell supersymmetric action, we need to fix $b_1=\frac{2}{n-2}$ and then extremize $S_\text{SUSY}$ with respect to $b_0,b_2,\dots,b_n$ as well
as the $D-n$ independent $\lambda_A$, subject to the constraint equation and flux quantization conditions in \eqref{tent_cc}.

\section{Toric geometry fibred over a spindle}\label{sec:toricoverspindle}

With the toric formalism of section \ref{sec:fibretoric} to hand, we now return to studying 
GK geometries $Y_{2n+1}$ of the fibred form
 \begin{align}\label{XoverYtoric}
X_{2n-1}\, \hookrightarrow\, Y_{2n+1} \, {\longrightarrow }\  \Sigma\, ,
\end{align}
introduced in section \ref{sec:fibred}, where now the fibres
$X_{2n-1}$ are toric Sasaki-Einstein manifolds and $\Sigma=\mathbb{WCP}^1_{[m_-,m_+]}$
is a spindle.
There is then a toric $U(1)^{n}$ action on $X_{2n-1}$, and when combined with the azimuthal rotation symmetry of the spindle
$\Sigma=\mathbb{WCP}^1_{[m_-,m_+]}$, this gives rise to a toric $U(1)^{n+1}$ action on $Y_{2n+1}$.
We will utilize the toric geometry summarized in the previous section both for $Y_{2n+1}$ and,
with suitable notational changes, for the fibres $X_{2n-1}$. Our main goal is to relate  
the quantities appearing in the geometric extremal problem 
for $Y_{2n+1}$ in section \ref{sec:extremal} to the master volume $\mathcal{V}_{2n-1}$ of the toric fibres $X_{2n-1}$. 

\subsection{Toric data for $Y_{2n+1}$}
As described in section \ref{sec:fibred}, the fibration of $\X$ over $\Sigma$ 
is specified by 
$n$ integers
 $(p_i)=(p_1,\dots, p_{n})$.\footnote{Setting $m_+=m_+=1$ is the case of the round $S^2$, which arises in \cite{Gauntlett:2018dpc} 
as a genus $g=0$ Riemann surface case. We should identify $p_i$ here with $-n_i$ there.} These are effectively Chern numbers for the associated $U(1)^n$ fibration 
 over $\Sigma$, and may be identified in terms of the local gluing data $\alpha_i^\pm$ of section~\ref{sec:fibred} 
 via \eqref{pidef}. Moreover, as explained in section~\ref{sec:volform}, we  have from \eqref{twistp}
\begin{align}\label{naughty}
p_1 = - (\sigma m_++ m_- )~,
\end{align}
where $\sigma=\pm1$ corresponds to the twist or anti-twist case, respectively:
\begin{align}
\sigma = \begin{cases}\ +1 & \mbox{twist}~, \\ \ -1  & \mbox{anti-twist}~.\end{cases}
\end{align}
Recall here that we are using a basis for the $U(1)^n$ action on $X_{2n-1}$ for which the holomorphic $(n,0)$-form $\Omega$
on the cone $C(X_{2n-1})$
is only charged with respect to the first $U(1)$. This $U(1)$, associated with the integer Chern number $p_1$ in \eqref{naughty}, may then be 
viewed as a fiducial choice of Reeb  action on the Calabi-Yau cone $C(X_{2n-1})$, which in the holographic context is dual to a fiducial choice of R-symmetry of the dual SCFT ({\it i.e.} not necessarily the 
superconformal R-symmetry). The remaining $U(1)^{n-1}$, associated with
the integer Chern numbers
\begin{align}
\vec p\, \equiv\,  (p_2,\dots, p_{n})\,,
\end{align} 
are dual to flavour symmetries.

The toric data for $C(Y_{2n+1})$ may be determined in terms of the toric data for the Calabi-Yau cone $C(X_{2n-1})$, together with the spindle data $m_\pm$, and the fibration data $p_i$, $\sigma$. 
To begin, let ${v}_{ai}$ with $a=1,\ldots,d$ and $i=1,\dots,n$ be the toric data for
$C(X_{2n-1})$, with $({v}_{ai})_{i=1}^n\in \Z^n$ for each $a$. In the basis for the $U(1)^n$ action described above, 
where the holomorphic $(n,0)$-form $\Omega$
on the cone $C(X_{2n-1})$ has charge 1 under the first $U(1)\subset U(1)^n$, and is uncharged 
under the remaining $U(1)^{n-1}$, we have 
$({v}_{ai})_{i=1}^n=(1,\vec{w}_a)$, with $\vec{w}_a\in\Z^{n-1}$ describing a convex 
lattice polytope. Following the discussion in section \ref{sec:fibred}, the toric data for the cone $C(Y_{2n+1})$ 
is then given by $D=d+2$ vectors $v_{A\mu}$ with $A=+,-,a$ and $\mu=0,1,\dots,n$ and 
$(v_{A\mu})_{\mu=0}^n \in\mathbb{Z}^{n+1}$ for each $A$.
Explicitly, in the basis of vector fields $\partial_{\varphi_\mu}$, $\mu=0,1,\ldots,n$, generating 
the $U(1)^{n+1}$ action introduced in section \ref{sec:fibred}, 
the $v_{A\mu}$ can be written  
\begin{align}\label{Y7data1}
{v}_{+\mu} &= (m_+,v_{+i})\,,\quad\qquad\qquad \mbox{where} \ v_{+i}\equiv  (1,-a_+ \vec{p} ) ~, \nn\\
{v}_{-\mu} &= (-\sigma m_-,v_{-i} )\,,\qquad\quad\,  \ \mbox{where} \ v_{-i}\equiv(1,-\sigma a_- \vec{p} )~, \nn\\
{v}_{a\mu}&= (0,{v}_{ai})\,,\qquad\qquad\qquad \ \mbox{where} \ v_{ai}\equiv (1,\vec{w}_a)~.
\end{align}
Here, as in \eqref{apm}, the
$a_\pm\in \Z$ satisfy
\begin{align}\label{bezout}
a_- m_++a_+ m_-   = 1~.
\end{align}
Recall that $\partial_{\varphi_0}$ is a vector field on $\Y$ that rotates the spindle direction, with 
the holomorphic $(n+1,0)$-form $\Psi$ on $C(\Y)$ having charge zero under  $\partial_{\varphi_0}$, while 
$\partial_{\varphi_i}$, $i=1,\ldots,n$, are vector fields that generate the 
toric $U(1)^n$ action on the fibres $\X$. This immediately leads to 
the result $v_{a\mu}=(0,v_{ai})$ in \eqref{Y7data1}. 
Geometrically, the $v_{ai}$ specify the $d$ $U(1)\subset U(1)^n$ 
subgroups that fix $d$ corresponding toric divisors in $C(\X)$, and 
in $C(\Y)$ there are then correspondingly $d$ toric divisors, 
fixed by $U(1)\subset U(1)^{n+1}$ and specified by $v_{a\mu}$; more precisely, 
these are the $d$ toric divisors of $C(\X)$ fibred over $\Sigma$.
On the other hand, in $C(\Y)$ there are two additional toric divisors, 
fixed by the $U(1)\subset U(1)^{n+1}$ subgroups specified by $v_{\pm \mu}$, respectively. Geometrically, these 
are cones over the fibres $X_\pm = \X/\Z_{m_\pm}$ over the
two poles of the spindle, and the vector 
fields $\zeta_+=(v_{+\mu})$, $\zeta_-=\sigma (v_{-\mu})$ were determined in \eqref{zetapm}, \eqref{vpmmu}. 

Recall that the integers $a_\pm$ in \eqref{bezout} are unique only up to shifts
\begin{align}\label{shiftbezout}
a_+ \, \mapsto \, a_+ -  \kappa\, m_+ \,,\qquad 
a_- \, \mapsto \, a_- + \kappa\,  m_-~,
\end{align}
where $\kappa\in \Z$ is arbitrary. The toric data $v_{A\mu}$ for $Y_{2n+1}$ given in \eqref{Y7data1} 
{\it a priori} depends on the choice of $a_\pm$, but recall that we are free to make 
$SL(n+1,\Z)$ transformations, corresponding to a change of basis for the torus $U(1)^{n+1}$. 
One can check that the matrix 
\begin{align}\label{sl4matrixexp}
M = \begin{pmatrix} 
1&0 & 0&\hdots& 0\\ 
0&1&0 &\hdots & 0   \\ 
\kappa p_2&0 &1 &\hdots & 0  \\ \vdots & \vdots&0 &\ddots & \vdots  \\ 
 \kappa p_{n} & 0 &0&\hdots & 1 \end{pmatrix}\, \in \, SL(n+1,\Z)\,,
\end{align}
acting on the $\mu$ index of $v_{A\mu}$, maps the vectors ${v}_{A\mu}$ in \eqref{Y7data1} to the same set of vectors with the replacements \eqref{shiftbezout}. 
The toric data we have presented is thus well-defined, with all choices of $a_\pm$ satisfying \eqref{bezout} describing the same toric geometry.
 \begin{figure}[h!]
	\centering
	\includegraphics[scale=0.4,trim=0 6cm 0 2cm]{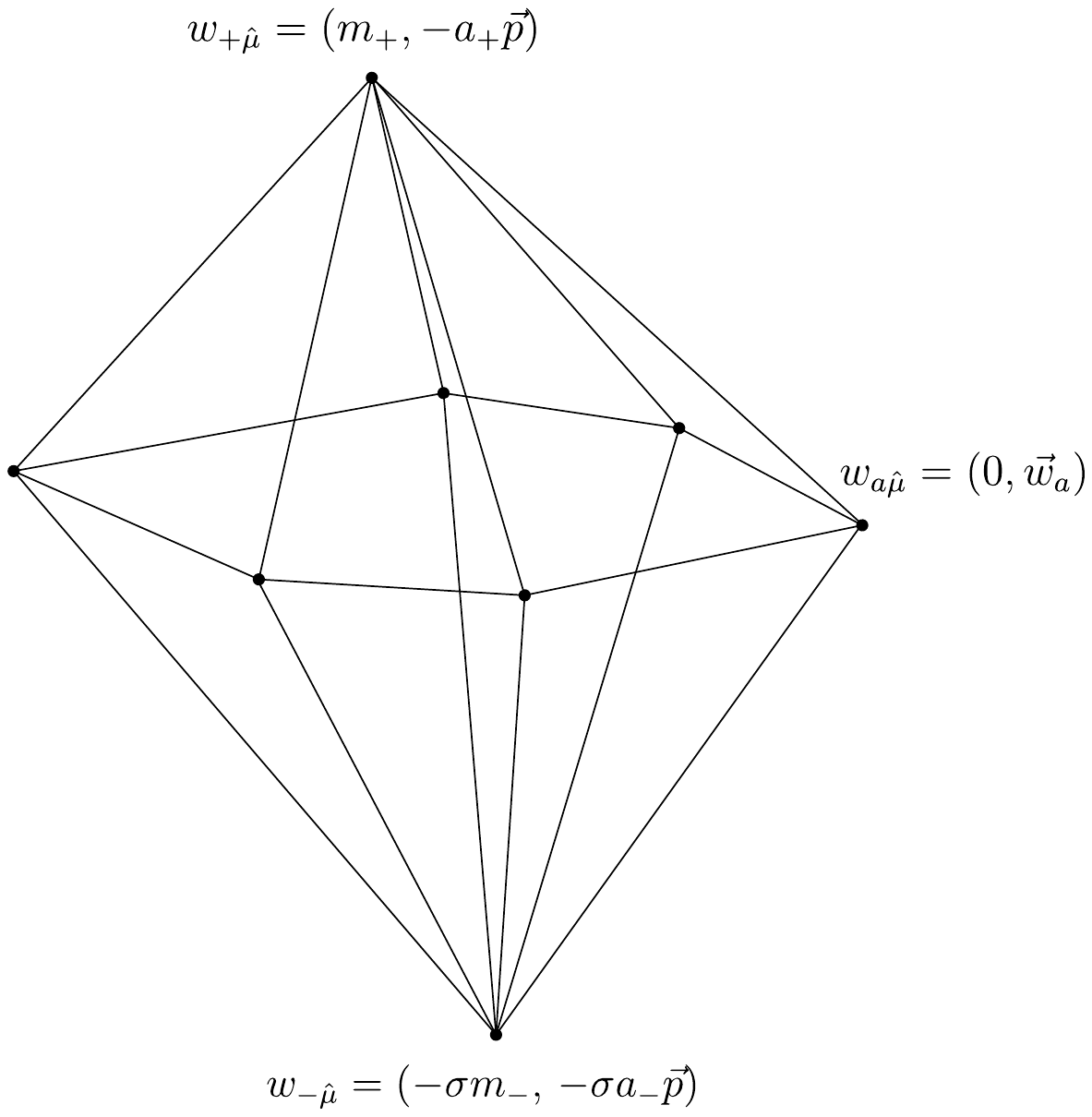}
	\caption{Toric data for a representative twist case ($\sigma=+1$) with $n=3$, $d=6$. The toric diagram is a suspension of the toric diagram for $X_5$, the two-dimensional polytope at height zero. For the anti-twist case ($\sigma=-1$) notice that the figure is no longer convex (recall $m_\pm\in\mathbb{N}$).}
	\label{suspensionnew}
\end{figure}

The toric data $v_{A\mu}$ for $Y_{2n+1}$ is given by a ``suspension" of the toric data for $X_{2n-1}$, 
at least in the twist case with $\sigma=+1$.
Dropping the $\mu=1$ component in \eqref{Y7data1} we can define truncated vectors 
\begin{align}\label{Y7data12}
{w}_{+\hat \mu} &= (m_+,-a_+ \vec{p})\,,\quad
{w}_{-\hat \mu} = (-\sigma m_-, -\sigma a_- \vec{p} )\,,\quad
{w}_{a\hat \mu}= (0,\vec{w}_a)\,,
\end{align}
with $\hat\mu=0,2,\dots, n$. In Figure \ref{suspensionnew} we have provided an illustration of the polytope that these vectors define for the twist case $\sigma=+1$, of dimension $n=3$, $d=7$. In particular the 
two-dimensional polytope at height zero is the ``toric diagram" for $X_5$, defined by the vectors $\vec{w}_a$; 
 we see that the toric diagram for $Y_7$ is the suspension of this, adding two 
 additional vectors ${w}_{\pm\hat \mu}$. For the twist case, when $\sigma=+1$, the polytope is convex, as illustrated in the figure,
 but it is not for the anti-twist case
 when $\sigma=-1$.

That the toric data \eqref{Y7data1} describes a fibration 
\eqref{XoverYtoric} directly follows from our discussion in section \ref{sec:fibred}.
An alternative point of view is provided by utilizing the Delzant construction (see {\it e.g.} \cite{Martelli:2004wu}), as discussed in appendix \ref{app:A}. 

   \subsection{Relating the master volumes of $Y_{2n+1}$ and $X_{2n-1}$}\label{relvolssec}

We now show that we can relate the master volume $\mathcal{V}_{2n+1}$ for $C(Y_{2n+1})$, or more precisely certain derivatives of it, to the master volume
$\mathcal{V}_{2n-1}$ associated with $C(X_{2n-1})$.  

Recall from \eqref{VEuc} that the master volume for $Y_{2n+1}$ can be written
\begin{align}\label{VEuc2}
\mathcal{V}_{2n+1} = \frac{(2\pi)^{n+1}}{|{b}|}\Vol(\mathcal{P})>0~,
\end{align}
where $|{b}|\equiv ({\sum_{\mu=0}^n b_\mu b_\mu})^{1/2}$ and $\Vol (\mathcal{P})$ is 
the Euclidean volume of the compact, convex $n$-dimensional polytope
\begin{align}\label{generalP}
\mathcal{P} & = \mathcal{P}(b_\mu;\lambda_a )\nn\\
&\, \equiv\,  \left\{(y^\mu)\in H(b_\mu) \ \mid \ 
\sum_{\mu=0}^n({y}^\mu-{y}^\mu_*)v_{A\mu} \geq \lambda_A~, \quad A=1,\ldots,D\right\}~,
\end{align}
with $(y^\mu_*)\in H=H(b_\mu)$ given by
\begin{align}\label{originP}
(y^\mu_*)_{\mu=0}^n =(0, y_*^i),\qquad y_*^i\equiv \left(\frac{1}{2b_1},0,\ldots ,0\right) ~.
\end{align}
The defining condition for the $(n+1)$-dimensional Reeb hyperplane $H$ given in
\eqref{ReebH} can then be written in the form
\begin{align}\label{reebhypcond2}
\sum_{\mu=0}^n y^\mu b_\mu = \tfrac{1}{2}
\quad\Leftrightarrow\quad   
y^0b_0+\sum_{i=1}^n({y}^i-{y}^i_*)b_i = 0\,.
\end{align}

\begin{figure}[h!]
	\centering
	\includegraphics[scale=0.3,trim=0 2cm 0 3cm]{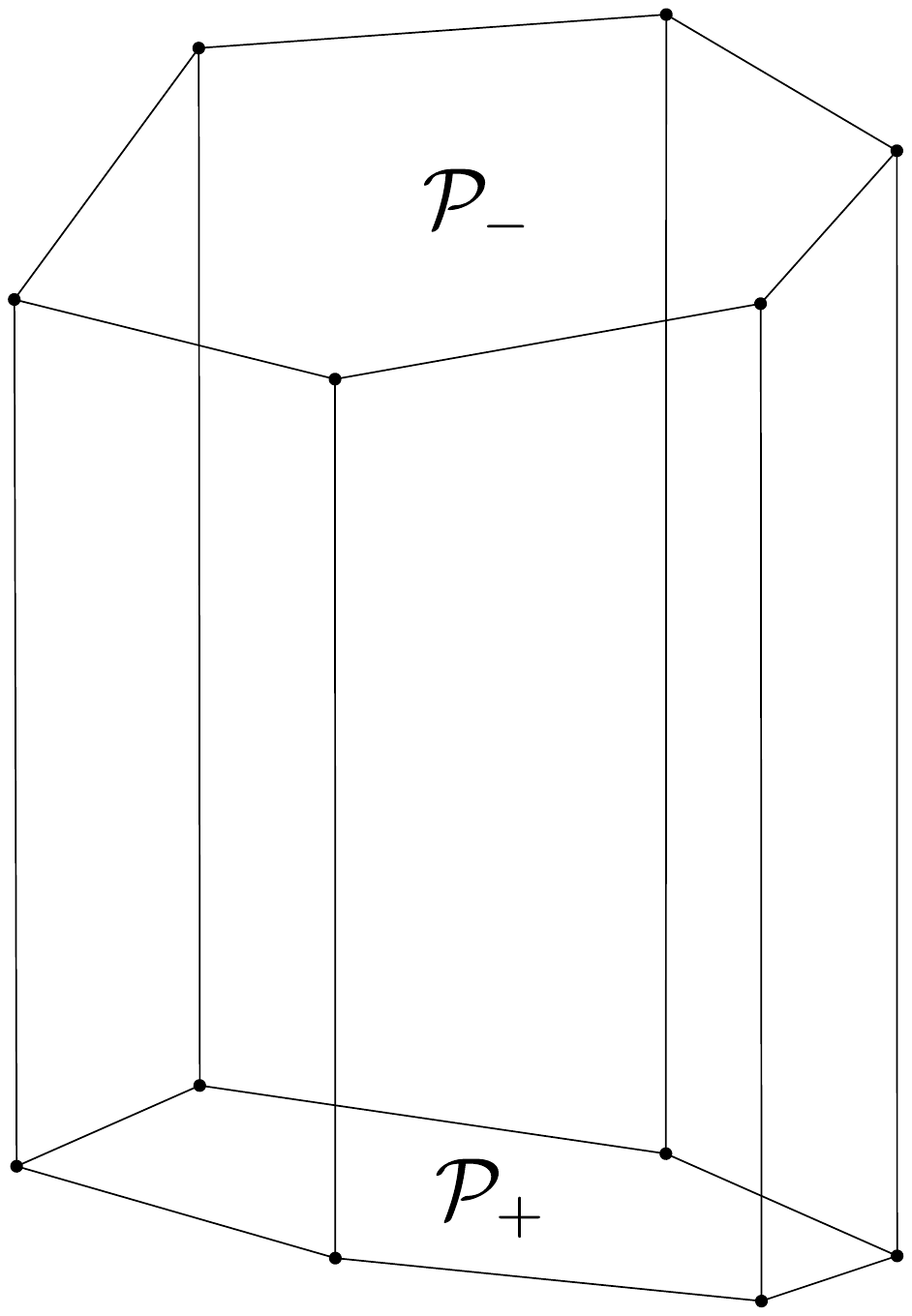}
	\caption{Toric polytope for a representative twist case with $n=3$, $d=6$.}
	\label{fig:prism}
\end{figure}

We now want to exploit the fact that the polytope 
$\mathcal{P}$ takes a special form, namely  a ``truncated prism'' with upper and lower faces, 
$\mathcal{P}_\pm$,
each of which is a polytope of one lower dimension specified by $d$ vectors; see Figure \ref{fig:prism}.
Specifically, these polytopes are defined by
\begin{align}
\mathcal{P}_\pm \, \equiv\, \mathcal{P} \cap \left\{\sum_{\mu=0}^n(y^\mu-{y}^\mu_*){v}_{\pm\mu} = \lambda_\pm\right\}~,
\end{align}
with ${v}_{\pm\mu}$ given in \eqref{Y7data1}.
Geometrically, these correspond to the $X_{2n-1}/\mathbb{Z}_{m_\pm}$ fibres
over the two poles of the spindle, associated with orbifold singularities $\mathbb{Z}_{m_\pm}$, respectively, as described in section \ref{sec:fibred}.
The faces $\mathcal{P}_\pm$ of the truncated prism are the images of toric divisors $\mathcal{D}_\pm$ in $C(\Y)$ under the moment map. 
We denote the corresponding $U(1)^{n+1}$ invariant $(2n-1)$-dimensional manifolds as 
$\StoX_\pm$,
which are copies of the fibre $\StoX_\pm \equiv X_{2n-1}/\mathbb{Z}_{m_\pm}$ over the poles of the spindle. 
These submanifolds were denoted $X_\pm$ in section \ref{sec:fibred}, and in particular in section \ref{sec:fixedpoints}, 
but as commented there we must be careful with orientations. 
From \eqref{SaVnew}, the volume of these manifolds are given by 
\begin{align}\label{tplusminus}
\mathrm{Vol}(\StoX_\pm)\equiv \int_{\StoX_\pm}\eta\wedge \tfrac{1}{(n-1)!}J^{n-1} = -\frac{1}{2\pi}\frac{\partial \mathcal{V}_{2n+1}}{\partial\lambda_\pm}~,
\end{align}
where the orientation on $T_\pm$ is induced from the complex structure
 and $\mathrm{Vol}(\StoX_\pm)>0$.
We shall see later that these satisfy the homology relation 
$m_+[\StoX_+]= \sigma m_-[\StoX_-]\in H_{2n-1}(\Y,\R)$, 
leading us to identify
\begin{align}\label{TtoX}
T_+ = X_+\, , \qquad T_-  = \sigma X_-\, .
\end{align}
In particular, $\mathrm{Vol}(T_\pm)$ should always be positive for a \emph{bona fide} GK geometry, 
where note that in section \ref{sec:fibred} (see also \eqref{signsofvols}) we then have $\mathrm{Vol}(X_-)=\sigma \mathrm{Vol}(T_-)$.

We now want to re-express $\mathrm{Vol}(\StoX_\pm)$ in terms of master volume $\mathcal{V}_{2n-1}$ for the fibres $\X$. 
First consider $\StoX_+$. We begin by
noting that the volume of $\StoX_+$ is obtained from the Euclidean volume of the $(n-1)$-dimensional
polytope $\mathcal{P}_+$, where one imposes
\begin{align}\label{Tpeqns}
\sum_{\mu=0}^n (y^\mu-{y}^\mu_*){v}_{+\mu}= \lambda_+~, \qquad \sum_{\mu=0}^n({y}^\mu-{y}^\mu_*){v}_{a\mu}\, \geq \, \lambda_a~.
\end{align}
Notice that since ${v}_{a\mu}=(0,{v}_{ai})$, this last inequality is exactly the same 
as the one for the $n$-dimensional polytope:
\begin{align}\label{ylambdaa}
\sum_{i=1}^n({y}^i-{y}^i_*){v}_{ai}\geq  \lambda_a~
\,.
\end{align}
On the other hand, the first equation in \eqref{Tpeqns} reads
\begin{align}\label{y4eqn}
y^0m_+ +\sum_{i=1}^n({y}^i-{y}^i_*){v}_{+i}\, = \ \lambda_+~.
\end{align}
We can solve this for $y^0$ and then substitute into \eqref{reebhypcond2}
to write the Reeb hyperplane condition in the form
\begin{align}\label{3dReebwithlambdap}
\sum_{i=1}^n({y}^i-{y}^i_*)\left({b}_i-\frac{b_0}{m_+}{v_{+i}}\right)= -\frac{b_0}{m_+}\lambda_+~.
\end{align}
When $\lambda_+=0$, notice this is an $n$-dimensional Reeb hyperplane equation, where we have shifted the $n$-dimensional Reeb vector
${b}_i\rightarrow {b}_i-\frac{b_0}{m_+}{v}_{+i}$, which was already introduced in \eqref{twistedblamfirst}.
We can also absorb the right hand side of \eqref{3dReebwithlambdap} into 
a new ${y}_*^i$, which recall is an arbitrary point in the $n$-dimensional Reeb hyperplane. 
It is then convenient to make the shift
\begin{align}
{y}^i_*\, \rightarrow \, {y}^i_*+\left(\frac{b_0}{m_+ b_1 - b_0}\lambda_+,0,\dots,0\right)  \equiv {y}_*^{(+)i}~,
\end{align}
to then find that the two conditions \eqref{3dReebwithlambdap} and \eqref{ylambdaa}, and hence
\eqref{Tpeqns} along with the Reeb hyperplane condition,
can be written in the equivalent form:
\begin{align}\label{lowerdimpoly}
\sum_{i=1}^n({y}^i-{y}_*^{(+)i}){b}^{(+)}_i&= 0~,\qquad\qquad
\sum_{i=1}^n ({y}^i-{y}_*^{(+)i})v_{ai}\geq \lambda_a^{(+)}~,\qquad\qquad
\end{align}
where we have also defined
\begin{align}
{b}^{(+)}_i&\equiv {b}_i-\frac{b_0}{m_+}{v}_{+i}\,,\qquad\qquad
\lambda^{(+)}_a\equiv \lambda_a + \frac{b_0}{m_+b^{(+)}_1}\lambda_+\,.
\end{align}

Now \eqref{lowerdimpoly} are precisely the conditions for 
specifying an $(n-1)$-dimensional polytope $\mathcal{P}_+$ 
associated with the fibre $X_{2n-1}/\mathbb{Z}_{m_+}$ sitting at the pole of the spindle with orbifold singularity $\mathbb{Z}_{m_+}$, as a function of the Reeb hyperplane vector 
${b}^{(+)}_i$ and K\"ahler class parameters $\lambda^{(+)}_a$. 
If $\mathcal{V}_{2n-1}(b_i;\lambda_a)$ 
denotes the master volume of $\X$ as a function of Reeb vector $b_i$ and K\"ahler class parameters $\lambda_a$, then 
the above analysis shows that 
\begin{align}\label{volSpmeqn}
\mathrm{Vol}(\StoX_+) = -\frac{1}{2\pi}\frac{\partial\mathcal{V}_{2n+1}}{\partial\lambda_+} = \frac{1}{m_+}\mathcal{V}^+_{2n-1}\, ,
\end{align}
where we have defined
\begin{align}
\mathcal{V}_{2n-1}^+\equiv\mathcal{V}_{2n-1}({b}^{(+)}_i;\lambda^{(+)}_a )\, .
\end{align}
Here the factor of $1/m_+$  on 
the right hand side of \eqref{volSpmeqn} is 
because the corresponding fibre 
is not $\X$, but rather $\StoX_+=\X/\Z_{m_+}$. 

A nearly identical analysis goes through for $\StoX_-$, and we can summarize the final results for both cases 
as
follows\footnote{
As already remarked, the polytope $\mathcal{P}$ is convex only in the twist case with $\sigma=1$, with the
anti-twist formulae with $\sigma=-1$ formally obtained from the twist formulae by replacing $m_-$ with $-m_-$ in the toric 
data \eqref{Y7data1}.}
\begin{align}\label{keyconn}
\mathcal{V}_{2n-1}^+\equiv\mathcal{V}_{2n-1}({b}^{(+)}_i;\lambda^{(+)}_a ) \, &= \, -\frac{m_+}{2\pi}\frac{\partial \mathcal{V}_{2n+1}}{\partial\lambda_+} = 
m_+ \mathrm{Vol}(T_+)~,\nn\\
\mathcal{V}_{2n-1}^-\equiv\mathcal{V}_{2n-1}({b}^{(-)}_i;\lambda^{(-)}_a)\,  &= -\frac{\sigma m_-}{2\pi}\frac{\partial \mathcal{V}_{2n+1}}{\partial\lambda_-} = 
\sigma m_- \mathrm{Vol}(T_-)~,
\end{align}
where we have defined the shifted vectors 
\begin{align}
{b}^{(+)}_i&={b}_i-\frac{b_0}{m_+}{v}_{+i}\,,\qquad\quad
\lambda^{(+)}_a=\lambda_a + \frac{b_0}{m_+b^{(+)}_1}\lambda_+\,,\nn\\
{b}^{(-)}_i&={b}_i+\frac{b_0}{\sigma m_-}{v}_{-i}\,,\quad\quad\ 
\lambda^{(-)}_a=\lambda_a - \frac{b_0}{\sigma m_-b_{1}^{(-)}}\lambda_-\,.
\label{twistedblam}
\end{align}
with ${b}^{(\pm)}_i$ already introduced in \eqref{twistedblamfirst}.
From \eqref{keyconn} we also immediately deduce that 
\begin{align}\label{secderivsv7}
\frac{\partial^2 \mathcal{V}_{2n+1}}{\partial\lambda_+\partial\lambda_a}&=-\frac{2\pi}{m_+}\frac{\partial\mathcal{V}_{2n-1}^+}{\partial \lambda_a}\,,\qquad\qquad\quad \ 
\frac{\partial^2 \mathcal{V}_{2n+1}}{\partial\lambda_-\partial\lambda_a}=-\frac{2\pi}{\sigma m_-}\frac{\partial\mathcal{V}_{2n-1}^-}{\partial \lambda_a}\,,\nn\\
\frac{\partial^2 \mathcal{V}_{2n+1}}{\partial\lambda_+^2}&=-\frac{2\pi b_0}{m_+^2b^{(+)}_1}\sum_{a=1}^d\frac{\partial\mathcal{V}_{2n-1}^+}{\partial \lambda_a}\,,\qquad
\frac{\partial^2 \mathcal{V}_{2n+1}}{\partial\lambda_-^2}
=
+\frac{2\pi b_0}{ m_-^2b^{(-)}_1}
\sum_{a=1}^d \frac{\partial\mathcal{V}_{2n-1}^-}{\partial \lambda_a} \,,\nn\\
\frac{\partial^2 \mathcal{V}_{2n+1}}{\partial\lambda_+\partial\lambda_-}&=0\,.
\end{align}
The last result is the statement that $\StoX_+$ and $\StoX_-$ do not intersect, which is obvious geometrically as they are distinct fibres of a fibration. 
Since $\mathrm{Vol}(T_\pm)>0$ we also note from \eqref{keyconn} that $\mathcal{V}_{2n-1}^+>0$ and $\sigma  \mathcal{V}_{2n-1}^->0$.

We next consider the
identity \eqref{id7dvol} satisfied by the master volume $\mathcal{V}_{2n+1}$.
The $\mu=1$ component of this vector equation is trivial. The $\mu=0$ component, which will play a key role in a moment, can be written as 
\begin{align}\label{keylemma}
	b_0\sum_{a=1}^d\frac{\partial\mathcal{V}_{2n+1}}{\partial\lambda_a}    = \bigg(m_+ {b_1}- b_0\bigg)\frac{\partial\mathcal{V}_{2n+1}}{\partial\lambda_+}-\bigg(\sigma m_-\,{b_1} + b_0\bigg)\frac{\partial\mathcal{V}_{2n+1}}{\partial\lambda_-} \,.
\end{align}
We do not explicitly write out the remaining components here.

\subsection{Geometric extremization for $X_{2n-1}\, \hookrightarrow\, Y_{2n+1} \, {\rightarrow}\, \Sigma$}\label{sec53}

We now have all the ingredients to translate the geometric extremization procedure for $Y_{2n+1}$, summarized in section \ref{sec:sumgeomext},
in terms of the shifted master volumes on the fibres $X_{2n-1}$. We first consider the supersymmetric action. From the definition
\eqref{tent_cc} and then using the identity satisfied by the master volume \eqref{keylemma}
we immediately have 
\begin{align}
S_{\text{SUSY}}=-\sum_{A=1}^D\frac{\partial \mathcal{V}_{2n+1}}{\partial \lambda_{A}} 
&=-\frac{b_1}{b_0} \bigg(m_+\,\frac{\partial\mathcal{V}_{2n+1}}{\partial\lambda_+} - \sigma m_-\,\frac{\partial\mathcal{V}_{2n+1}}{\partial\lambda_-}\bigg)\nn\\
&= \frac{2\pi b_1}{b_0}\left[m_+\mathrm{Vol}(\StoX_+)-\sigma m_-\mathrm{Vol}(\StoX_-)\right]\,,
\end{align}
and we notice that the last expression is in exact alignment with the general result
\eqref{generalblock} that we proved earlier,  
taking into account the orientations in \eqref{TtoX}.
Hence, using \eqref{keyconn}, we obtain the key result 
\begin{align}\label{keygravresult}
	\boxed{
	S_{\text{SUSY}}	= 2\pi\frac{b_1}{b_0} \left(
		\mathcal{V}_{2n-1}^+ - \mathcal{V}_{2n-1}^- \right) 
		}
\end{align}
and recall that $S_{\text{SUSY}}>0$.

We next consider the fluxes on $\Y$. We denote these by $\NAS_A=(\NAS_+,\NAS_-,\Nas_a)$, which are the fluxes over the 
corresponding toric codimension two submanifolds $T_A\subset \Y$, where $T_A=(T_+,T_-,T_a)$. In particular 
\begin{align}
\NAS_\pm \, \equiv \, \frac{1}{\nu_n}\int_{\StoX_\pm} \eta\wedge \rho \wedge \frac{J^{n-2}}{(n-2)!}\, .
\end{align}
From \eqref{TtoX} we take
\begin{align}\label{Npmchange}
\NAS_+ = \Np\, , \qquad \NAS_- = \sigma \Nm\, ,
\end{align}
where $\Npm>0$ were the fluxes introduced in section \ref{sec:fibred}.
The $T_a$ may be identified with 
$\Sigma_a$ in \eqref{Sigmaa}, and are the total spaces of toric codimension two submanifolds $S_a\subset \X$, fibred 
over the spindle $\Sigma$, with $a=1,\ldots,d$. 

  We first note that
the linear relations satisfied by these fluxes given in \eqref{constraintfluxes} can be written in the equivalent form
\begin{align}\label{homfluxes}
m_+ \NAS_+ & = \sigma m_- \NAS_-\equiv N~,\nn\\
\sum_{a=1}^d \Nas_a&=-\frac{\sigma m_+ + m_-}{ m_+  m_-}N\,,\nn\\
\sum_{a=1}^d \vec{w}_{a}\Nas_{a}&= \frac{N}{ m_+  m_-}\vec{p}\,,
\end{align}
and these include the constraint equation $\sum_A \NAS_A=0$.
The first line implies $m_+[\StoX_+]= \sigma m_-[\StoX_-]\in H_{2n-1}(\Y,\R)$ that we mentioned earlier.
As an aside notice that for the twist case, $\sigma=+1$, the second line implies that some of the $\Nas_a<0$.
We can also easily find expressions for the fluxes $\NAS_A$ in terms of $\mathcal{V}_{2n-1}$. Indeed from the expression for the $\NAS_A$ in terms of derivatives of $S_{\text{SUSY}}$ given in \eqref{tent_cc}
we immediately deduce
\begin{empheq}[box=\fbox]{align}\label{N7exps}
m_+\NAS_+&=-\frac{1}{\nu_n}\frac{b_1}{b^{(+)}_1}\sum_{a=1}^d\frac{\partial\mathcal{V}_{2n-1}^+}{\partial \lambda_a}\, ,\nn\\
\sigma m_- \NAS_-&=-\frac{1}{\nu_n}\frac{b_1}{b^{(-)}_1}\sum_{a=1}^d\frac{\partial\mathcal{V}_{2n-1}^-}{\partial \lambda_a}\, , \nn\\
\Nas_a&=-\frac{1}{\nu_n}\frac{b_1}{b_0}\bigg(
\frac{\partial\mathcal{V}_{2n-1}^+}{\partial \lambda_a} - \frac{\partial\mathcal{V}_{2n-1}^-}{\partial \lambda_a}\bigg)
\end{empheq}
The constraint equation can be recast in the form
\begin{align}\label{N7exps2con}
\boxed{
\frac{1}{b_1^{(+)}}\sum_{a=1}^d\frac{\partial\mathcal{V}_{2n-1}^+}{\partial \lambda_a}=
\frac{1}{b_1^{(-)}}\sum_{a=1}^d \frac{\partial\mathcal{V}_{2n-1}^-}{\partial \lambda_a}
}
\end{align}
which ensures $m_+ \NAS_+  = \sigma m_- \NAS_-\equiv N$. 

Due to the constraints \eqref{homfluxes}, the set of toric fluxes $\NAS_A=(\NAS_+,\NAS_-,\Nas_a)$ is overdetermined, 
and it is sometimes convenient to instead work with a minimal set of unconstrained quantized fluxes $\mathcal{N}_\alpha$, 
as originally introduced in equation \eqref{fluxquantize}.
 We may make contact between the two descriptions 
using the results of appendix \ref{app:A}. Here the index on $\mathcal{N}_\alpha$ may be identified 
as $\alpha=\hat{I}=0,1,\ldots,d-n$, labelling a basis for the free part of $H_{2n-1}(\Y,\Z)$,
while recall that $A=1,\ldots,D=d+2$. We then have the general 
homology relation in toric geometry (see {\it e.g.} \cite{Franco:2005sm})
\begin{align}\label{refeqn}
{\NAS}_{A} = \sum_{I=0}^{d-n} Q^A_{\hat{I}}\,  \mathcal{N}_{\hat{I}}\, .
\end{align}
Written out explicitly using the charge matrix $Q^A_{\hat{I}}$, 
satisfying $\sum_{A=1}^D Q^A_{\hat{I}}\, {v}_{A\mu} = 0$, and deduced in appendix \ref{app:A}, this reads
\begin{align}\label{toricfluxrelations}
\NAS_+ & = m_-\,  \mathcal{N}_0\, , \qquad \NAS_- = \sigma m_+\,  \mathcal{N}_0\, ,\qquad 
\Nas_a =  q_0^a \, \mathcal{N}_0 + \sum_{I=1}^{d-n}q_I^a\,  \mathcal{N}_I\, .
\end{align}
Here $q_I^a$ is the corresponding charge matrix for the toric fibres $\X$, 
satisfying $\sum_{a=1}^d q^a_{{I}}\, {v}_{ai} = 0$,
and 
$q_0^a$ satisfies $\sum_{a=1}^d q_0^a\, v_{ai} = p_i$, where recall the integers
$p_i$ determine the fibration of $\X$ over $\Sigma$. Notice that the $q_0^a$ are not unique, in general,  
and can be replaced with $q_0^a\to \sum_{I=1}^{d-n} c_Iq^a_I$ where $c_I$ are constants. 
In turn this shifts the ``baryonic fluxes'' $\mathcal{N}_I$, via $\mathcal{N}_I\to c_I\mathcal{N}_0 $
showing that they are not unique either, and we comment on this further in section \ref{bhsec9}.

Given that in \eqref{homfluxes} 
we identify $m_+\NAS_+=\sigma m_- \NAS_- = N$, where recall $\sigma^2=1$, the first two equations in \eqref{toricfluxrelations} allow us to identify
\begin{align}\label{expfornzero}
\mathcal{N}_ 0 = \frac{N}{m_+m_-}\, .
\end{align}
Since $\mathcal{N}_0$ is required to be a positive integer \eqref{fluxquantize}, note this shows that $N$ is divisible by $m_+m_-$, 
as commented earlier, below \eqref{defN}. 
The toric flux numbers $\Nas_a$ thus satisfy the last equation in \eqref{toricfluxrelations}. 
In appendix~\ref{app:A} we further introduce the (non-unique) integers $\alpha_a^j$ that satisfy 
\begin{align}\label{alphadef}
\sum_{a=1}^dv_{ai} \alpha_a^j = \delta_{i}^j\,.
\end{align}
In terms of the gauged linear sigma model description, 
$\alpha_a^j$ is the charge of the $a$th coordinate on $\C^d$ under the $j$th $U(1)$ in the toric $U(1)^n$
action on the fibres $\X$. Using this, we may then write $q_0^a=\sum_{i=1}^n p_i \alpha_a^i$, and 
hence the last equation in \eqref{toricfluxrelations} reads
\begin{align}\label{yetanotherflux}
\Nas_a = \frac{N}{m_+m_-} \sum_{i=1}^n {p_i} \alpha_a^i  + \sum_{I=1}^{d-n}q_I^a\,  \mathcal{N}_I\, .
\end{align}
This expresses the toric flux numbers $\Nas_a$ directly in terms of the ``flavour fluxes'' $p_i$, 
and 
a  set of 
``baryonic fluxes'' $\mathcal{N}_I$, one for each independent cycle, together with $N$ and the toric data. 
 Although as we comment further in section \ref{bhsec9}, the $\mathcal{N}_I$ depend  on 
an (arbitrary) choice of the $\alpha_a^j$ satisfying \eqref{alphadef}, and different choices lead to 
different~$\mathcal{N}_I$. 

Although we shall not work through the details, the flavour twist case of section \ref{nobar} is obtained by imposing the condition \eqref{kahlerclass}, which in the toric setting amounts to set all the $\lambda_a\equiv \lambda$ to be equal in a particular gauge. 
To see this, from \eqref{omegalambda} and \eqref{rhoca} we can write the transverse K\"ahler class for $\Y$ in the form
$[J] = -2\pi( \lambda_+c_++\lambda_+c_-+\lambda\sum_ac_a)$, while the first Chern class of the foliation is 
$[\rho] = 2\pi( c_++c_-+\sum_ac_a)$. In particular it is clear that \eqref{kahlerclass} is satisfied on the toric fibres at the poles.

\subsection{R-charges of baryonic operators}\label{rchgetoricsec}

We now provide an expression for the geometric R-charges that are associated with certain supersymmetric cycles $T_{\pm a}$, $a=1,\dots, d$, of codimension four on $Y_{2n+1}$, which were introduced in section \ref{gravblockformrchgetoo}. Recall that these are defined as copies of $U(1)^n$ invariant 
codimension two submanifolds
$S_a\subset\X$, whose cones are divisors in the Calabi-Yau cone $\X$, over each
of the two poles of the spindle. 
In the notation of section \ref{sec:fibred} we have 
\begin{align}\label{TpmtoS}
T_{+a} = S_a^+\, , \qquad T_{-a} = \sigma S_a^-\, , 
\end{align}
precisely as in \eqref{TtoX}. 
In the case of $n=3,4$ the geometric R-charges are dual to the R-charges of baryonic operators associated with D3, M5-branes wrapping these cycles, respectively.

In section \ref{sec:master} we introduced $c_A\in H^2_B(\mathcal{F}_\xi)$ to be basic representatives 
of integral classes in $H^2(Y_{2n+1},\mathbb{Z})$, which are Poincar\'e dual to 
the $D$ toric divisors $\mathcal{D}_A$ on $C(Y_{2n+1})$. We write $c_A=\{c_\pm,c_a\}$, with
$c_\pm$ Poincar\'e dual to the divisors $\mathcal{D}_\pm$. The codimension four supersymmetric cycles
$T_{\pm a}$ are then Poincar\'e dual to $c_\pm\wedge c_a$. 
Then using the result \eqref{UBBnew} in the definition of the R-charges \eqref{Rpmdef0} we deduce\footnote{Note that the normalization of the geometric R-charges here is slightly different to what has appeared before in the literature. For example, setting $m_\pm=1$, $\sigma=+1$ and $b_0=0$, we have the set-up associated with
a Sasaki-Einstein space fibred over a sphere. In this case $\NAS_\pm=N$ and the geometric R-charges here differ from those in \cite{Gauntlett:2018dpc,Gauntlett:2019roi} by a factor of $N$.}
  \begin{align}\label{Rpmdef}
  R_a^\pm &\equiv \frac{4 \pi }{\nu_n \NAS_\pm} \, \int_{T_{\pm a}} \eta \wedge \frac{J^{n-2}}{(n-2)!} \,\nn\\
  &=\frac{1}{\NAS_\pm}\frac{2}{\nu_n}\frac{1}{(2\pi)}\frac{\partial^2\mathcal{V}_{2n+1}}{\partial\lambda_\pm \partial\lambda_a}~.
    \end{align}
 Notice that these agree with the expressions defined in \eqref{Rpmdef0}, with the different choices 
   of orientation in the cycles in \eqref{TpmtoS} cancelling due to the division by 
   $\NAS_\pm$ in \eqref{Rpmdef}, rather than $\Npm$ in \eqref{Rpmdef0},  where these are similarly related via \eqref{Npmchange}.
Since $\NAS_+>0$ and $\sigma \NAS_->0$ from the first line of \eqref{Rpmdef} we have
   \begin{align}
   \boxed{
     R_a^+ >0,\qquad \sigma R_a^->0
     }
     \end{align}
    We will prove in a moment that these R-charges satisfy the identities
\begin{align}\label{Rpmidentities1}
\boxed{\sum_{a=1}^d R^+_a {v}_{a\mu} \,  =\, \frac{2}{b_1}\left(b_\mu - \frac{b_0}{m_+}{v}_{+\mu}\right),\qquad
\sum_{a=1}^d R^-_a {v}_{a\mu} \,  =\, \frac{2}{b_1}\left(b_\mu + \frac{b_0}{\sigma m_-}{v}_{-\mu}\right)}
\end{align}
  
 Using the results in the previous subsection we can now express the geometric R-charges in terms of $\mathcal{V}_{2n-1}$. Indeed using \eqref{secderivsv7} we have
\begin{align}
\label{Rapmequiv}
\boxed{
R_a^\pm 
=  -\frac{1}{N}\frac{2}{\nu_n}\frac{\partial\mathcal{V}_{2n-1}^\pm}{\partial \lambda_a}
}
\end{align}
In addition, using the expressions for the $\NAS_A$ given in \eqref{N7exps}
we immediately derive the result 
\begin{align}\label{Rpmidentities12}
\boxed{
\sum_{a=1}^d R^+_a  =\, 2- \frac{2b_0}{b_1m_+}~,\qquad
\sum_{a=1}^d R^-_a  =\, 2 + \frac{2b_0}{b_1\sigma m_-}}
\end{align}
and we also note that
\begin{align}\label{dconstraints}
\boxed{R^+_a-R^-_a=\frac{2b_0}{b_1}\frac{\Nas_a}{N}}
\end{align}
agreeing with the general formula \eqref{fluxes} derived earlier.
It is also interesting to highlight that in the toric case, from 
\eqref{Rpmidentities12} we can write
\begin{align}\label{Rpmidentities123}
\boxed{\frac{1}{2}\sum_{a=1}^d \left(R_a^++ R_a^-\right) = 2 - \frac{m_--\sigma m_+}{m_+m_-}\frac{b_0}{b_1}}
\end{align}
Note that for the toric anti-twist case, with $\sigma=-1$, this expression is the same as what we saw
in the universal anti-twist case in \eqref{Nauniversaltwist3}.

Observe that \eqref{Rpmidentities12} is actually the $\mu=1$ component of \eqref{Rpmidentities1}.
Notice that the $\mu=0$ component of \eqref{Rpmidentities1} is trivially satisfied.
Thus, to complete the proof of \eqref{Rpmidentities1} we therefore just need to check the remaining components which read
\begin{align}\label{Rpmidentities3}
\sum_{a=1}^d R^+_a \vec{v}_a  & = \frac{2}{b_1}\vec{b}^{(+)}~,\qquad
\sum_{a=1}^d R^-_a \vec{v}_a   = \frac{2}{b_1}\vec{b}^{(-)}~.
\end{align}
Using the expressions for $R^\pm_a$ given in
\eqref{Rapmequiv}, we find that these are identities after using the fact that the $(2n-1)$-dimensional master volumes
$\mathcal{V}_{2n-1}(\vec{b}^{(\pm)};\lambda^{(\pm)}_a)$ satisfy the identity 
\begin{align}
\sum_{a=1}^d \bigg(\vec{v}_a-\frac{\vec{b}^{(\pm)}}{b_1^{(\pm)}}\bigg)\frac{\partial \mathcal{V}^\pm_{(2n-1)}}{\partial\lambda_a} = 0~,
\end{align}
which is
the $(2n-1)$-dimensional version of \eqref{id7dvol}.

\subsection{The limit when $b_0=0$, $m_\pm=1$}\label{mpmeqonecase}

It is interesting to consider taking the limit $b_0\to 0$ in the expressions for the 
geometric extremization problem above. In particular, if we set $m_\pm=1$, and $\sigma=+1$
the spindle is then a two-sphere and we could expect to recover the results for GK geometry fibred over a Riemann surface of genus $g=0$ with a topological twist. In particular, in this set-up there is no mixing of the R-symmetry
vector with the $U(1)$ action on the two sphere and hence $b_0=0$.

We first note that setting $m_\pm =1$ and taking the $b_0\to 0$ limit of the off-shell supersymmetric action
\eqref{keygravresult} we find
\begin{align}\label{tent_cc2}
	\lim_{b_0\to 0}S_\text{SUSY}&=
	-2\pi b_1\sum_{i=1}^n(-{p}_i)\frac{\partial\mathcal{V}_{2n-1}}{\partial b_i}
	-{A}\sum_{a=1}^d\frac{\partial \mathcal{V}_{2n-1}}{\partial\lambda_a}\,,		\end{align}
	with $-p_1=+2$, and 
where we have defined
\begin{equation}
	A \, \equiv\,  -2\pi \left({\lambda_+}+{\lambda_-}\right) .
\end{equation}
Note that in these expressions we should take derivatives before setting $b_1=2/(n-2)$.
In addition, from \eqref{N7exps} for the fluxes, 
in the $b_0\rightarrow 0$ limit we obtain
\begin{align}
	 N&=-\frac{1}{\nu_n}\sum_{a=1}^d \frac{\partial\mathcal{V}_{2n-1}}{\partial \lambda_a}\,,\nn\\
	 \Nas_a &=  \frac{1}{\nu_n}\left[
	\frac{A}{2\pi}\sum_{b=1}^d\frac{\partial^2\mathcal{V}_{2n-1}}{\partial\lambda_a\partial\lambda_b}+
	b_1\sum_{i=1}^n (-{p}_i)\frac{\partial^2\mathcal{V}_{2n-1}}{\partial\lambda_a\partial b_i}
	\right]\,,
\end{align}
and hence the constraint equation $\sum_A \NAS_A=0$ becomes in this limit 
\begin{align}
	0=
-2\pi (-{p}_1)	\sum_{a=1}^d\frac{\partial\mathcal{V}_{2n-1}}{\partial \lambda_a}+
	{A}\sum_{a,b=1}^d\frac{\partial^2\mathcal{V}_{2n-1}}{\partial\lambda_a\partial\lambda_b}+
	2\pi b_1\sum_{a=1}^d\sum_{i=1}^n (-{p}_i)\frac{\partial^2\mathcal{V}_{2n-1}}{\partial\lambda_a\partial b_i} \,.
\end{align}
Finally, the R-charges \eqref{Rapmequiv} are given by
\begin{align}
\label{Rapmequiv2}
\lim_{b_0\to 0} R_a^\pm =  -\frac{1}{N}\frac{2}{\nu_n}\frac{\partial\mathcal{V}_{2n-1}}{\partial \lambda_a}\,.
\end{align}
These are precisely the formulae derived in
\cite{Gauntlett:2018dpc,Gauntlett:2019roi,Gauntlett:2019pqg}, 
associated with a Sasaki-Einstein space fibred over a Riemann surface of genus $g=0$,
after taking into account that the R-charges differ by an extra factor of $N$ as noted before, as well
as identifying $-p_i$ here with $n_i$ there.

We expect that setting $m_\pm=1$ implies that we must have $b_0=0$ (which we assumed above). For the case
$n=3$ we can explicitly show that this is true in section~\ref{sec:nvs}.

\section{Matching AdS$_3\times Y_7$ solutions with field theory}\label{secftmatch}

This section will focus on the toric set up of the previous section with $n=3$. This is 
associated with AdS$_3\times Y_7$ solutions of type IIB supergravity, where $Y_7$ is a toric GK geometry consisting of 
a fibration of a toric Sasaki-Einstein space $X_5$ over a spindle. These solutions can be interpreted as being dual to the $\mathcal{N}=1$, $d=4$ SCFT, which is dual to AdS$_5\times X_5$, that is then compactified on the spindle with magnetic fluxes switched on.
We determine the explicit map between the field theory variables involved in $c$-extremization and those appearing in
the extremization of the GK geometry. We also illustrate some of our formalism explicitly by considering
the examples when $X_5=S^5 $ and also $T^{1,1}$.

\subsection{$\mathcal{N}=1$, $d=4$ SCFT on $\mathbb{R}^{1,1}\times \Sigma$}\label{secqftcalc}

Consider the four-dimensional $\mathcal{N}=1$ SCFT that is associated with $N$ D3-branes sitting at the apex of the Calabi-Yau 3-fold cone 
$C(X_5)$, with toric vectors $v_{ai}$, $a=1,\dots, d$ and $i=1,\dots3$. Recall that the SCFT has, generically\footnote{In some cases this
is enlarged to a non-Abelian symmetry.} , $U(1)^d$ symmetry with
$U(1)^{d-3}$ baryonic symmetry associated with the number of independent three cycles $d-3=\dim\,  H_3(X_5,\R)$ on the Sasaki-Einstein manifold $X_5$.
We then place the SCFT on a spindle with background magnetic fluxes for the $U(1)^d$ symmetry, associated with either a twist or an anti-twist to preserve supersymmetry. Assuming that the resulting field theory flows to a $d=2$, $\mathcal{N}=(0,2)$ SCFT in the IR, we can extract the central charge, as well as the R-charges of  the baryonic operators, using $c$-extremization \cite{Benini:2012cz}. The extremization can be carried out in two stages by first extremizing over the baryonic directions first.
The resulting central charge after the first stage, which is still off-shell, is what can be matched with the off-shell gravitational computation, as we shall show in detail.

Let  $\Delta_a$ be the trial R-charges of the fields that are associated with the toric divisors of $C(X_5)$  -- see, for example, \cite{Franco:2005sm}, \cite{Butti:2005vn}
for a general discussion of quiver gauge theories dual to toric $C(X_5)$, and  this trial R-charge assignment.
 The $\Delta_a$ 
satisfy the constraint
\begin{align}\label{Deltacon1}
\sum_{a=1}^d \Delta_a = 2~,
\end{align}
which is equivalent to the statement that the superpotential of the theory has R-charge 2. 
We also turn on background magnetic fluxes for the $U(1)^d$ symmetry, 
with field strengths $F_a$, $a=1,\ldots,d$,
given by 
\begin{align}
\kf_a = \frac{1}{2\pi}\int_\Sigma F_a\, ,
\end{align}
where the $\kf_a$ are integers divided by $(m_-m_+)$.
These may be viewed 
as twisting the associated fields into sections of $\mathcal{O}(\kf_a)$ over the spindle, where to preserve supersymmetry we have the constraint 
\begin{align} \label{constraint_ft1}
\sum_{a=1}^d \kf_a= -\frac{\sigma m_++ m_-}{m_+ m_-}~,
\end{align}
associated with the twist and the anti-twist case when $\sigma=\pm 1$, respectively.

We next define the field theory quantities\footnote{These were labelled $\Delta_a^{(1)}$, $\Delta_a^{(2)}$ in \cite{Hosseini:2021fge}.}
\begin{align}\label{Deltapm1}
\Delta_a^+ & \, \equiv \,  \Delta_a + \frac{1}{2}\varepsilon \left(\kf_a - \frac{r_a}{2}\frac{m_- -\sigma m_+}{m_- m_+}\right)\,,\nn\\
\Delta_a^-& \, \equiv \, \Delta_a -\frac{1}{2}\varepsilon \left(\kf_a + \frac{r_a}{2}\frac{m_- -\sigma m_+}{m_- m_+}\right)\,,
\end{align}
with $r_a$ a set of arbitrary variables satisfying the constraint 
\begin{align}
\sum_{a=1}^d  r_a = 2~.
\end{align}
Different choices of $r_a$ give different gauges, and 
so will drop out of final results.\footnote{In various papers the results for 
$\Delta_a$ in different gauges have been reported.} To see this, notice that the freedom we have is
\begin{align}
r_a \, \mapsto\, r_a + \delta r_a~,\qquad \mbox{with}\qquad \sum_{a=1}^d \delta r_a = 0~.
\end{align}
On the other hand, from \eqref{Deltapm1} we see that such a shift may be absorbed into the trial R-charges $\Delta_a$ via 
\begin{align}
\Delta_a \, \mapsto\, \Delta_a +\frac{\delta r_a}{2}\frac{m_- -\sigma m_+}{m_- m_+}\varepsilon~.
\end{align}
and, moreover, this shift preserves the constraint \eqref{Deltacon1}. Notice that
\eqref{Deltacon1}--\eqref{Deltapm1} imply that $\Delta^\pm_a$ satisfy 
\begin{align}\label{delpmids}
\sum_{a=1}^d \Delta^+_a = 2 - \frac{\varepsilon}{m_+}~, \qquad \sum_{a=1}^d \Delta^-_a = 2 + \frac{\varepsilon}{\sigma m_-}~.
\end{align}

Using anomaly polynomials, it has been shown that the off-shell trial central charge in the large $N$ limit is given by \cite{Hosseini:2021fge}
\begin{align} \label{ftexp}
c_{\mathrm{trial}}= \frac{3}{\varepsilon}\left(\sum_{a<b<c} (\vec v_a,\vec v_b,\vec v_c) \Delta^+_a\Delta^+_b\Delta^+_c -\sum_{a<b<c} (\vec v_a,\vec v_b,\vec v_c) \Delta^-_a\Delta^-_b\Delta^-_c\right)N^2~,
\end{align}
where $(\vec v_a,\vec v_b,\vec v_c)\equiv \det (\vec v_a,\vec v_b,\vec v_c)$.
Associated with the $d-3$ baryonic directions, we can first impose a partial set of extremization conditions
\begin{align}\label{baryonicmixing}
\sum_{a=1}^d q^a_I\, \frac{\partial c_{\mathrm{trial}}}{\partial \Delta_a} =0~, \qquad I=1,\ldots,d-3~,
\end{align}
where we recall that $q^a_I$ are the kernel vectors for the toric data for $C(X_5)$ satisfying  
\begin{align}\label{kernut}
\sum_{a=1}^d q^a_{{I}}\, {v}_{ai} = 0\, .
\end{align}
which specifies the embedding $U(1)^{d-3}\subset U(1)^d$ (see appendix \ref{app:A}). 
After solving these conditions, and substituting back into $c_{\mathrm{trial}}$, one obtains an off-shell function,
$c_{\mathrm{trial}}|_{\mathrm{baryonic}}$, of 
$d-(d-3)=3$ variables: we start with  the $d+1$ variables $\Delta_1,\ldots,\Delta_d,\varepsilon$, with one constraint \eqref{Deltacon1}, and end up 
with (say) $\Delta_1,\Delta_2,\varepsilon$, where we have eliminated $\Delta_3,\ldots, \Delta_d$ using the constraints 
\eqref{baryonicmixing}. 
After extremizing over the baryonic directions, the resulting $\Delta^\pm_a|_{\mathrm{baryonic}}$ are then also functions of trial R-charges 
$\Delta_1,\Delta_2,\varepsilon$. The on-shell results are then obtained by further extremizing over $\Delta_1,\Delta_2,\varepsilon$.
This procedure is carried out for fixed magnetic fluxes $\kf_a$.

It is interesting to highlight that the baryon mixing constraints \eqref{baryonicmixing} can be recast as
an expression that is linear in either $\Delta_a$ and $\varepsilon$. We prove this in appendix \ref{app:matching} and, for example, we
can write the conditions as
\begin{equation}\label{appbfinaleqquesttext}
	\sum_{b=1}^d \sum_{a=1}^b \sum_{c=1}^d q_I^a (\vec v_{a}, \vec v_{b}, \vec{v}_c) \left( \mathfrak{p}_b \Delta_c^+ +\mathfrak{p}_c \Delta_b^+ + \varepsilon\, \mathfrak{p}_b \mathfrak{p}_c\right)=0 \:.
\end{equation}

\subsection{Matching with GK geometry}\label{sec:match}

Recalling \eqref{homfluxes}, the rank of the field theory gauge group, $N$, is identified on the gravity side via
\begin{align}\label{match1}
N\equiv m_+ \NAS_+ & = \sigma m_- \NAS_-\,.
\end{align}
Furthermore the background magnetic fluxes for the SCFT $\kf_a$ are identified on the gravity side via 
\begin{align}\label{fluxident}
\kf_a\, \equiv \, \frac{\Nas_a}{N}\,,
\end{align}
where $\Nas_a$ characterize the quantized five-form flux through the $d$ toric five-cycles on $Y_7$ associated with the toric divisors
$\mathcal{D}_a$. Importantly, the second condition in \eqref{homfluxes} on the $\Nas_a$
implies that this identification is consistent with the field theory constraint \eqref{constraint_ft1} on the $\kf_a$.
Note also from the third condition in \eqref{homfluxes}, for consistency, we deduce that the background magnetic fluxes of the SCFT
are related to the geometric data $\vec{p}$ which specifies the fibration of the $SE_5$ over the spindle, via
\begin{align}\label{wppdictionary}
\sum_a\vec{w}_a\mathfrak{p}_a=\frac{1}{m_- m_+}\vec{p}\,.
\end{align}

This particular relation may be understood in field theory as follows. Recall that 
turning on the magnetic flux $\kf_a$, the corresponding fields $Z_a$ become sections of 
$\mathcal{O}(\kf_a)$ over the spindle $\Sigma$. On the other hand, in the gauged linear sigma 
model description of the fibres $C(X_5)$, the $Z_a$ are also sections of non-trivial bundles over the fibres $X_5$, 
 as they are (typically) charged 
under the baryonic $U(1)^{d-3}$ symmetries. 
 However, from \eqref{kernut} 
one can verify that $\prod_{a=1}^d Z_a^{v_{a i}}$ are sections of trivial bundles on the fibres $X_5$, 
that moreover have charge $\delta_{ij}$ under the $j$th toric $U(1)\subset U(1)^3$. 
The variables $\vec{p}$  then precisely describe the twisting of this 
$i$th flavour direction over the spindle $\Sigma$, for $i=2,3$, which is the equality 
\eqref{wppdictionary}. A more detailed discussion of this may be found in 
appendix B of \cite{Gauntlett:2019roi}, albeit in the case that $\Sigma=\Sigma_g$ is a 
smooth Riemann surface of genus $g$.

We also make the identification 
\begin{align}\label{bnoughtep}
\varepsilon=b_0 ~,
\end{align}
as well as 
\begin{align}\label{matchRcharges}
 \left.\Delta_a^+\right|_{\mathrm{baryonic}}=R^+_a ~,\qquad \left.\Delta_a^-\right|_{\mathrm{baryonic}}=R^-_a ~,
\end{align}
where the notation on the left hand side means that we have extremized the trial $c$-function in field theory over the baryonic 
directions, as discussed in the last subsection. We immediately notice that the conditions on the $ \left.\Delta_a^\pm \right|_{\mathrm{baryonic}}$
arising from \eqref{delpmids} are consistent with the conditions on the $R^\pm_a$ 
in \eqref{Rpmidentities12}.
Observe that $R^\pm_a$ are functions of the trial R-symmetry vector 
$(b_0, b_2,b_3)$ (with $b_1=2$, as required for a GK geometry) while the $\Delta^\pm_a|_{\mathrm{baryonic}}$ are functions of trial R-charges 
$\Delta_1,\Delta_2,\varepsilon$, after extremizing over the baryonic directions. 

The claim is that \eqref{matchRcharges} gives the change of variables between field theory and GK geometry. Notice this 
is {\it a priori} over-determined, as there are $2d$ equations in 
\eqref{matchRcharges}, but only three independent variables. However, we can prove that the identification is consistent; the proof is somewhat involved so we have presented it in appendix \ref{app:matching}. We will also show there, using the key results of the previous section, that the off-shell trial central charge in gravity \eqref{keygravresult} can be expressed in the form
\begin{equation}\label{ZexpressionRtext}
		\mathscr{Z}	= \frac{3 N^2b_1^2}{4b_0}  \sum_{a < b < c}  (\vec v_{a}, \vec v_{b}, \vec{v}_c) \left(R_a^+ R_b^+ R_c^+ - R_a^- R_b^- R_c^-\right).
\end{equation}
With the above dictionary, this is then in exact agreement with the field theory result~\eqref{ftexp}.

\subsection{New variables}\label{sec:nvs}

In field theory, instead of using the variables $\Delta_a$ subject to the constraints \eqref{Deltacon1} and \eqref{baryonicmixing}
we can use a slightly different set of variables $\varphi_a$ defined via
\begin{align}
	\varphi_a \equiv  \frac{1}{2} \, (\Delta_a^+ + \Delta_a^-) =\Delta_a+\frac{\varepsilon}{4}\frac{\sigma m_+-m_-}{ m_+ m_-} r_a\,,
\end{align}
where we used \eqref{Deltapm1} and we recall the gauge parameters satisfy $\sum_ar_a=2$. The inverse relation is $\Delta_a^\pm = \varphi_a \pm \frac{\varepsilon}{2}\mathfrak{p}_a$.
The constraints \eqref{Deltacon1}, \eqref{baryonicmixing} then become the following constraints on the $\varphi_a$:
\begin{align}\label{constr1}
	\sum_{a=1}^d \varphi_a 
	= 2 +  \frac{\varepsilon}{2} \frac{\sigma m_+-m_-}{ m_+ m_-} \,,\nn\\
	\sum_{b=1}^d \sum_{a=1}^b \sum_{c=1}^d q_I^a (\vec v_{a}, \vec v_{b}, \vec{v}_c) \left( \varphi_b \mathfrak{p}_c + \varphi_c \mathfrak{p}_b\right) =0 \,,
\end{align}
where we used \eqref{kernelconstr_simp1} to get the second expression. We also highlight that while $(\vec v_{a}, \vec v_{b}, \vec{v}_c)$ is antisymmetric in $b,c$ and $\left( \varphi_b \mathfrak{p}_c + \varphi_c \mathfrak{p}_b\right)$ is symmetric, the sum is not trivial due to
middle summation running from $a=1,\dots,b$.
The off-shell trial central charge \eqref{ftexp} can be written in terms of the $\varphi_a$ as:
\begin{equation}\label{ctrial_newvar}
		c_{\mathrm{trial}}	= 3 N^2 \sum_{a < b < c} (\vec v_{a}, \vec v_{b}, \vec{v}_c) \left(\mathfrak{p}_a \varphi_b \varphi_c + \mathfrak{p}_b \varphi_a \varphi_c + \mathfrak{p}_c \varphi_a \varphi_b + \frac{\varepsilon^2}{4}\,\mathfrak{p}_a\mathfrak{p}_b\mathfrak{p}_c\right)\,.
\end{equation}

These formulae generalize those of \cite{Hosseini:2021fge,Faedo:2021nub}, who studied the case of $X_5=S^5$, to arbitrary $X_5$. 
Indeed for the special case of $X_5=S^5$, which we will study in the next subsection, there is no baryonic symmetry 
and we just have the first constraint in \eqref{constr1}. Furthermore,
the trial central charge \eqref{ctrial_newvar} reads
\begin{equation}\label{ctrial_newvars5}
		c_{\mathrm{trial}}	= 3 N^2 \left(\mathfrak{p}_1 \varphi_2 \varphi_3 + \mathfrak{p}_2 \varphi_1 \varphi_3 + \mathfrak{p}_3 \varphi_1 \varphi_2 + \frac{\varepsilon^2}{4}\,\mathfrak{p}_1\mathfrak{p}_2\mathfrak{p}_3\right),
\end{equation}
and we have recovered the result (5.26) of \cite{Faedo:2021nub}. 

For general $X_5$, using the $\varphi_a$ variables it is straightforward to show that when the spindle becomes a two-sphere, then necessarily $\varepsilon=0$
(recall that in section \ref{mpmeqonecase} we assumed $\varepsilon=0$). 
Indeed, setting $m_\pm=1$ and $\sigma= +1$, both of the 
constraints \eqref{constr1} are independent of $\varepsilon$. Thus, the only $\varepsilon$ dependence in $c_{\mathrm{trial}}$ is the quadratic dependence in the last term in \eqref{ctrial_newvar} and hence the extremal point will necessarily have $\varepsilon=0$.
 
\subsection{Examples}
We finish this section by discussing two toric examples when $X_5=S^5$ and $T^{1,1}$, making the relation between the quantities appearing in   field theory and GK geometry that was discussed in section \ref{sec:match} very explicit. One can also check that the general formula for the associated toric $Y_7$ GK geometry discussed in section \ref{sec:fibretoric}, based on the master volume $\mathcal{V}_7$, agrees with the formula that we derived in 
section \ref{sec:toricoverspindle} using $\mathcal{V}^\pm_5$. In fact, studying these examples was very helpful in elucidating
the general results of section~\ref{sec:toricoverspindle}.
For the $S^5$ example, we also recover all of the results of 
section \ref{sec3s5ex}, where we analysed this case as an example of an $X_5$ with no baryonic symmetries.

\subsubsection{$S^5$ fibred over a spindle}\label{torics5example}
The toric data for $S^5$ can be specified by $d=3$ inward pointing normal vectors given by
\begin{align}\label{toricC31}
{v}_{1i} = (1,\vec{w}_1)~, \qquad {v}_{2i} = (1,\vec{w}_2)~, \qquad {v}_{3i} = (1,\vec{w}_3)\,,
\end{align}
where 
\begin{align}\label{toricC32}
\vec{w}_1 = (0,0)~, \qquad \vec{w}_2 = (1,0)~, \qquad \vec{w}_3= (0,1)\,.
\end{align}
There are no baryonic directions associated with $S^5$ and the kernel $q^a_I$ for the toric data is trivial.

For $S^5$ fibred over a spindle, the toric data for the associated toric GK geometry $Y_7$  is then given by $D=5$ inward pointing normal vectors given in \eqref{Y7data1}:
\begin{align}\label{s5toricvecs}
{v}_{+\mu}=(m_+,1,-a_+\vec{p})\,,\quad
{v}_{-\mu}=(-\sigma m_-,1,- \sigma a_-\vec{p})\,,\quad
{v}_{a\mu}=(0,1,\vec{w}_a)\,,
\end{align}
where $\vec{p}=(p_2,p_3)\in\mathbb{Z}^2$ and 
$a_\pm\in \Z$ satisfy $a_- m_++a_+ m_-   = 1$. The three integers
$p_i=(p_1,p_2,p_3)$, with $p_1 = - (m_- +\sigma m_+)$, specify the fibration.
There is a one-dimensional kernel $Q^A_0$, satisfying
$\sum_{A=1}^5 Q^A_0\, {v}_{A\mu} = 0$ (see appendix \ref{app:A}), given by 
\begin{align} \label{ker_S5}
Q^A_0\, & = \, (m_-,\sigma m_+, - (m_- +\sigma m_+)-p_2-p_3,p_2,p_3)~.
\end{align}
Using the vectors \eqref{s5toricvecs} we can now obtain an explicit, and lengthy, expression for the
master volume $\mathcal{V}_7$ for $Y_7$ using the formula given in (2.27) of \cite{Gauntlett:2019roi}. 

We can now carry out the geometric extremization using the procedure summarized in section \ref{sec:sumgeomext}. 
The master volume $\mathcal{V}_7$ depends on just $5-3=2$ of the 5 K\"ahler class parameters $\lambda_A$. 
We therefore solve the constraint equation 
and one of the five-form flux quantization conditions in \eqref{tent_cc} for two of the $\lambda_A$. For example, we can solve
the constraint equation and the expression for $\NAS_+$ for $(\lambda_+,\lambda_-)$. It must be the case, and indeed it
is, that the remaining K\"ahler class parameters $\lambda_1,\lambda_2,\lambda_3$ then drop out of any final formula.
The flux vector $\NAS_A$ in 
\eqref{tent_cc} is now expressed in terms of $\NAS_+, p_2, p_3$ and, in particular, we find 
$\sigma m_- \NAS_-=m_+ \NAS_+\equiv N$.  

Instead of using these three variables, it is illuminating to
express the flux vector in terms of $\NAS_1,\NAS_2,\NAS_3$ to get 
 \begin{align}\label{fluxesS5}
\NAS_A=\left \{ \NAS_+, \frac{\sigma m_+}{m_-} \NAS_+, \NAS_1,\NAS_2,\NAS_3\right\}~,\qquad
 \NAS_+= -\frac{m_-}{m_-+ \sigma m_+}\sum_{a=1}^3 \Nas_a\,.
 \end{align}
We can further rescale the $\Nas_a$ by a factor of $N= \sigma m_-\NAS_-=m_+\NAS_+$ and introduce $\kf_a$, which are to be identified 
 with the background magnetic fluxes of the SCFT shortly, via 
 \begin{align}\label{constonpss5}
\kf_a\equiv \frac{ \Nas_a}{N}~, \qquad \sum_{a=1}^3 \kf_a= -\frac{m_-+\sigma m_+}{m_+ m_-}\,.
 \end{align}
Note from \eqref{wppdictionary}, the fibration data $\vec{p}=(p_2,p_3)$ is related to the $\kf_a$ via 
 \begin{align}\label{ppdics5}
p_2=m_- m_+\mathfrak{p}_2\,,\qquad
p_3=m_- m_+\mathfrak{p}_3\,.
 \end{align}
Also, importantly, \eqref{constonpss5} implies 
\begin{align}
\mathfrak{p}_1=-(\mathfrak{p}_2+\mathfrak{p}_3)-\frac{m_-+\sigma m_+}{m_-m_+}
=\frac{p_1-p_2-p_3}{m_-m_+}\,,
 \end{align}
where recall 
$p_1 = - (m_- +\sigma m_+)$. Observe that there is a $\mathbb{Z}_3$ symmetry permuting the $\kf_a$ but
not $(p_1,p_2,p_3)$: the latter is a consequence of the fact that we singled out the $p_1$ direction in constructing the fibration.

 At this point we can now obtain an expression for the off-shell central charge $\mathscr{Z}$ from 
 \eqref{offshellz} and the expression for $S_\text{SUSY}$ in \eqref{tent_cc}. This is expressed in terms of $m_\pm$,
 $a_\pm$, $ \kf_a$, $b_0,b_1,b_2,b_3$ and we should set $b_1=2$. 
 The expression for $\mathscr{Z}$ is quadratic in $b_0,b_2,b_3$ and after extremizing
 over these variables we find that the on-shell central charge, $c_{\mathrm{sugra}}$, as calculated from the GK geometry, can be expressed 
 as
  \begin{align}\label{cC3}
c_{\mathrm{sugra}}\equiv \mathscr{Z}_\text{os} =
\frac{6m_-^2m_+^2 \kf_1 \kf_2 \kf_3}{m_-^2+m_+^2-m_-^2m_+^2(\kf_1^2+\kf_2^2+\kf_3^2) }N^2\,,
   \end{align}
 exactly as in section \ref{sec3s5ex}.
 Notice that $a_\pm$ have dropped out of the final expression.  With the master volume $\mathcal{V}_7$ in hand, it is also straightforward to obtain explicit expressions for both the off-shell and on-shell 
 R-charges $R^\pm _a$ using \eqref{Rpmdef}. The on-shell expressions can be written
  \begin{align}
   R^+_a=&\ 
-Cm_+\left(\kf_1[\sigma +m_- \kf_1],\kf_2[\sigma +m_- \kf_2],\kf_3[\sigma +m_- \kf_3]\right)\,,\nn\\
 R^-_a=&\ -Cm_- \left(\kf_1[1+m_+\kf_1],\kf_2[1+ m_+\kf_2],\kf_3[1+m_+\kf_3]\right)\,,
  \end{align}
 where
 \begin{align}
 C=\frac{2m_-m_+}{m_-^2+m_+^2-m_-^2m_+^2(\kf_1^2+\kf_2^2+\kf_3^2)}\,,
 \end{align}
 again in agreement with section \ref{sec3s5ex}.
Demanding $c_{\mathrm{sugra}}>0$ as well $ R_a^+ >0$ and $\sigma R_a^->0$
gives the restrictions on the parameters in alignment with the explicit supergravity solutions
as discussed in section \ref{sec3s5ex}.
  
  We can now compare these results with the corresponding calculations in field theory using anomaly polynomials and $c$-extremization. 
  In fact these calculations were carried out for the case of the anti-twist
 already in \cite{Ferrero:2020laf,Boido:2021szx,Ferrero:2021etw}. We consider $\mathcal{N}=4$ SYM theory dual to AdS$_5\times S^5$, 
 with $SU(N)$ gauge group.  We place the theory on a spindle with background magnetic fluxes, parametrized by $\kf_a$, for the $U(1)^3\subset SU(4)$ global symmetry with $\sum_{a=1}^3 \kf_a= -\frac{m_-+\sigma m_+}{m_+ m_-}$.
 The trial field theory central charge, $c_\text{trial}$, is parametrized by $\Delta_a$, satisfying $\sum_{a=1}^3\Delta_a=2$.
 From \eqref{ftexp} we find that $c_\text{trial}$ is explicitly given by
    \begin{align} \label{N4ctrial}
  	&c_\text{trial}= \Bigg\{ 
  	3(\Delta_1\Delta_2 \kf_3+\Delta_1\Delta_3 \kf_2+\Delta_2\Delta_3 \kf_1\nn)\\
  	&\quad -\epsilon \frac{3(m_--\sigma m_+)[(r_3\Delta_2 + r_2\Delta_3)\kf_1+(r_3\Delta_1 + r_1\Delta_3)\kf_2+(r_2\Delta_1 + r_1\Delta_2)\kf_3]}{4m_-m_+}        \nn\\
  	&\quad +3\epsilon^2\left[\frac{(r_2 r_3\, \kf_1+r_1 r_3\,\kf_2+r_1 r_2\, \kf_3)(m_--\sigma m_+)^2+9\kf_1\kf_2\kf_3 m_-^2 m_+^2)}{16m_-^2m_+^2}\right]
  	\Bigg\}N^2\,.
  \end{align}
  
We then find that $c_\text{trial}$ as a function of the three independent variables $\Delta_2, \Delta_3,\varepsilon$ (say) exactly agrees
with the off-shell gravitational charge 
$\mathscr{Z}$ as a function of
$b_0, b_2, b_3$ (with $b_1=2$), provided that we utilise the dictionary given in section \ref{sec:match}:
\begin{align}\label{Deltapm1two0}
b_0&=\varepsilon\,,\nn\\
\kf_a\quad&\leftrightarrow\quad \frac{\Nas_a}{N}\,, \nn\\
\Delta_a^\pm \,   \equiv \, \Delta_a \pm \frac{1}{2}\varepsilon \left(\kf_a \mp \frac{r_a}{2}\frac{m_- -\sigma m_+}{m_- m_+}\right)
\quad&\leftrightarrow\quad
R^\pm_a\, ,
\end{align}
as well as identifying $N$ on each side.
Explicitly, for this example the dictionary reads
\begin{align}\label{beedeltas5}
b_0&=\varepsilon\,,\nn\\
b_2&=\Delta_2+\frac{1}{2}\mathfrak{p}_2\varepsilon - \frac{r_2}{4}\left(\frac{m_--\sigma m_+}{m_- m_+}\right)\varepsilon-a_+\mathfrak{p}_2m_-\varepsilon\,,\nn\\
b_3&=\Delta_3+\frac{1}{2}\mathfrak{p}_3\varepsilon - \frac{r_3}{4}\left(\frac{m_--\sigma m_+}{m_- m_+}\right)\varepsilon-a_+\mathfrak{p}_3m_-\varepsilon\,.
\end{align}
Notice that the dictionary involves the parameters $a_+$. Recall that $a_+$ is only defined up the transformation given in
\eqref{shiftbezout}. Making this shift in \eqref{beedeltas5}, and using \eqref{ppdics5},
induces a transformation on the vector $(b_0,b_1,b_2,b_3)$ which
is precisely given by the $SL(4,\mathbb{Z})$ transformation in \eqref{sl4matrixexp} with $b_2\to b_2+\kappa p_2 b_0$ and
$b_3\to b_2+\kappa p_3 b_0$. 
 
\subsubsection{$T^{1,1}$ fibred over a spindle}

The toric data for $T^{1,1}$ can be specified by $d=4$ inward pointing normal vectors given by
\begin{align}\label{toricT111}
{v}_{1i} = (1,\vec{w}_1)~, \qquad {v}_{2i} = (1,\vec{w}_2)~, \qquad {v}_{3i} = (1,\vec{w}_3)\,,\qquad {v}_{4i} = (1,\vec{w}_4)\,,
\end{align}
where 
\begin{align}\label{toricT112}
\vec{w}_1 = (0,0)~, \qquad \vec{w}_2 = (1,0)~, \qquad \vec{w}_3= (1,1)\,,
\qquad \vec{w}_4= (0,1)\,.
\end{align}
There is a one-dimensional kernel $q^a_1$, satisfying $\sum_{a=1}^4 q^a_{{1}}\, {v}_{ai} = 0$ (see appendix \ref{app:A}), given by
\begin{align}\label{t11kernel}
q^a_1 & =  (-1,1,-1,1)~.
\end{align}

The toric data for $T^{1,1}$ fibred over a spindle, $Y_7$, is then given by $D=6$ inward pointing normal vectors
given in \eqref{Y7data1}:
\begin{align}\label{t11toricvecs}
{v}_{+\mu}=(m_+,1,-a_+\vec{p})\,,\quad
{v}_{-\mu}=(-\sigma m_-,1,-\sigma a_-\vec{p})\,,\quad
{v}_{a\mu}=(0,1,\vec{w}_a)\,,
\end{align}
where $\vec{p}=(p_2,p_3)\in\mathbb{Z}^2$ and 
$a_\pm\in \Z$ satisfy $a_- m_++a_+ m_-   = 1$. The three integers
$p_i=(p_1,p_2,p_3)$, with $p_1 = - (m_- +\sigma m_+)$, specify the fibration.
There is a two-dimensional kernel $Q^A_{\hat{I}}=(Q^A_0, Q^A_1)$, satisfying
$\sum_{A=1}^6 Q^A_{\hat{I}}\, {v}_{A\mu} = 0$ (see appendix \ref{app:A}), given by
\begin{align} \label{ker_T11}
Q^A_0\, & = \, (m_-,\sigma m_+, - (m_- +\sigma m_+)-p_2-p_3,p_2,0,p_3)~,\nn\\
Q^A_1\, & = \, (0,0,1,-1,1,-1)=(0,0,q_1^1,q_1^2,q_1^3,q_1^4)~.
\end{align}
Using the vectors \eqref{s5toricvecs} we can now obtain an explicit, and lengthy, expression for the
master volume $\mathcal{V}_7$ for $Y_7$ using the formula given in (2.27) of \cite{Gauntlett:2019roi}. 

The master volume $\mathcal{V}_7$ depends on just $6-3=3$ of the 6 K\"ahler class parameters $\lambda^A$. 
We therefore solve the constraint equation 
and two of the five-form flux quantization conditions in \eqref{tent_cc} for three of the $\lambda^A$. For example, we can solve
the constraint equation and the expressions for $\NAS_+$ and $\NAS_1$ for $(\lambda_+,\lambda_-,\lambda_1)$. It must be the case, and indeed it
is, that the remaining K\"ahler class parameters $\lambda_2,\lambda_3,\lambda_4$ then drop out of any final formula.
The flux vector $\NAS_A$ in 
\eqref{tent_cc} is now expressed in terms of $\NAS_+, \NAS_1, p_2, p_3$ and, in particular, we find $ m_+\NAS_+=\sigma m_-\NAS_-\equiv N$.  
Instead of using these four variables, it is most illuminating to
express the flux vector in terms of $\NAS_1,\NAS_2,\NAS_3,\NAS_4$ to obtain
 \begin{align}\label{fluxesT11}
\NAS_A=\left \{ \NAS_+, \frac{\sigma m_+}{m_-} \NAS_+, \NAS_1,\NAS_2,\NAS_3,\NAS_4\right\}~,\qquad
 \NAS_+= -\frac{m_-}{m_-+ \sigma m_+}\sum_{a=1}^4 \Nas_a\,.
 \end{align}
 We can further rescale the $\Nas_a$ by a factor of $N= \sigma m_-\NAS_-=m_+\NAS_+$ and introduce $\kf_a$, which are to be identified 
 with the background magnetic fluxes of the SCFT shortly, via
 \begin{align}\label{constonps}
\kf_a\equiv \frac{ \Nas_a}{N}~, \qquad \sum_{a=1}^4 \kf_a= -\frac{m_-+\sigma m_+}{m_+ m_-}\,,
 \end{align}
and also, from \eqref{wppdictionary}, we have the fibration data $\vec{p}=(p_2,p_3)$ is related to the $\kf_a$ via
 \begin{align}\label{ppdict11}
p_2=m_- m_+(\mathfrak{p}_2+\mathfrak{p}_3)\,,\qquad
p_3=m_- m_+(\mathfrak{p}_3+\mathfrak{p}_4)\,.
 \end{align}
 
We can now obtain an expression for the off-shell central charge $\mathscr{Z}$ from 
 \eqref{offshellz} and the expression for $S_\text{SUSY}$ in \eqref{tent_cc}. This is expressed in terms of $m_\pm$,
 $a_\pm$, $ \kf_a$, $b_0,b_1,b_2,b_3$ and we should set $b_1=2$. It is quadratic in $b_0,b_2,b_3$ and after extremizing
 $\mathscr{Z}$ over these variables we find that the on-shell central charge as calculated from gravity can be expressed 
 as\footnote{This corrects some typos in \cite{Hosseini:2021fge} who analysed this from field theory as explained below.}
  \begin{align}
c_{\mathrm{sugra}}& \equiv \mathscr{Z}_\text{os} \nonumber\\
 &=  \frac{3(m_-+\sigma m_+)^2 \sum_{a<b, c\ne a,b}\kf_a \kf_b \kf_c^2\sum_{a< b< c} \kf_a \kf_b \kf_c}{(m_-^2-\sigma m_-  m_++ m_+^2)\prod_{a<b}(\kf_a+\kf_b)-\sigma m_+ m_-\Theta_{KW}}N^2\,,
 \end{align}
 where 
 \begin{align}
 \Theta_{KW}=\sum_{a<b, c\ne a,b}\kf _a \kf_b \kf_c^4-2\sum_{a<b}\kf_a \kf_b\prod_c \kf_c\,.
 \end{align}
 Notice that $a_\pm$ have dropped out of the final expression.
The extremal values $b_0,b_2,b_3$ are lengthy and we don't give them here explicitly.
 With the master volume in hand, it is also straightforward to obtain explicit expressions for both the off-shell and on-shell 
 R-charges $R^\pm _a$ using \eqref{Rpmdef}. The on-shell expressions are expressed in terms of $m_\pm$ and $\kf_a$ as expected.
One then determine the ranges of allowed parameters by imposing, on-shell,
 $c_{\mathrm{sugra}}>0$, $R^+_a>0$ and $\sigma R^-_a>0$.
 
Now $H^2(T^{1,1},\mathbb{Z})\cong \mathbb{Z}$, associated with there being a single baryonic $U(1)$ symmetry in the 
Klebanov-Witten field theory dual to AdS$_5\times T^{1,1}$. Thus, $H_{5}(Y_7,\Z)=\mathbb{Z}^2$ and there are two independent fluxes $\mathcal{N}_0, \mathcal{N}_1$. 
We can therefore express $\NAS_A$ in terms of $\mathcal{N}_0, \mathcal{N}_1$, given the spindle data $m_\pm, \sigma$ and 
the fibration data $p_i$. Recall from \eqref{expfornzero}
we have $\mathcal{N}_ 0 = \frac{N}{m_+m_-}$. Also, from \eqref{ker_T11} we have $q_0^a=(-(m_-+\sigma m_+)-p_2-p_3,p_2,0,p_3)$ and 
$q_1^a=(-1,1,-1,1)$. 
Hence from \eqref{toricfluxrelations} we can write 
\begin{align}\label{t11nacalNs}
(\NAS_1,\NAS_2,\NAS_3,\NAS_4) & = \frac{N}{m_+m_-}(-(m_-+\sigma m_+)-p_2-p_3,p_2,0,p_3) \nn\\
& \quad + \mathcal{N}_1(1,-1,1,-1)\, .
\end{align}
and $\mathcal{N}_1$ is the baryonic charge. 
As we emphasized earlier, the definition of baryonic charge is not unique.
In particular, for a given $M_a$, we could also write \eqref{toricfluxrelations} in the form
 $\Nas_a =  (q_0^a+q_1^a \mathcal{N}_1/\mathcal{N}_0)\mathcal{N}_0$ and then using 
 $\tilde q_0^a\equiv q_0^a+q_1^a \mathcal{N}_1/\mathcal{N}_0$ instead of $q_0^a$ to define the baryonic charge, 
 which still satisfies $\sum_{a=1}^d \tilde q_0^a\, v_{ai} = p_i$,
 we would conclude that
 the baryonic charge vanishes.
 
  It is interesting to examine the special sub-class associated with the flavour twist that we discussed in sections
 \ref{nobar} and \ref{someexamplessec}. We first recall that the
Sasakian volume of $T^{1,1}$ is given by
 \begin{equation}\label{sasvolT11}
 	\Vol_S(T^{1,1}) = \frac{\pi^3 b_1}{b_2 b_3 (b_1 - b_2) (b_1 - b_3)} \:,
 \end{equation}
while the Sasakian volume of the three-dimensional supersymmetric submanifolds are
\begin{align}
	\Vol_S(S_1) &= \frac{2\pi^2}{b_2 b_3} \:, \qquad\qquad\qquad\qquad
	\Vol_S(S_2) = \frac{2\pi^2}{b_3 (b_1 - b_2)} \:, \nn\\
	\Vol_S(S_3) &= \frac{2\pi^2}{ (b_1 - b_2) (b_1 - b_3)} \:, \qquad\,
	\Vol_S(S_4) = \frac{2\pi^2}{b_2 (b_1 - b_3)} \:. 
\end{align}
The Sasaki volume \eqref{sasvolT11} is extremized for $b_2 = b_3 = \frac{b_1}{2}$. Setting $b_1 = 3$ we get the Sasaki-Einstein volumes which are $\Vol_{SE}(T^{1,1})=\frac{16\pi^3}{27}$ and $\Vol_{SE}(S_a) = \frac{8\pi^2}{9}$. 

 The flavour twist is obtained by demanding the condition \eqref{kahlerclass}, which in the toric setting amounts to set all the $\lambda_a$ to be equal in a particular gauge. Using the gauge transformations \eqref{lamgt}, we can fix $\gamma_2$ and $\gamma_3$ in such a way that three of the four $\lambda_a$ are equal. In order to fix the last $\lambda_a$, we need to fix one of the fluxes $\Nas_a$, {\it e.g.}
 \begin{equation}\label{N1T11}
 	\NAS_3 = \frac{\pi N}{2 b_0} \left[\frac{\Vol_S(S_3)}{\Vol_S(T^{1,1})}\bigg|_{\vec{b}^{(+)}} - \frac{\Vol_S(S_3)}{\Vol_S(T^{1,1})}\bigg|_{\vec{b}^{(-)}}\right]\, ,
 \end{equation}
 ({\it c.f.} \eqref{t11nacalNs}).
All the other fluxes are fixed by \eqref{homfluxes}, and likewise they read
 \begin{equation}
	\Nas_a = \frac{\pi N}{2 b_0} \left[\frac{\Vol_S(S_a)}{\Vol_S(T^{1,1})}\bigg|_{\vec{b}^{(+)}} - \frac{\Vol_S(S_a)}{\Vol_S(T^{1,1})}\bigg|_{\vec{b}^{(-)}}\right]\,,
\end{equation}
 exactly as we derived earlier in \eqref{Naformula}.
 
The universal anti-twist case 
 is obtained by setting \eqref{kahlerclass2} as well as \eqref{unitwistcond}. The latter combined with \eqref{newvariableri} tells us that the on-shell twisting parameters are fixed to be, with $\sigma=-1$,
 \begin{equation}
 	p_2 = p_3 = \frac{p_1}{2} = \frac{1}{2} \left(m_+ - m_-\right).
 \end{equation}
Furthermore, the condition \eqref{N1T11} implies 
 \begin{equation}\label{t11utwistcase}
	\mathfrak{p}_1 =\mathfrak{p}_2 =\mathfrak{p}_3 =\mathfrak{p}_4 = \frac{m_+ - m_-}{4m_+ m_-} \:.
\end{equation}
We then recover, as expected, all the general results of section \ref{unitwist} for the universal anti-twist, in particular the on-shell results
for $\mathscr{Z}$ given in \eqref{offshellz3} and $R^\pm_a$ given in \eqref{rpmads3caseos}
with 
\begin{equation}
	a_{4\mathrm{d}} = \frac{27N^2}{64} \:, \qquad\qquad	R_a^{4\mathrm{d}} = \frac{1}{2} \:, \quad a=1,2,3,4 \:.
\end{equation}

 We can now compare with field theory. For the case of the anti-twist the relevant field theory calculations were carried out 
 already in \cite{Hosseini:2021fge}; here we also include the twist case. We consider the quiver gauge theory dual to AdS$_5\times T^{1,1}$, with rank $N$ gauge groups. This gauge theory has $U(1)^3\times U(1)_B$ symmetry where
 $U(1)_B$ is a baryonic symmetry associated with the kernel \eqref{t11kernel}. 
 We now put this gauge theory on a spindle with background magnetic fluxes, parametrized by $\kf_a$, for the $U(1)^3\times U(1)_B$ symmetry with $\sum_{a=1}^4 \kf_a= -\frac{m_-+\sigma m_+}{m_+ m_-}$.
 The trial central charge is parametrized by $\Delta_a$, satisfying $\sum_a\Delta_a=2$ and from the general result
\eqref{ftexp} we find 
(setting here $r_1=r_2=r_3=r_4=\tfrac{1}{2}$ for simplicity)
 \begin{align} \label{KWctrial}
 c_\text{trial}&={N^2} \Big\{ 
 3[\kf_1 (\Delta_2 \Delta_3+\Delta_2 \Delta_4+\Delta_3 \Delta_4)+\kf_2 (\Delta_1 \Delta_3+\Delta_1 \Delta_4+\Delta_3 \Delta_4)\nn\\
 &\qquad+\kf_3 (\Delta_1 \Delta_2+\Delta_1 \Delta_4+\Delta_2 \Delta_4)+\kf_4 (\Delta_1 \Delta_2+\Delta_1 \Delta_3+\Delta_2 \Delta_3)]\nn\\
& -\varepsilon\frac{3(m_--\sigma m_+)}{4m_- m_+} (\kf_1(2-\Delta_1) +\kf_2(2-\Delta_2) + \kf_3(2-\Delta_3)+\kf_4(2-\Delta_4) )\nn\\
&+\varepsilon^2[-\frac{9(m_-+\sigma m_+)(m_--\sigma m_+)^2}{64m_-^3 m_+^3}         \nn\\       
& +\frac{3}{4} (\kf_1 \kf_2 \kf_3+\kf_1 \kf_2 \kf_4+\kf_1 \kf_3 \kf_4+\kf_2 \kf_3 \kf_4)]
 \Big\}N^2\,.
 \end{align}

The on-shell central charge calculated in field theory is obtained by extremizing with respect to $\Delta_a$ and $\epsilon$, subject to $\sum_a\Delta_a=2$. We can do this in two steps, 
first extremizing over the baryonic direction defined by $q^a_1 \Delta_a=\Delta_1-\Delta_2+\Delta_3-\Delta_4$, and then over the remaining three independent variables, which can taken to be, for example, $\Delta_1, \Delta_2,\varepsilon$. After the first step
we get $c_\text{trial}|_{\text{baryonic}}$ as a function of $\Delta_1, \Delta_2,\varepsilon$ and we find that there is an exact agreement with
the off-shell gravitational charge 
$\mathscr{Z}$, as a function of
$b_0, b_2, b_3$. The dictionary between the two variables is as described in section \ref{sec:match}:
\begin{align}\label{Deltapm1two}
b_0 \, & = \,\varepsilon\,,\nn\\
\kf_a\quad&\leftrightarrow\quad \frac{\Nas_a}{N}\,,\nn\\
\Delta_a^\pm|_{\mathrm{baryonic}}  \equiv  \Delta_a \pm \frac{1}{2}\varepsilon \left(\kf_a \mp \frac{r_a}{2}\frac{m_- -\sigma m_+}{m_- m_+}\right)
\quad&\leftrightarrow\quad
R^\pm_a\,,
\end{align}
as well as identifying $N$ on each side.
Explicitly, for this example we find
\begin{align}\label{reebcoc}
	b_0 \, & = \,\varepsilon\,,\nn\\
	b_2 &  = \frac{1}{\mathfrak{p}_1+\mathfrak{p}_2}\left[2\mathfrak{p}_2 - (\mathfrak{p}_2+\mathfrak{p}_3)\,\Delta_1 + (\mathfrak{p}_1+\mathfrak{p}_4)\,\Delta_2\right] + \frac{1}{2} \,\varepsilon (\mathfrak{p}_2+\mathfrak{p}_3) 
	\nn\\
	& \quad \frac{1}{4}\varepsilon\left(\frac{m_--\sigma m_+ }{m_+ m_-}\right) 
	\frac{(-r_2\mathfrak{p}_1+(r_1-2) \mathfrak{p}_2 +r_1\mathfrak{p}_3 - r_2\mathfrak{p}_4)}{\mathfrak{p}_1 + \mathfrak{p}_2} -a_+ m_
	- \varepsilon  (\mathfrak{p}_2+\mathfrak{p}_3)~,\nn\\
	b_3& = 2-\Delta_1-\Delta_2 + \frac{1}{2} \,\varepsilon (\mathfrak{p}_3+\mathfrak{p}_4) - \frac{1}{4}\,\varepsilon(r_3 + r_4) \left(\frac{m_--\sigma m_+}{m_+ m_-}\right) \nn\\ & \quad -a_+m_- \varepsilon  (\mathfrak{p}_3+\mathfrak{p}_4)~.
\end{align}
Notice that the dictionary involves the parameters $a_+$. Recall that $a_+$ is only defined up the transformation given in
\eqref{shiftbezout}. Making this shift in \eqref{reebcoc}, and using \eqref{ppdict11},
induces a transformation on the vector $(b_0,b_1,b_2,b_3)$ which
is precisely given by the $SL(4,\mathbb{Z})$ transformation in \eqref{sl4matrixexp} with $b_2\to b_2+\kappa p_2 b_0$ and
$b_3\to b_2+\kappa p_3 b_0$.  
 
 \section{Black holes in AdS$_4\times SE_7$}\label{bhsec9}

For $n=4$ we expect the AdS$_2\times Y_9$ solutions, with $Y_9$ fibred as in \eqref{fibrationformintro},
can arise as the near horizon limit of supersymmetric, accelerating black holes that carry magnetic charge and
asymptotically approach AdS$_4\times X_7$, with $X_7$ a Sasaki-Einstein manifold. 
Such solutions, and our associated entropy function \eqref{ads2entess}, have recently been discussed in 
\cite{Boido:2022iye}. In this section we make contact with that work, along with some related observations.

As explained in \cite{Boido:2022iye}, from a four-dimensional perspective the near horizon black hole solutions 
take the form AdS$_2\times \Sigma$, with the conical deficits of the spindle horizon $\Sigma$ 
being related to a non-zero acceleration parameter for the black hole \cite{Ferrero:2020twa}.
Such black holes can carry two types of magnetic charge, associated to massless gauge fields in AdS$_4$ 
with a different $D=11$ origin: one type arise from isometries of $X_7$ (``flavour symmetries''), while the other 
type arise from homology cycles of $X_7$ (``baryonic symmetries''). The flavour magnetic charges 
can immediately be identified with the twisting parameters $p_i\in \Z$: this follows directly from their 
definition in \eqref{pis}, where the $A_i$ are connection one-forms for the twisting of the
$U(1)^s$ action on $X_7$ over the spindle $\Sigma$. As already alluded to in section \ref{sec53}, the baryonic magnetic 
charges, or equivalently ``baryonic fluxes'' $\mathcal{N}_I$,  instead  require 
(arbitrary) choices to define unambiguously. This is a reflection of the well-known mixing between 
baryonic and flavour symmetries in field theory, and can also be seen directly at the level 
of Kaluza-Klein reduction. 

Specifically, as described in \cite{Boido:2022iye}, in considering 
a linear Kaluza-Klein reduction of $D=11$ supergravity on an AdS$_4\times X_7$ background, one 
has the perturbation of the six-form potential
\begin{align}\label{deltaC6}
\delta C_6 = \left[ \sum_{i=1}^s A_i\wedge \omega_i + \sum_{I=1}^{b_5(X_7)} A_I\wedge \omega_I\right]\frac{\nu_4 L^6}{2\pi} N\, .
\end{align}
Here both $\omega_i$ and $\omega_I$ are co-closed five-forms on $X_7$, but 
$\omega_I$ is closed while $\diff\omega_i=\partial_{\varphi_i}\lrcorner\,  \mathrm{vol}_{X_7}$, with 
$\mathrm{vol}_{X_7}$ a suitably normalized volume form \cite{Benvenuti:2006xg} (see also \cite{Bah:2019rgq,Bah:2020jas}).  
We are then free to shift $\omega_i\rightarrow\omega_i + \sum_{I} c_i^I\, \omega_I$, 
for arbitrary constants $c_i^I$, which is the freedom to mix baryonic symmetries into 
flavour symmetries. This is the same freedom discussed after equation \eqref{toricfluxrelations}, 
or equivalently the freedom to shift $\alpha_a^i\rightarrow \alpha_a^i +\sum_I c_i^I q_I^a$.  
This correspondingly shifts the four-dimensional gauge field as $A_I\rightarrow A_I - \sum_{i=1}^s c_i^I A_i$, 
and hence also the flux $\int_\Sigma \diff A_I/2\pi$ through the spindle horizon $\Sigma$. Although one 
might (naively) define this as the ``baryonic flux'', as shown in appendix \ref{app:KK}, this is \emph{not}
 in general the same as the ``baryonic flux'' $\mathcal{N}_I$ already defined. 

Recall that the original flux integrals $\mathcal{N}_\alpha$, defined in 
\eqref{fluxquantize}, depend only on the homology classes of the seven-cycles 
$\Sigma_\alpha \subset Y_9$. The ambiguity in correspondingly 
defining the fluxes $\mathcal{N}_I$, with associated homology cycles 
labelled by $I=1,\ldots,b_5(X_7)$ in the fibre $X_7$, arises precisely because 
of the twisting. To explain this, pick representatives $\mathcal{C}_I\subset X_7$ of the five-cycles in the fibre, 
which are invariant under the $U(1)^s$ action that we twist by. The $\mathcal{C}_I$ 
then fibre over the spindle $\Sigma$ to give seven-cycles in $Y_9$, with the 
seven-form flux  through these cycles \eqref{fluxquantize} then defining a set of fluxes. 
However, when the flavour magnetic charges/twisting parameters $p_i$ are non-zero, 
the homology classes of these seven-cycles in general depend on the representative 
of $\mathcal{C}_I$: specifically, two representatives of the same cycle in $X_7$, 
but with different $U(1)^s$ actions, will twist differently, resulting in different 
seven-form fluxes.\footnote{For a concrete illustration of this in the AdS$_3$ case, 
see the formula \eqref{t11nacalNs} for the fluxes $(\Nas_1,\Nas_2,\Nas_3,\Nas_4)$ 
for $T^{1,1}$ internal space. In this case there is only one internal three-cycle, 
with the four toric representatives all being $\pm 1$ times this cycle in homology in $T^{1,1}$. But this 
flux vector depends explicitly on the twisting variables $p_i$, showing that the homology 
classes of the corresponding four five-cycles in $Y_7$ (the fibration of $T^{1,1}$ over $\Sigma$) are distinct
{\it i.e.} in \eqref{t11nacalNs} we see that the four $M_a$ are not simply related by $\pm $ signs, but instead
are different linear combinations of $\mathcal{N}_0$ and $\mathcal{N}_1$.}
This is just another manifestation of the baryonic/flavour 
mixing problem described in the previous paragraph.   
A precise relation 
between the Kaluza-Klein point of view and fluxes of the AdS$_2$ solution, and in 
particular equation \eqref{yetanotherflux}, is spelled out in appendix \ref{app:KK}.

Rather than trying to separate out a set of  ``baryonic magnetic charges'', it is 
therefore preferable to work directly with the seven-form fluxes, with a chosen basis of 
seven-cycles, which 
are then defined unambiguously. Or, even better, in the toric case to work with the 
toric fluxes $\Nas_a$, given by \eqref{toricfluxrelations} and which are subject to the constraints
in the last two lines of \eqref{homfluxes}. It is the latter that 
most directly describe how the field theory is twisted over $\R\times \Sigma$, 
on the conformal boundary of the AdS$_4$ black hole solutions. Indeed, 
more generally it is presumably possible to define a set of ``equivariant'' fluxes, 
utilizing a form of equivariant cohomology that generalizes toric geometry appropriately, 
so that the flavour and baryonic charges may be combined naturally. 
We leave further development of this for future work, anticipating that the results of \cite{Bah:2019rgq,Bah:2020jas}
could play an important role.

In the case of the universal anti-twist such black hole solutions have been explicitly constructed in $D=4$ minimal gauged supergravity\footnote{Note that $N_{SE}$ in {\it e.g.} (4.46) of \cite{Ferrero:2020twa} is the same as $N$ in this paper.} 
and then uplifted on Sasaki-Einstein $X_7$ to obtain black hole solutions in $D=11$ \cite{Ferrero:2020twa}. 
As discussed in section~\ref{nobar}, the universal anti-twist is a special case of the flavour twist. 
In fact a richer class of black hole solutions were constructed in \cite{Ferrero:2020twa} that have non-vanishing rotation and also electric flavour charge. The entropy of these black holes was computed in \cite{Ferrero:2020twa} and, for vanishing rotation and electric charge, precisely agrees with the result \eqref{exuatads2} obtained using the formalism of this paper. However, a curious feature of these 
black hole solutions is that while setting the electric charge and rotation to zero
leads to a regular AdS$_2\times Y_9$ horizon, the $\mathbb{R}\times \Sigma$ conformal boundary degenerates into
a singular geometry. 

It is possible that this is a feature of purely magnetically charged black holes
in the anti-twist class more generally\footnote{Indeed, it is
 challenging to solve the conformal Killing spinor equation on the boundary \cite{Ferrero:2020twa,Arav:2022lzo}.}, 
 {\it i.e.} not just in the universal anti-twist class. 
 If this is the case, then we expect that they should be viewed as limiting, degenerate cases of larger families of black hole solutions that also have non-vanishing electric charge and angular momentum. On the other hand we do not expect such restrictions in the twist class, 
and we expect, generically, that one will be able to construct purely magnetic charged and accelerating black holes with regular $\mathbb{R}\times \Sigma$ conformal boundaries and regular AdS$_2\times Y_9$ horizons.

The results of this paper allow us to calculate the entropy of supersymmetric, accelerating and purely magnetically charged black holes. 
In the flavour twist class, the results of section \ref{ads2solsexamples} for the $X_7=S^7$ case precisely agree with the results for explicit black hole solutions constructed using STU gauged supergravity \cite{Ferrero:2021ovq}. The results for $V^{5,2}$ provide a precise new prediction for the entropy. A much richer family of cases arises when $X_7$ is toric and the techniques we developed in section \ref{sec:toricoverspindle} allow one to obtain the entropy very explicitly\footnote{ Obtaining the final result in closed form can be challenging since the extremization procedure requires finding roots of polynomials.} by solving a system of algebraic equations. It would be very interesting to recover these results from a field theory calculation and hence obtain a microstate counting interpretation of the black hole entropy. 

A similar discussion to the above can be made for AdS$_3\times Y_7$ solutions with $Y_7$ fibred as in \eqref{fibrationformintro}. 
One now expects these to arise as the near horizon limit of supersymmetric accelerating black strings in AdS$_5$, carrying flavour and baryonic magnetic charges. Many of the above points for black holes are also applicable to black strings, but there are some differences.
For example, less is known about explicit solutions: indeed accelerating black strings in $D=5$ minimal gauged supergravity are not known, which would be the analogue of the $D=4$ accelerating black holes of \cite{Ferrero:2020twa}. Another important difference is that one cannot add electric charge (flavour or baryonic) and preserve the AdS$_3\times \Sigma$ horizon, which could be particularly relevant for anti-twist constructions.
On the other hand, within the toric case one can make a precise connection with a field theory computation using $c$-extremization and anomaly polynomials, as we discussed in section \ref{secftmatch}. 

 \section{Discussion}\label{sec:disc}

In this paper we have studied various aspects of GK geometry, consisting of
a Sasaki-Einstein space fibred over a spindle. 
One of our main results is the formula \eqref{generalblock} which,
in particular, gives rise to the gravitational block form for the supersymmetric action as given in \eqref{actiongb}. 
By setting $m_\pm=1$, this formula is also valid if the spindle is replaced with a two-sphere, giving a new perspective
on some of the results on GK geometry presented in \cite{Gauntlett:2018dpc,Hosseini:2019use,Hosseini:2019ddy,Gauntlett:2019roi,Gauntlett:2019pqg}. These results provide a concrete demonstration that gravitational blocks appear very generally in the context of M2 and D3-branes reduced on two-dimensional
spaces with an azimuthal symmetry.

Since the gravitational block formula \eqref{generalblock} receives contributions from the two poles of the spindle, one might 
 imagine it arises by applying an appropriate
  fixed point formula, with the associated action generated by azimuthal rotations of the spindle (or sphere). A technical  
   observation is that there are some (implicit) signs in the gravitational block formula, 
  related to the choice of whether we are in the twist or anti-twist class. As shown in
 \cite{Ferrero:2021etw}, this  is in turn related to the chirality of the Killing spinor at the fixed poles of the spindle. 
 Thus, if a standard fixed point formula,
   such as the Berline-Vergne formula, can be used to derive the gravitational block result, 
   the equivariant forms must be sensitive to these signs, related to spinor chirality. 

 On the other hand, in the specific case of the universal
 anti-twist, the supersymmetric action/entropy function given by \eqref{osutwist} 
  has already been related to the localization result  in minimal gauged supergravity \cite{BenettiGenolini:2019jdz}
  in reference \cite{Cassani:2021dwa}. However, this relation is for now somewhat indirect,  as 
  in the present paper we have shown by direct computation 
  that the supersymmetric action for this class of AdS$_2$ GK geometries is equal to the 
  on-shell action of a corresponding class of supersymmetric accelerating 
  black holes in AdS$_4$, for which the former arise as the near horizon limits
   of the latter. It is then more specifically the holographically renormalized on-shell action 
   of these black holes that  the localization formula 
   of  \cite{BenettiGenolini:2019jdz} applies to, as described in  \cite{Cassani:2021dwa}. 
   This intriguing relation between these two approaches to black hole entropy functions is discussed further in \cite{Boido:2022iye}.
   
More generally one might imagine that conjectured gravitational block formulas 
 for other classes of solutions, in different theories and in different dimensions, 
could be derived using a similar approach to this paper, and/or by an appropriate 
fixed point theorem. For example, the various black holes in AdS$_4$ and AdS$_5$ discussed in 
\cite{Hosseini:2019iad}, the class of branes wrapped on spindles and higher-dimensional orbifolds in  
 \cite{Faedo:2021nub,Giri:2021xta,Cheung:2022ilc,Suh:2022olh,Couzens:2022lvg,Faedo:2022rqx}, and the general higher derivative gravitational block formula conjectured in  \cite{Hristov:2021qsw} 
for black holes in AdS$_4$ and asymptotically flat space. We note that the latter formula was 
inspired by the localization formula in minimal gauged supergravity in \cite{BenettiGenolini:2019jdz}. 
These generalizations might also include adding various internal fluxes to the 
GK geometries focused on in the present paper, for example the solutions constructed 
recently in \cite{Arav:2022lzo}, 
as well the more general extensions of GK geometry considered in
\cite{Couzens:2020jgx,Couzens:2022agr}. Another possibility would be to 
consider including higher derivative terms in the effective action.
The geometric structures imposed by supersymmetry in different theories and different dimensions 
are often quite distinct, but the universality of this class of gravitational block/black hole entropy functions
suggest there is a more universal approach to deriving these formulas, indeed perhaps utilizing 
an appropriate fixed point formula that is largely insensitive to the detailed geometry. 
It will be very interesting to pursue this in future work.

For the AdS$_3\times Y_7$ class of GK geometries in 
type IIB string theory discussed in section \ref{secftmatch}, we have obtained a very general 
proof that our off-shell supersymmetric action 
agrees with the off-shell trial $c$-function in field theory, 
thus obtaining a very general exact result 
in AdS/CFT.  Here $Y_7$ is an arbitrary fibration 
of a toric Sasaki-Einstein five-manifold $X_5$ over a spindle~$\Sigma$. 
However, a key assumption on the gravity side is that 
the corresponding extremal AdS$_3$ supergravity solutions actually exist -- 
provided they do, the extremal value of the supersymmetric action 
 computes the central charge, but there might be obstructions to existence. 
 Ultimately one needs an existence result for the corresponding PDE 
-- see, for example, the discussion in \cite{Couzens:2018wnk,Gauntlett:2018dpc}. 
 However, it is natural to conjecture that this existence is guaranteed if and only if 
the R-charges $R_a^\pm$ satisfy the positivity conditions $R_a^+>0$, $\sigma R_a^->0$
(in our conventions). 
 That this is a necessary condition 
 can be seen from the first equality in \ref{Rpmdef}: 
 when $n=3$ these R-charges are proportional to the  K\"ahler class 
 integrated over a basis of toric submanifolds in the fibres over the two poles of the spindle, and these
 should all be positive.  On the field theory side this question is related to whether 
 the $d=4$, $\mathcal{N}=1$ SCFT, dual to the AdS$_5\times X_5$ solution, 
 indeed flows to a $d=2$, $(0,2)$ SCFT in the IR when the theory is wrapped 
 on a spindle. 


\section*{Acknowledgements}

\noindent 
We thank Chris Couzens for discussions.
This work was supported in part by STFC grants  ST/T000791/1 and 
ST/T000864/1.
JPG is supported as a Visiting Fellow at the Perimeter Institute. 
 
 \appendix
 
 \section{Delzant construction}\label{app:A}

To see that the toric data \eqref{Y7data1} indeed describes a fibration 
\eqref{XoverYtoric}, we can utilize the Delzant construction for toric cones (as summarized, for example, in \cite{Martelli:2004wu}). 

We first recall this construction
for $C(X_{2n-1})$ and, for simplicity, we assume that $X_{2n-1}$ is simply-connected. There is a linear map
\begin{align}
\mathcal{A}^{(n)}: \R^d\rightarrow \R^{n}~, \qquad \mathcal{A}^{(n)}({e}_{a})= {v}_{a}\in \R^{n}~,
\end{align}
where $\{e_{a}\}$ denotes the standard orthonormal basis for $\R^d$, with components 
$e_{ab}=\delta_{ab}$, and $v_{a}=(v_{ai})_{i=1}^n$ are the toric data for $\X$, where in what follows 
it will be convenient to suppress the vector index $i$ on $v_a$ (which are the components of $v_a$ in a basis for the Lie algebra of $U(1)^n$).
 Since $\mathcal{A}^{(n)}$ maps $\Z^d$ to $\Z^{n}$, there is an induced 
map of tori $U(1)^d=\R^d/\Z^d\rightarrow \R^n/\, \mathrm{span}_\Z\{{v}_{a}\}$, 
with a kernel $U(1)^{d-n}$.\footnote{When $X_{2n-1}$ is not simply-connected
there is also a finite group as part of this kernel.} The latter is generated by an integer $d\times (d-n)$ matrix $q_I^a$, $I=1,\ldots,d-n$, satisfying
\begin{align}\label{3kernel}
\sum_{a=1}^d q_I^a\, {v}_{ai}= 0\,,\qquad i=1,\dots,n\,,
\end{align}
which specifies the embedding $U(1)^{d-n}\subset U(1)^d$. The toric $U(1)^n$ action on $C(X_{2n-1})$ is then via the quotient $U(1)^n=U(1)^d/U(1)^{d-n}$. In physics language, the above 
construction describes a gauged linear sigma model (GLSM) with $d$ complex fields and $U(1)^{d-n}$ charges 
specified by $q_I^a$, with $C(X_{2n-1})$ being the vacuum moduli space (see {\it e.g.} \cite{Martelli:2004wu}).

For a toric $C(Y_{2n+1})$ there is then a similar construction, with linear map
\begin{align}
\mathcal{A}^{(n+1)}: \R^{D}\rightarrow \R^{n+1}~, \qquad \mathcal{A}^{(n+1)}(e_{A})= v_{A}\in \R^{n+1}~,
\end{align}
with standard orthonormal basis $\{e_A\}$ of $\R^D$, with components $e_{AB}=\delta_{AB}$, 
toric data $v_A=(v_{A\mu})_{\mu=0}^n$ for $C(\Y)$, 
and
with a kernel specified by an integer $D\times (D-n-1)$ matrix 
$Q^A_{\hat{I}}$, $\hat I=0,1,\dots,d-n$, satisfying
\begin{align}\label{4kernel}
\sum_{A=1}^D Q^A_{\hat{I}}\, {v}_{A\mu} = 0\,,\qquad \mu=0,1,\dots, n\,.
\end{align}

From \eqref{Y7data1}, which recall relates the toric data for $C(\Y)$ in terms of that for $C(\X)$, 
the spindle data $m_\pm$, and the twisting variables $p_i$, $a_\pm$,  we can immediately identify part 
of this kernel:
\begin{align}\label{Qa}
Q_I= (0,0,q^1_I,\ldots,q^d_I)~, \qquad I=1,\ldots,d\, .
\end{align}
Note here that this satisfies \eqref{4kernel} by virtue of \eqref{3kernel}, where ${v}_{a}=(1,\vec{w}_a)$. 
Since $\hat I$ takes $D-n-1=d+1-n$ values and $I=1,\dots,d-n$, there is one more kernel vector to identify, which we label
as the $\hat{I}=0$ component of $Q^A_{\hat{I}}$. We write this charge vector as
\begin{align}\label{Q0}
Q_0 = (m_-,\sigma m_+,q_0^1,\ldots,q_0^d)~.
\end{align}
Notice here that the first two entries are fixed by the zero'th components of the vectors in \eqref{Y7data1}. 
From \eqref{Y7data1} and \eqref{4kernel} 
we see that $Q_0$ is a kernel vector provided that the vector $q_0^a$ satisfies
\begin{align}\label{twistcond}
\sum_{a=1}^d q_0^a \, {v}_{ai}\, & = \, p_i~,\qquad i=1,\dots, n\,,
\end{align}
where we recall from \eqref{naughty} that $p_1 = - (m_- + \sigma m_+)$, with $\sigma=+1$ being the twist,
 and $\sigma=-1$ being the anti-twist.
In fact the above equations are precisely describing the fibration \eqref{XoverYtoric} in a two step process, 
as we now describe. 

To begin, recall that to fibre $C(X_{2n-1})$ over a spindle $\Sigma=\mathbb{WCP}^1_{[m_-,m_+]}$, we may 
first fibre $\C^d$ over $\Sigma$. To do so we must first lift the $U(1)^n$ action on 
$C(X_{2n-1})$ to $\C^d$, which means specifying  $n$ vectors $\alpha^i = (\alpha_a^{i})_{a=1}^d \in\Z^d$, $i=1,\ldots,n$ satisfying
\begin{align}\label{lift}
\mathcal{A}^{(n)}(\alpha^{i})= {e}_{i}\in \Z^n~,
\end{align}
where $\{{e}_i\}$ is the standard orthonormal basis for $\R^n$. In components this condition 
reads
\begin{align}\label{valpha}
\sum_{a=1}^d v_{ai}\, \alpha_a^{j}= \delta_{i}^{j}~.
\end{align}
Of course the choice of each ${\alpha}^{i}\in\Z^d$, $i=1,\dots, n$, is unique only up to the kernel of 
$\mathcal{A}^{(n)}$, generated by $q_I^a$. Geometrically, this is because 
$C(X_{2n-1})$ is precisely a K\"ahler quotient of $\C^d$ via the torus $U(1)^{d-n}$ generated by
this kernel. Put more simply, the charge of the $a$th coordinate $z^a$ of $\C^d$ under the $i$th $U(1)\subset U(1)^n$
is precisely $\alpha_a^{i}$. 

With this in hand, we now re-examine \eqref{twistcond} and the kernel 
$Q^A_{\hat{I}}$. We first consider the twist case with $\sigma=+1$. The GLSM for $C(Y_{2n+1})$ begins with $\C^{D}=\C^{d+2}$. 
Taking the quotient by $U(1)$ with charge vector $Q_0$ given by \eqref{Q0}
then describes the total space of a $\C^d$ bundle over $\Sigma=\mathbb{WCP}^1_{[m_-,m_+]}$, 
where the charge vector $q_0^a$ describes the twisting. That is, as a first step we construct
the $(d+1)$-dimensional space
\begin{align}\label{Xdef}
Z\, \equiv\,  \mathcal{O}({q}^a_0)_{\Sigma}\times_{U(1)^d} \C^d~.
\end{align}
Here the notation means we twist the $a$th coordinate of $\C^d$ by the line bundle $O(q_0^a)_\Sigma$. 
The zeroth component of the condition \eqref{twistcond} then simply says that $Z$ is a Calabi-Yau $(d+1)$-fold. Notice 
here that $q_0^a<0$ is necessary for the convex toric geometry description we have given to be 
applicable, although the space $Z$ defined by \eqref{Xdef} exists as a complex manifold with zero first Chern class 
irrespective of the signs of the charges.

In the anti-twist case, with $\sigma=-1$, the charge of the second coordinate 
on $\C^{D}=\C^{d+2}$ under $Q_0$ in \eqref{Q0} is negative. 
Taking the K\"ahler quotient by this $U(1)$ then does not result 
in a space with the topology given in 
\eqref{Xdef}. Formally 
we may complex conjugate the second coordinate on $\C^D$, to see that 
topologically $Z$ given by \eqref{Xdef} is a partial resolution 
of the conical geometry in the anti-twist case. However, the space $Z$ is then  no longer Calabi-Yau, since 
$\sum_{a=1}^d q_0^a$ is not equal to $-(m_-+m_+)$ when $\sigma=-1$.\footnote{This fact was discussed in some detail in \cite{Ferrero:2021etw} for various explicit  supergravity 
solutions where $d=n$ and $C(\X)=\C^{n}$.}  
This lack of any clear relation to a partially resolved Calabi-Yau 
geometry, with a blown up copy of $\Sigma$ on which the branes may wrap, is one reason why the anti-twist solutions are more difficult 
to interpret physically.

Returning to the twist case with $\sigma=+1$, 
at this stage we have quotiented by $U(1)$ out of $U(1)^{d+1-n}$ to obtain a Calabi-Yau $(d+1)$-fold $Z$, which 
is a $\C^d$ fibration over the spindle $\Sigma$. Quotienting by the remaining $U(1)^{d-n}=U(1)^{d+1-n}/U(1)$ 
then precisely turns the fibre $\C^d$ into $C(X_{2n-1})$ -- this is clear from \eqref{Qa}, which only 
acts on the $\C^d$ fibre direction of $Z$.  Since $\C^d$ is fibred non-trivially over the spindle,
 the $C(X_{2n-1})$ will also be fibred, and equation \eqref{twistcond} is telling us how. Specifically, 
 this condition is solved by writing
\begin{align}\label{qn}
q_0^a= \sum_{i=1}^n p_i\, \alpha_a^{i}~.
\end{align}
 The fact this solves \eqref{twistcond} follows 
from \eqref{valpha}:
\begin{align}
\sum_{a=1}^d q_0^a\, v_{ai}= \sum_{a=1}^d\sum_{j=1}^{n} p_j\, \alpha_a^{j}\, v_{ai}= p_i~,\qquad i=1,\dots, n\,.
\end{align}
Geometrically, recall that $\alpha_a^{i}$ gives the charge of the $a$th coordinate on $\C^d$ under 
the $i$th $U(1)$ in $U(1)^n$. Thus $\vec{p}=(p_2,\dots, p_{n})$ precisely has the interpretation of 
twisting $\C^d$ by the line bundle $O(\vec{p})_\Sigma$, so that we may also write
\begin{align}
Z= O(\vec{p})_{\Sigma}\times_{U(1)^n} \C^d~,
\end{align}
where $q_0$ in \eqref{Xdef} is specified by \eqref{qn}.
The vacuum moduli space with the full quotient is then
\begin{align}\label{appfibre}
O(\vec{p})_\Sigma \times_{U(1)^n} C(\X)~.
\end{align}
In the anti-twist case this 
last quotient of $\C^d$ by $U(1)^{d-n}$ to obtain 
the Calabi-Yau cone fibres $C(\X)$ is still valid, but the 
space in \eqref{appfibre}, while being 
a partial resolution of $C(\Y)$, is no longer Calabi-Yau. 

This concludes our proof that the toric data \eqref{Y7data1} describes the fibration 
\eqref{XoverY}. 

\section{Matching GK geometry for AdS$_3\times Y_7$ with field theory}\label{app:matching}

Here we consider GK geometry for toric $Y_7$ comprised of toric $X_5$ fibred over a spindle.
We show that the off-shell central charge $\mathscr{Z}$ for the  AdS$_3\times Y_7$ solution matches
with the off-shell central charge $c_{\mathrm{trial}}|_{\mathrm{baryonic}}$ in the field theory, where the extremization over
the baryonic directions has been carried out.  The key results of sections \ref{sec53} and
\ref{rchgetoricsec}, are central to the proof and we also use some results of \cite{Hosseini:2019use}.

It will be convenient to utilise different choices of gauge for the K\"ahler class parameters $\lambda^A$.
Recall that the master volume for toric $Y_7$, $\mathcal{V}_7$, is invariant under the gauge transformations
given in \eqref{lamgt}:
\begin{align}\label{lamgt2}
\lambda_A\ \to\ \lambda_A+\sum_{\mu =0}^{3}\gamma^\mu(v_{A\mu} b_1-b_\mu)\,.
\end{align}
Although $\gamma^\mu$ has four components only three of them yield a non-trivial transformation since 
the dependence on $\gamma^1$ drops out. Any function which is homogeneous in $\lambda_A$ is invariant under \eqref{lamgt2} and, in particular, $\mathcal{V}_{7}$, $\mathcal{V}_{5}^\pm$, $R_a^\pm$, and $\mathscr{Z}$ are all gauge-invariant quantities.

We will use three different gauge choices.
We first note that the transformation of $\lambda_a^{(+)}$ only depends on $\gamma^2,\gamma^3$. We can therefore choose 
$\gamma^2,\gamma^3$ to set two of the $\lambda_a^{(+)}$ to vanish, for example
	\begin{equation}\label{gaugeplus}
\textit{Plus gauge:}\qquad		\lambda_1^{(+)} = \lambda_2^{(+)} = 0 \:.
	\end{equation}
Similarly, we can instead choose $\gamma^2,\gamma^3$ to set two of the $\lambda_a^{(-)}$ to vanish, for example
	\begin{equation}\label{gaugeminus}
\textit{Minus gauge:}\qquad		\lambda_1^{(-)} = \lambda_2^{(-)} = 0 \:.
	\end{equation}
Finally, noting that the transformation of $\lambda^\pm$ depends on $\gamma^0,\gamma^2,\gamma^3$ we can choose
$\gamma^2,\gamma^3$ (say) to obtain
\begin{equation}\label{gaugesym}
\textit{Symmetric gauge:}\qquad		\lambda_+ = \lambda_- = 0 \quad \Rightarrow  \quad \lambda_a^{(\pm)} = \lambda_a \:.
	\end{equation}
Note that these three gauge choices leave a residual one-parameter gauge invariance parametrised by $\gamma^0$, which we will not
need to fix. 

\subsection{$\mathscr{Z}$ as a function of $R_a^\pm$}
The master volume for toric $X_5$, $\mathcal{V}_{5}$, is homogeneous of degree 2 in the $\lambda_a$ and hence we can write
\begin{equation}\label{V5identity}
	\mathcal{V}_{5}^\pm = \frac{1}{2} \sum_{a=1}^d \frac{\partial \mathcal{V}_{5}^\pm}{\partial \lambda_a} \, \lambda_a^{(\pm)} = - \frac{N\nu_3}{4}  \sum_{a=1}^d R_a^\pm\, \lambda_a^{(\pm)} \:,
\end{equation}
where we used the expressions for the shifted R-charges, $R_a^\pm$, given in \eqref{Rapmequiv}. We now show that it is possible to express $\lambda_a^{(\pm)}$ in \eqref{V5identity} in terms of $R_a^\pm$ and hence 
obtain an expression for the supersymmetric action $S_{\text{SUSY}}$, and hence the off-shell central charge $\mathscr{Z}$,
in terms of $R_a^\pm$ using \eqref{keygravresult} and \eqref{offshellz}. We will
use different gauges to treat the $\pm$ cases, but the final results are gauge invariant.

It is helpful to recall the explicit formula for $\mathcal{V}_{5}=\mathcal{V}_5(\vec{b};\{\lambda_a\})$  given in (1.3) of \cite{Gauntlett:2018dpc}:
\begin{align}\label{volform}
\mathcal{V}_5
&  =   \frac{(2\pi)^3}{2}\sum_{a=1}^d \lambda_a \frac{\lambda_{a-1}({\vec{v}}_a,{\vec{v}}_{a+1},\vec{b}) - \lambda_a ({\vec{v}}_{a-1},{\vec{v}}_{a+1},\vec{b})+\lambda_{a+1}({\vec{v}}_{a-1},{\vec{v}}_a,\vec{b})}{({\vec{v}}_{a-1},{\vec{v}}_a,\vec{b})({\vec{v}}_{a},{\vec{v}}_{a+1},\vec{b})}~.
\end{align}
We now define
\begin{equation}
	J_{ab} \equiv  \frac{\partial \mathcal{V}_{5}}{\partial \lambda_a \partial \lambda_b} \:, \qquad  J_{ab}^\pm \equiv  J_{ab}\big|_{\vec b = \vec b^{(\pm)}} \:,
\end{equation}
with $J_{ab}$ independent of $\lambda_a$ and $\vec{b}=(b_1,b_2,b_3)$.
The only non-vanishing components of the matrix $J_{ab}$ are $J_{aa}$ and $J_{a,a+1} = J_{a+1,a}$ 
given by
\begin{align}\label{Jsexp}
J_{aa}=-(2\pi)^3\frac{(\vec v_{a-1},\vec v_{a+1},\vec b)}{(\vec v_{a-1},\vec v_a,\vec b)(\vec v_a,\vec v_{a+1},\vec b)}\,,\qquad
J_{a,a+1} =(2\pi)^3\frac{1}{(\vec v_a,\vec v_{a+1},\vec b)}\,.
\end{align}
The homogeneity of the master volume $\mathcal{V}_{5}$ implies that we can recast $R_a^\pm$ in \eqref{Rapmequiv} as
\begin{align}\label{Rpm}
	R_a^\pm &= -\frac{2}{N\nu_3} \, \sum_{b=1}^d J_{ab}^\pm\, \lambda_b^{(\pm)}\,,\nn\\
		&= -\frac{16\pi^3}{N\nu_3}\bigg[\frac{\lambda_{a-1}}{(\vec{v}_{a-1}, \vec{v}_{a} , \vec{b}^{(\pm)})} - \frac{(\vec{v}_{a-1}, \vec{v}_{a+1} , \vec{b}^{(\pm)}) \,\lambda_a}{(\vec{v}_{a-1}, \vec{v}_{a} , \vec{b}^{(\pm)})(\vec{v}_{a}, \vec{v}_{a+1} , \vec{b}^{(\pm)})} + \frac{\lambda_{a+1}}{(\vec{v}_{a}, \vec{v}_{a+1} , \vec{b}^{(\pm)})}\bigg] \:.
\end{align} 

We now focus on the $+$ case and choose the ``plus gauge" \eqref{gaugeplus}, $\lambda_1^{(+)} = \lambda_2^{(+)} = 0$.
In this gauge we have
\begin{align}\label{explicitRp}
		&R_1^+ = -\frac{2}{N\nu_3}\left(J_{1d}^+\, \lambda_d^{(+)}\right) , \nn\\
		&R_2^+ = -\frac{2}{N\nu_3}\left(J_{23}^+\, \lambda_3^{(+)}\right) , \nn\\
		&R_3^+ = -\frac{2}{N\nu_3}\left(J_{33}^+\, \lambda_3^{(+)} + J_{34}^+\, \lambda_4^{(+)}\right) , \nn \\
		&R_4^+ = -\frac{2}{N\nu_3}\left(J_{43}^+\, \lambda_3^{(+)} + J_{44}^+\, \lambda_4^{(+)} + J_{45}^+\, \lambda_5^{(+)}\right) ,  \nn\\
		&\dots \quad ,
\end{align}
and these equations can be solved recursively for the $\lambda_a^{(+)}$ to get (as in \cite{Hosseini:2019use})
\begin{equation}\label{lambda}
	\lambda_{a}^{(+)}=-\frac{N\nu_3}{16 \pi^{3}}\sum_{b=2}^{a}(\vec v_{b}, \vec v_{a}, \vec{b}^{(+)}) \, R_{b}^+, \qquad a=3, \ldots, d\,.
\end{equation}
The identity for $R_{a}^+$ given in \eqref{Rpmidentities3} ensures that the expression for $\lambda_d^{(+)}$ we get from the first equation in \eqref{explicitRp} is consistent with \eqref{lambda}. 
If we now substitute \eqref{lambda} back into \eqref{V5identity} we get
\begin{equation}
	\mathcal{V}_{5}^+ =  \frac{N^2\nu_3^2}{64\pi^3} \sum_{a=3}^d \sum_{b=2}^a (\vec v_{b}, \vec v_{a}, \vec{b}^{(+)}) R_a^+\, R_b^+ \:,
\end{equation}
and then after again using the identity \eqref{Rpmidentities3} we can write
\begin{align}
		\mathcal{V}_{5}^+ &=  \frac{N^2\nu_3^2 b_1}{128\pi^3}  \sum_{a=3}^d \sum_{b=2}^a \sum_{c=1}^d (\vec v_{b}, \vec v_{a}, \vec{v}_c) R_a^+ R_b^+ R_c^+  \nn\\
		&= 
		\frac{N^2\nu_3^2 b_1}{128\pi^3} \sum_{ a < b < c} (\vec v_{a}, \vec v_{b}, \vec{v}_c)\, R_a^+ R_b^+ R_c^+ \:.
\end{align}
This expression does not depend on the gauge choice that we used and hence it holds in all gauges.

Analogously, for the $-$ case we can utilise the ``minus gauge" \eqref{gaugeminus}, and obtain an equivalent 
expression for $\mathcal{V}_{5}^-$ with $R^+_a$ replaced with $R^-_a$.  
Putting these results together, we can write the supersymmetric action \eqref{keygravresult}
and hence the trial central charge $\mathscr{Z}$ defined in \eqref{offshellz} in the form
\begin{equation}\label{ZexpressionR}
		\mathscr{Z}	= \frac{3 N^2b_1^2}{4b_0}  \sum_{a < b < c} (\vec v_{a}, \vec v_{b}, \vec{v}_c) \left(R_a^+ R_b^+ R_c^+ - R_a^- R_b^- R_c^-\right).
\end{equation}
In this expression we note that $R_a^\pm$ can be viewed as functions of $b_\mu$ as well as the five-form fluxes $\NAS_A$ (after eliminating the $\lambda$'s using the second two lines of \eqref{tent_cc} or, equivalently, \eqref{N7exps} and \eqref{N7exps2con}).

We now observe that after setting $b_1 = 2$, which is required in the GK extremization procedure, the expression for
$\mathscr{Z}$ in \eqref{ZexpressionR} has precisely the same form as the field theory expression for the central charge \eqref{ftexp} 
using the dictionary between field theory and geometric quantities discussed in section \ref{sec:match}:
\begin{align}\label{bnoughtep2}
\varepsilon&=b_0 ~,\nn\\
 \left.\Delta_a^+\right|_{\mathrm{baryonic}}&=R^+_a ~,\qquad \left.\Delta_a^-\right|_{\mathrm{baryonic}}=R^-_a ~.
\end{align}
We also identify the rank of the field theory gauge group, $N$, with the flux $N\equiv m_+ \NAS_+  = \sigma m_- \NAS_-$ and
the background magnetic fluxes for the SCFT $\kf_a$ with the five-from fluxes via $\kf_a\, \equiv \, \frac{\Nas_a}{N}$
as in \eqref{fluxident}\eqref{bnoughtep}.

We now argue that the identification \eqref{bnoughtep2} is possible. On the GK geometry side we have three independent variables, $b_0, b_2,b_3$. We also have three independent variables on the field theory side, the $d+1$ variables 
$\varepsilon, \Delta_a$ satisfying the $d-2$ constraints
\begin{align}\label{consdeltaapp}
\sum_{a=1}^d \Delta_a = 2\,,\qquad\qquad
\sum_{a=1}^d q^a_I\, \frac{\partial c_{\mathrm{trial}}}{\partial \Delta_a} =0~, \qquad I=1,\ldots,d-3~.
\end{align}
Interestingly, we will show below (see \eqref{appbfinaleqquest}) that the latter constraints are actually $d-2$ linear constraints in terms of the variables $\Delta_a,\varepsilon$. We will also show in section \ref{linearity} that the $R_a^\pm$, after eliminating the $\lambda_a$ variables, are linear functions of $b_2, b_3$ and $b_0$.
Thus, \eqref{bnoughtep2} is $2d+1$ linear equations for 3 variables and hence the system seems to be overdetermined. However, only 3 of the equations are independent, and the other $2d-2$ can be obtained as linear combinations of these. 

To see this, we introduce the notation
\begin{equation}
	E_a^\pm \equiv \left.\Delta_a^\pm\right|_{\mathrm{baryonic}} - R^\pm_a \: ,
\end{equation}
so that the $2d$ linear equations in the second line of \eqref{bnoughtep2} are simply $E_a^\pm = 0$. Then, combining the identities \eqref{dconstraints} and the definition \eqref{Deltapm1}, we conclude
\begin{equation}
	E_a^+ - E_a^- = 0\,,
\end{equation}
which eliminates $d$ equations, say those encoded by $E_a^-$, from the original system. Another equation can be eliminated because
\begin{equation}
	\sum_{a=1}^d E_a^+ = 0 \:,
\end{equation}
where we used \eqref{Rpmidentities12} and \eqref{delpmids}. Finally, the last set of $d-3$ linear constraints comes from the baryonic extremization condition \eqref{consdeltaapp}. 

To see this we will prove in the next subsection that 
\begin{equation}\label{kernelconstr}
	\sum_{a=1}^d q_I^a \, \frac{\partial c_{\text{trial}}}{\partial \Delta_a} \bigg|_{\Delta_a^\pm = R_a^\pm (\vec B, \NAS_A)}= 0 \:, \qquad I=1,\dots, d-3 \:.
\end{equation}
Furthermore, to do so we will also prove that the baryonic extremization condition \eqref{consdeltaapp} can be re-expressed in the equivalent form
\begin{equation}\label{bar_constr_intermediate}
	\sum_{b=1}^d \sum_{a=1}^b \sum_{c=1}^d q_I^a (\vec v_{a}, \vec v_{b}, \vec{v}_c) \left( \Delta_b^+ \Delta_c^+ -  \Delta_b^- \Delta_c^-\right)=0 \:.
\end{equation}
Writing 
\eqref{bar_constr_intermediate} purely in terms of $\Delta_a^+$, say, yields
\begin{equation}\label{appbfinaleqquest}
	\sum_{b=1}^d \sum_{a=1}^b \sum_{c=1}^d q_I^a (\vec v_{a}, \vec v_{b}, \vec{v}_c) \left( \mathfrak{p}_b \Delta_c^+ +\mathfrak{p}_c \Delta_b^+ + \varepsilon\, \mathfrak{p}_b \mathfrak{p}_c\right)=0 \:,
\end{equation}
 from which we see that the baryonic constraints are linear in the $\Delta_a$ and $\varepsilon$ as noted above. Hence, given \eqref{kernelconstr}, 
 the baryonic extremization condition implies that the system of linear equations $E_a^+$ is, indeed, also subject to $d-3$ linear constraints
\begin{equation}
	\sum_{b=1}^d \sum_{a=1}^b \sum_{c=1}^d q_I^a (\vec v_{a}, \vec v_{b}, \vec{v}_c) \left( \mathfrak{p}_b E_c^+ +\mathfrak{p}_c E_b^+ \right)=0 \:.
\end{equation}

Thus, the linear identifications \eqref{bnoughtep2} are actually not overdetermined. In practice we can take the identification  to be, for example,
\begin{equation}\label{finalsystem}
		\varepsilon=b_0 ~,\qquad	\left.\Delta_1^+\right|_{\mathrm{baryonic}}=R^+_1\,, \qquad \left.\Delta_2^+\right|_{\mathrm{baryonic}}=R^+_2\,. 
	\end{equation}

\subsection{Proof of baryonic extremization conditions}\label{proof}
We now prove \eqref{kernelconstr} and \eqref{bar_constr_intermediate}. We begin with the latter and show that we can 
rewrite the field theory baryon mixing condition
\begin{align}\label{baryonicmixingappb}
\sum_{a=1}^d q^a_I\, \frac{\partial c_{\mathrm{trial}}}{\partial \Delta_a} =0~, \qquad I=1,\ldots,d-3~,
\end{align}
in the form
\begin{equation}\label{kernelconstr_simp1}
	\sum_{b=1}^d \sum_{a=1}^b \sum_{c=1}^d q_I^a (\vec v_{a}, \vec v_{b}, \vec{v}_c) \left( \Delta_b^+ \Delta_c^+ -  \Delta_b^- \Delta_c^-\right)=0 \:.
\end{equation}

To see this, we begin by using the expression for $c_\text{trial}$ in \eqref{ftexp} to write
\begin{equation}\label{sum}
 \sum_{a=1}^d q_I^a 	\frac{\partial c_{\text{trial}}}{\partial \Delta_a} =
	 \frac{3}{\varepsilon}\sum_{e<b<c} (\vec v_{e}, \vec v_{b}, \vec{v}_c)\left[q_I^e \Delta_b^+ \Delta_c^+ + q_I^b \Delta_e^+ \Delta_c^+ + q_I^c \Delta_b^+ \Delta_e^+ - (+\leftrightarrow -)\right]\,.
\end{equation}
We can rewrite the triple sum in the following way
\begin{equation}
	\sum_{e<b<c} = \sum_{e=1}^d \sum_{b=e}^d \sum_{c=b}^d = \sum_{b=1}^d \sum_{e=1}^b \sum_{c=b}^d \:,
\end{equation}
where due to the asymmetry of the determinant the sum gets no contribution if we set any two of $e,b,c$ equal.
Using \eqref{3kernel}, we can rewrite the third term in \eqref{sum} as
\begin{equation}
	-\sum_{b=1}^d \sum_{e=1}^b \sum_{c=1}^b (\vec v_{e}, \vec v_{b}, \vec{v}_c) q_I^c \Delta_b^+ \Delta_e^+ = \sum_{b=1}^d \sum_{e=1}^b \sum_{c=1}^b (\vec v_{e}, \vec v_{b}, \vec{v}_c) q_I^e \Delta_b^+ \Delta_c^+ \:,
\end{equation}
where we  swapped the indices $e$ and $c$ to get the second expression. 
Hence, summing the first and the third term of \eqref{sum} yields
\begin{equation}
	\sum_{b=1}^d \sum_{e=1}^b \sum_{c=1}^d (\vec v_{e}, \vec v_{b}, \vec{v}_c) q_I^e \Delta_b^+ \Delta_c^+ \:.
\end{equation}
For the second term in \eqref{sum}, we can exploit the skew-symmetry of the summand to extend the range of $c$ and then massage it via some relabelling and again using \eqref{3kernel} to find 
\begin{align}
		\sum_{b=1}^d \sum_{e=1}^b & \sum_{c=b}^d (\vec v_{e}, \vec v_{b}, \vec{v}_c) q_I^b \Delta_e^+ \Delta_c^+ = \sum_{b=1}^d \sum_{e=1}^b \sum_{c=1}^d (\vec v_{e}, \vec v_{b}, \vec{v}_c) q_I^b \Delta_e^+ \Delta_c^+ \nn\\
		&= \sum_{e=1}^d \sum_{b=e}^d \sum_{c=1}^d (\vec v_{e}, \vec v_{b}, \vec{v}_c) q_I^b \Delta_e^+ \Delta_c^+ = -\sum_{e=1}^d \sum_{b=1}^e \sum_{c=1}^d (\vec v_{e}, \vec v_{b}, \vec{v}_c) q_I^b \Delta_e^+ \Delta_c^+\nn \\
		&= \sum_{b=1}^d \sum_{e=1}^b \sum_{c=1}^d (\vec v_{e}, \vec v_{b}, \vec{v}_c) q_I^e \Delta_b^+ \Delta_c^+ \:.
\end{align}
Hence we find that the second term in \eqref{sum} is equal to the sum of the first and the third term. We
can obtain a similar result for the terms involving $\Delta_a^-$ and hence we have proven \eqref{kernelconstr_simp1}.

We now prove \eqref{kernelconstr}:
\begin{equation}\label{kernelconstr2}
	 \sum_{a=1}^d q_I^a \, \frac{\partial c_{\text{trial}}}{\partial \Delta_a} \bigg|_{\Delta_a^\pm = R_a^\pm (\vec B, \NAS_A)}= 0 \:, \qquad I=1,\dots, d-3 \:.
\end{equation}
Using what we just proved in \eqref{kernelconstr_simp1}, as well as the conditions satisfied by $R^\pm_c$ given in \eqref{Rpmidentities3} to carry out the sum over $c$, this condition
can be written equivalently as
\begin{equation}\label{sum2}
W\equiv 	\sum_{a=1}^d \sum_{b=a}^d q_I^a \left[(\vec{v}_a, \vec{v}_b, \vec b^{(+)}) R_b^+ - (\vec{v}_a, \vec{v}_b, \vec b^{(-)}) R_b^-\right] =0\:,
\end{equation}
where we have also swapped the sums over $a$ and $b$. 
To proceed we exploit the determinant identity $(\vec{a},\vec{b},\vec{c}) = \vec{a}\cdot(\vec{b}\wedge\vec{c})$
to write $W$ in the form
\begin{equation}
\label{defomega}
W = \sum_{a=1}^d q_I^a \, \vec{v}_a \cdot \vec{u}_a\,,\qquad 
	\vec{u}_a \equiv  \sum_{b=a}^d  \left[(\vec{v}_b \wedge\vec b^{(+)}) R_b^+ - (\vec{v}_b \wedge \vec b^{(-)}) R_b^-\right]. 
\end{equation}
Note that the expression for $W$ is unchanged under the following shift of the vectors $\vec{u}_a$:
\begin{equation}\label{upzexp}
	\vec{u}_a\to\vec{u}'_a = \vec{u}_a + \vec{v}_a \wedge \vec z_a \:,
\end{equation}
for an arbitrary vector $\vec z_a$. The strategy is to show that we can choose $\vec z_a$ in such a way that all of the $\vec u'_a$ are actually the same vector $\vec{u}$, since if this is the case we then have
\begin{equation}
	W = \sum_{a=1}^d q_I^a \, \vec{v}_a \cdot \vec{u}'_a = \sum_{a=1}^d q_I^a \, \vec{v}_a \cdot \vec{u} 
	= 0 \:,
\end{equation}
where in the last step we again used \eqref{3kernel}.

We now show that such a choice of $\vec z_a$ is possible. We first compute the difference between two adjacent $\vec u'_a$ vectors:
\begin{align}
		\vec u'_{a+1}- \vec u'_a &= -(\vec{v}_a \wedge\vec b^{(+)}) R_a^+ + (\vec{v}_a \wedge \vec b^{(-)}) R_a^- + \vec{v}_{a+1} \wedge \vec z_{a+1} - \vec{v}_a \wedge \vec z_a\,.
\end{align}
We thus want to choose the variables $\vec z_a$, $a=1,\dots, d$ to solve the system of equations
\begin{equation}\label{system}
	-\vec{v}_a \wedge \left[ \vec b^{(+)} R_a^+ - \vec b^{(-)} R_a^- + \vec z_a\right] + \vec{v}_{a+1} \wedge \vec z_{a+1} = 0 \:, \qquad a=1,\dots, d \:,
\end{equation}
where, as usual, the indices are identified cyclically {\it i.e.} $d+1 = 1$. 
Now since $\vec{v}_a$ and $\vec{v}_{a+1}$ are not parallel vectors the only way to solve this is if each of the two terms
are proportional to $\vec{v}_a\wedge\vec{v}_{a+1}$ which implies that we can write
\begin{align}
		\vec b^{(+)} R_a^+ - \vec b^{(-)} R_a^- + \vec z_a &= k_a \vec{v}_{a+1} + k'_a \vec v_a\,,\nn \\
		\vec{z}_{a+1} &= - k_a \vec{v}_a + k''_{a+1} \vec{v}_{a+1}\,,
\end{align}
for some real constants $k_a$, $k'_a$, and $k''_a$, $a=1, \dots, d$. This is equivalent to
\begin{align}\label{twozexps}
		\vec{z}_a &= k_a \vec{v}_{a+1} + k'_a \vec v_a -\vec b^{(+)} R_a^+ + \vec b^{(-)} R_a^-\,,\nn \\
		\vec{z}_a &= -k_{a-1} \vec{v}_{a-1} + k''_{a} \vec{v}_{a}\,.
\end{align}
Now recall from \eqref{upzexp} that $\vec{z}_a$ can be shifted by a vector proportional to $\vec{v}_a$ without changing $\vec{u}'_a$.
Hence, we can reabsorb, for example, the term $k''_{a} \vec{v}_{a}$ inside $\vec{y}_a$ with an associated redefinition of 
$k_a'$. Thus, consistency of the two expressions for each $\vec{z}_a$ in \eqref{twozexps} amounts to solving
the following system of equations for $k_a$ and $k'_a$
\begin{equation}\label{system2}
	k_a \vec{v}_{a+1} + k'_a \vec{v}_a+ k_{a-1} \vec{v}_{a-1} =\vec b^{(+)} R_a^+ - \vec b^{(-)} R_a^-  \:, \qquad a=1,\dots, d \:.
\end{equation}

This is $3(d-1)$ independent equations for $2d$ variables $k_a, k'_a$. A solution can exist provided that a set of $d-3$ constraints are satisfied. Let us introduce the notation
\begin{equation}
	\vec b^{(+)} R_a^+ - \vec b^{(-)} R_a^- \equiv (\beta_a^1, \, \vec{\beta}_a) \:,
\end{equation}
where $\vec \beta_a$ is a two dimensional vector. Then the first component of \eqref{system2} can be used to
solve for $k_a'$ as follows
\begin{equation}
	k'_a = \beta_a^1 - k_a - k_{a-1} \:.
\end{equation}
Substituting this back into \eqref{system2} we get the set of two-dimensional equations
\begin{equation}\label{system2d}
	k_a (\vec{w}_{a+1} - \vec{w}_a) - k_{a-1} (\vec{w}_{a} - \vec{w}_{a-1}) = \vec{\beta}_a - \beta_a^1 \vec{w}_a \:.
\end{equation}
Now if we project this two-dimensional equation onto vectors orthogonal to $(\vec{w}_{a+1} - \vec{w}_a)$ and $ (\vec{w}_{a} - \vec{w}_{a-1})$, we get two simple linear equations for $k_{a+1}$ and $k_a$ respectively.\footnote{Note that we can do this since by construction $(\vec{w}_{a+1} - \vec{w}_a)$ and $ (\vec{w}_{a} - \vec{w}_{a-1})$ are linearly independent.} 
We therefore deduce
\begin{align}
		k_a \left[(\vec{w}_{a+1}, \vec{w}_a) - (\vec{w}_{a+1}, \vec{w}_{a-1}) + (\vec{w}_{a}, \vec{w}_{a-1})\right] &= (\vec \beta_a, \vec{w}_a) - (\vec \beta_a, \vec{w}_{a-1}) + \beta_a^1(\vec w_a, \vec{w}_{a-1})\,,\nn\\
		-k_{a-1} \left[(\vec{w}_{a}, \vec{w}_{a+1}) - (\vec{w}_{a-1}, \vec{w}_{a+1}) + (\vec{w}_{a-1}, \vec{w}_{a})\right] &= (\vec \beta_a, \vec{w}_{a+1}) - (\vec \beta_a, \vec{w}_{a}) - \beta_a^1(\vec w_{a}, \vec{w}_{a+1}) \,,
\end{align}
where here $ (\vec{a}, \vec{c}) $ denotes the determinant of the 2$\times$2 matrix built with the vectors $\vec{a}$ and $\vec{c}$.
So, we obtain two different expressions for $k_a$ and the constraint is that they must be equal
\begin{align}\label{geetingthereeq}
&\frac{(\vec \beta_a, \vec{w}_a) - (\vec \beta_a, \vec{w}_{a-1}) + \beta_a^1(\vec w_a, \vec{w}_{a-1})}{(\vec{w}_{a+1}, \vec{w}_a) - (\vec{w}_{a+1}, \vec{w}_{a-1}) + (\vec{w}_{a}, \vec{w}_{a-1})} \nn\\
&\qquad\qquad\qquad\qquad \qquad= - \frac{(\vec \beta_{a+1}, \vec{w}_{a+2}) - (\vec \beta_{a+1}, \vec{w}_{a+1}) - \beta_{a+1}^1(\vec w_{a+1}, \vec{w}_{a+2})}{(\vec{w}_{a+1}, \vec{w}_{a+2}) - (\vec{w}_{a}, \vec{w}_{a+2}) + (\vec{w}_{a}, \vec{w}_{a+1})}	\:.
\end{align}

We are left to show that $\beta_a^1$ and $\vec{\beta}_a$ are such that \eqref{geetingthereeq} is satisfied. Notice that this equation can be rewritten in terms of 3$\times$3 determinants as
\begin{equation}\label{consistency}
	\frac{(\vec{v}_a,\vec{v}_{a-1},\vec{b}^{(+)})R_a^+ - (\vec{v}_a,\vec{v}_{a-1},\vec{b}^{(-)})R_a^-}{(\vec{v}_{a+1},\vec{v}_a,\vec{v}_{a-1})} = \frac{(\vec{v}_{a+2},\vec{v}_{a+1},\vec{b}^{(+)})R_{a+1}^+ - (\vec{v}_{a+2},\vec{v}_{a+1},\vec{b}^{(-)})R_{a+1}^-}{(\vec{v}_{a+2},\vec{v}_{a+1},\vec{v}_{a})} \:.
\end{equation}
This equation is gauge invariant, so we can choose a convenient gauge to show that it is satisfied. In particular it is convenient to pick the ``symmetric gauge" \eqref{gaugesym}:
\begin{equation}
	\lambda_+ = \lambda_- = 0 \quad \Rightarrow \quad \lambda_a^{(\pm)} = \lambda_a \:.
\end{equation}
We substitute the expression \eqref{Rpm} for $R_a^\pm$
into \eqref{consistency}. The terms involving $\lambda_{a-1}$ and $\lambda_{a+2}$ cancel and
we are left with an expression involving $\lambda_a$ and $\lambda_{a+1}$ given by\footnote{We could in principle have exploited the residual gauge invariance to set {\it e.g.} $\lambda_{a+1}=0$ but this is not necessary to conclude the proof.}
\begin{align}
\label{finaltoprove}
		&\lambda_a \bigg[\frac{(\vec{v}_{a-1}, \vec{v}_{a+1} , \vec{b}^{(+)})(\vec{v}_{a}, \vec{v}_{a+1} , \vec{b}^{(-)}) - (\vec{v}_{a-1}, \vec{v}_{a+1} , \vec{b}^{(-)})(\vec{v}_{a}, \vec{v}_{a+1} , \vec{b}^{(+)})}{(\vec{v}_{a+1}, \vec{v}_{a} , \vec{v}_{a-1})}\nn \\
		 &\qquad - \frac{(\vec{v}_{a+2}, \vec{v}_{a+1} , \vec{b}^{(+)})(\vec{v}_{a}, \vec{v}_{a+1} , \vec{b}^{(-)}) - (\vec{v}_{a+2}, \vec{v}_{a+1} , \vec{b}^{(-)})(\vec{v}_{a}, \vec{v}_{a+1} , \vec{b}^{(+)})}{(\vec{v}_{a+2}, \vec{v}_{a+1} , \vec{v}_{a})}\bigg] \nn\\
		 +&\lambda_{a+1} \bigg[\frac{(\vec{v}_{a}, \vec{v}_{a-1} , \vec{b}^{(+)})(\vec{v}_{a}, \vec{v}_{a+1} , \vec{b}^{(-)}) - (\vec{v}_{a}, \vec{v}_{a-1} , \vec{b}^{(-)})(\vec{v}_{a}, \vec{v}_{a+1} , \vec{b}^{(+)})}{(\vec{v}_{a+1}, \vec{v}_{a} , \vec{v}_{a-1})} \nn\\
		 &\qquad - \frac{(\vec{v}_{a}, \vec{v}_{a+2} , \vec{b}^{(+)})(\vec{v}_{a}, \vec{v}_{a+1} , \vec{b}^{(-)}) - (\vec{v}_{a}, \vec{v}_{a+2} , \vec{b}^{(-)})(\vec{v}_{a}, \vec{v}_{a+1} , \vec{b}^{(+)})}{(\vec{v}_{a+2}, \vec{v}_{a+1} , \vec{v}_{a})}\bigg]=0\:.
\end{align}
We can show that both the terms inside the square brackets vanish independently as a result of the vector quadruple product identity.
Specifically for any four vectors $\vec{a},\vec{b},\vec{c},\vec{d}$ in $\R^3$ we have
\begin{align}
 \vec{a} (\vec{b},\vec{c},\vec{d}) -\vec{b} (\vec{c},\vec{d},\vec{a})  +\vec{c} (\vec{d},\vec{a},\vec{b}) - \vec{d} (\vec{a},\vec{b},\vec{c}) =0 \,,
\end{align}
which immediately implies the following identity involving products of determinants:
\begin{equation}
( \vec{a},\vec{d},\vec{e}) (\vec{b},\vec{d},\vec{c}) + (\vec{b}, \vec{d},\vec{e}) (\vec{c},\vec{d},\vec{a})  +(\vec{c},\vec{d},\vec{e})  
(\vec{a},\vec{d},\vec{b})  =0 \:.
\end{equation}
Using this identity we can write
\begin{align}
(\vec{v}_{a-1}, \vec{v}_{a+1} , \vec{b}^{(+)})(\vec{v}_{a}, \vec{v}_{a+1} , \vec{b}^{(-)}) - (\vec{v}_{a-1}, &\vec{v}_{a+1} , \vec{b}^{(-)})(\vec{v}_{a}, \vec{v}_{a+1} , \vec{b}^{(+)})\nn\\ &= (\vec{v}_{a+1}, \vec{v}_{a} , \vec{v}_{a-1}) (\vec{v}_{a+1}, \vec{b}^{(-)} , \vec{b}^{(+)}) \:,\nn\\
(\vec{v}_{a+2}, \vec{v}_{a+1} , \vec{b}^{(+)})(\vec{v}_{a}, \vec{v}_{a+1} , \vec{b}^{(-)}) - (\vec{v}_{a+2}, &\vec{v}_{a+1} , \vec{b}^{(-)})(\vec{v}_{a}, \vec{v}_{a+1} , \vec{b}^{(+)}) \nn\\ &= (\vec{v}_{a+2}, \vec{v}_{a+1} , \vec{v}_{a})(\vec{v}_{a+1}, \vec{b}^{(-)} , \vec{b}^{(+)}) \,,
\end{align}
which implies that the coefficient of $\lambda_a$ in \eqref{finaltoprove} vanishes. A similar argument shows that
the coefficient of $\lambda_{a+1}$ in \eqref{finaltoprove} also vanishes and this concludes the proof of
\eqref{kernelconstr2}.

\subsection{$R_a^\pm$ as functions of $b_2$, $b_3$}\label{linearity}
Here we show that $R_a^\pm$, after eliminating the $\lambda_a$ variables, are linear functions of $b_2, b_3$ and $b_0$.
This shows that the second line in \eqref{bnoughtep2} are linear equations in these variables and also that \eqref{appbfinaleqquest}, after substituting 
$\Delta_a^+\to R_a^+$, are linear constraints. 

The key equation from the previous subsection is \eqref{consistency}. 
After eliminating $R_a^-$ via \eqref{dconstraints} and using also \eqref{diffbpbm}, we find \eqref{consistency} can be rewritten as
\begin{equation}
	\frac{(\vec{v}_a,\vec{v}_{a-1},\vec{p})\frac{R_a^+}{m_+ m_-} + (\vec{v}_a,\vec{v}_{a-1},\vec{b}^{(-)}) \frac{\Nas_a}{N}}{(\vec{v}_{a+1},\vec{v}_a,\vec{v}_{a-1})} = \frac{(\vec{v}_{a+2},\vec{v}_{a+1},\vec{p}) \frac{R_{a+1}^+}{m_+ m_-} + (\vec{v}_{a+2},\vec{v}_{a+1},\vec{b}^{(-)})\frac{\Nas_{a+1}}{N}}{(\vec{v}_{a+2},\vec{v}_{a+1},\vec{v}_{a})} \:.
\end{equation}
Schematically we can write this in the form
\begin{equation}\label{explicitRa}
	R_a^+ = R_{a+1}^+ A_a +B_a \:,
\end{equation}
where $A_a$ is independent of $b_2, b_3$ and $b_0$, while $B_a$ is linear in $b_2, b_3$ and $b_0$.
By recursively substituting the expression for $R_{a+1}^+$ into the expression for $R_a^+$, we can eventually express all $R_a^+$ in terms of the last one, $R_d^+$, as 
\begin{equation}
	R_a^+ = R_{d}^+ A'_a +B'_a \:,
\end{equation}
with $A'_a$, $B'_a$ having the same properties. Next we can use \eqref{Rpmidentities12} to find
\begin{equation}\label{explicitRd}
	R_d^+ = \bigg(\sum_{a=1}^{d} A'_a\bigg)^{-1} \left[2-\frac{b_0}{m_+} - \sum_{a=1}^{d} B'_a\right]\, ,
\end{equation}
to conclude that $R_d^+$ is a linear function of $b_2, b_3$ and $b_0$. From \eqref{explicitRa} we find that all of
the $R_a^+$ are linear functions of $b_2, b_3$ and $b_0$ and, recalling that $R_a^-$ are related to $R_a^+$ via \eqref{fluxes}, the
same is true for $R_a^-$. 
 
 \section{Fluxes from Kaluza-Klein reduction}\label{app:KK}

In this appendix we expand further on how flavour and baryonic fluxes 
arise from Kaluza-Klein reduction in AdS$_4\times X_7$/AdS$_5\times X_5$ solutions with $n=4$, $n=3$, respectively, 
and in particular analyse the relation \eqref{yetanotherflux} for the toric fluxes $\Nas_a$ from this point 
of view. Essentially this expands on the discussion in \cite{Benvenuti:2006xg} (using some of their results), 
to make contact with some flux formulas that appear in the present paper. Notice that here 
we are relating fluxes in a linearized Kaluza-Klein analysis around AdS$_4\times X_7$/AdS$_5\times X_5$ solutions, 
to fluxes in near horizon AdS$_2\times Y_9$/AdS$_3\times Y_7$ solutions -- while one cannot literally compare the solutions 
at this level, the fluxes are quantized, and hence one should be able to make such a matching; 
and indeed, we find this leads to a consistent picture.

We work in general dimension $n$, with $n=3$ and $n=4$ relevant to the AdS$_5\times X_5$ and AdS$_4\times X_7$ solutions, 
respectively. We may then consider Kaluza-Klein reduction of, respectively, Type IIB and $D=11$ supergravity on a Sasaki-Einstein manifold $X_{2n-1}$ with $U(1)^s $ isometry. 
As explained in \cite{Benvenuti:2006xg}, we obtain a supergravity theory with 
an AdS$_5$/AdS$_4$ vacuum, 
which has massless gauge fields $A_a$, where the index $a$ runs from 1 to $d \equiv s + b_{2n-3}$, where $ b_{2n-3}$ is the $(2n-3)$-th Betti number, {\it i.e.} $b_{2n-3}\equiv \dim H_{2n-3}(X_{2n-1},\R)$. Such gauge fields appear in the expansion of the $(2n-2)$-form potential 
\begin{equation}\label{Cexpansion}
	C_{2n-2} = \left[ \frac{2\pi}{V} \,\vol(X_{2n-1}) + \sum_{a=1}^{d} A_a \wedge \omega_a  + \cdots\right] \frac{\nu_n L^{2n-2} }{(5-n)2\pi}N\, .
\end{equation}
Here $V \equiv \Vol_{SE}(X_{2n-1})$ and $\vol(X_{2n-1})$ is the volume form on $X_{2n-1}$, coming from the AdS background, 
and the  terms in the $\cdots$ in \eqref{Cexpansion} are not relevant for our discussion. The remaining term $\sum_{a=1}^d A_a\wedge \omega_a$ 
is the Kaluza-Klein fluctuation term (called 
$\delta C_6$ in \eqref{deltaC6} with $n=4$),  
where $\omega_a$ are $(2n-3)$-forms satisfying  \cite{Benvenuti:2006xg}
\begin{align}\label{domegaa}
	\diff \omega_a &= \sum_{i=1}^{s} v_{ai} \, \partial_{\varphi_i} \lrcorner\, \vol(X_{2n-1}) \:, \\
	\diff *_{2n-1} \omega_a &= 0 \, .
\end{align}
 The $v_{ai}$ are constants, but 
in the case that $X_{2n-1}$ is toric, {\it i.e.} $s=n$, the $v_{ai}$ are  precisely the toric data \cite{Benvenuti:2006xg}. While we expect everything in this appendix 
to be true in general, for concreteness we assume that $X_{2n-1}$ is toric so as to be able to make use of various toric formulae that appear in the main text, and appendix \ref{app:A}. 

In general, quantities labelled by the index $a$ can be split into linear combinations of quantities labelled by an index $i$, $i=1, \dots, s$, and those labelled by an index $I$, $I=1,\dots, b_{2n-3}$. 
We thus begin by writing 
\begin{align}
	A_a &= \sum_{i=1}^{s} \alpha_a^i \,A_i + \sum_{I=1}^{b_{2n-3}} q_I^a A_I \:, \label{Aasplit} \\
	\omega_a &= \sum_{i=1}^{s} v_{ai}\, \omega_i + \sum_{I=1}^{b_{2n-3}} m_I^a\,\omega_I \:.\label{omegaasplit}
\end{align}
Here in \eqref{Aasplit} we take $\alpha_a^i$ and the baryonic charge matrix 
$q_I^a$ to be the quantities introduced in appendix \ref{app:A}. This is then precisely what one means by a splitting into ``flavour'' and ``baryonic'' 
symmetries: by definition $\alpha_a^i$ is the charge of the $a$th coordinate $z^a$ of 
the GLSM $\C^d$ under the $i$th $U(1)\subset U(1)^s$, while $q_I^a$ is the 
charge of this coordinate under the $I$th baryonic symmetry. By definition, these then
couple to the flavour gauge field $A_i$ and baryonic gauge field $A_I$, respectively, 
as in  \eqref{Aasplit}. Similarly, in \eqref{omegaasplit} we take the 
$\omega_I$ to be harmonic $(2n-3)$-forms, constituting a basis of $H^{2n-3}(X_{2n-1},\R)$ and such that $\int_{\mathcal{C}_I} \omega_J = \delta_{IJ}$ for $\mathcal{C}_I$  a basis 
for the free part of $H_{2n-3}(X_{2n-1},\Z)$, while $\omega_i$ satisfy
\begin{equation}
	\diff \omega_i = \partial_{\varphi_i} \lrcorner\, \vol(X_{2n-1}) \label{omegai} \:.
\end{equation}
This then ensures that \eqref{domegaa} holds. Moreover, taking the $m_I^a$ to satisfy the relations
\begin{align}
	\sum_{a=1}^d v_{ai}\,\alpha_a^j &= \delta_i^j \:,\qquad\qquad\;\;\, \sum_{a=1}^d v_{ai}\,q_I^a = 0 \:, \label{toricv}\\
	\sum_{a=1}^{d} q_I^a\, m_J^a &= \delta_{IJ} \:, \qquad\qquad \sum_{a=1}^d \alpha_a^i\, m_I^a = 0 \:, \label{toricm}
\end{align}
then gives the projections
\begin{align}\label{projectstuff}
	A_i &= \sum_{a=1}^{d} v_{ai} \, A_a \:, \qquad\qquad  A_I = \sum_{a=1}^{d} m_I^a  A_a \:,\\
	\omega_i &= \sum_{a=1}^{d} \alpha_a^i \, \omega_a \:, \qquad\qquad\;
	\omega_I = \sum_{a=1}^{d} q_I^a \, \omega_a \: .\label{omegaiI}
\end{align}
The forms $\omega_i$ and $\omega_I$ are then precisely the projections onto flavour and baryonic directions, respectively. 
The relations in \eqref{toricv} are precisely those in appendix \ref{app:A}, where $v_{ai}$ give the toric data for $X_{2n-1}$. 
In particular, recall that $\alpha_a^j$ are not unique, precisely due to the kernel 
generated by $q_I^a$. On the other hand, in the first equation in \eqref{toricm} the baryonic charges $q_I^a$, $I=1,\ldots,b_{2n-3}$, are by definition linearly independent, 
and $m_J^a$ is simply a choice of inverse. Due to the second equation in \eqref{toricv} we are free to shift $m_I^a\rightarrow m_I^a - \sum_{i=1}^s c_i^I v_{ai}$, 
and using this freedom one can then always impose the second equation in \eqref{toricm}. 

Notice, however, that we are still free to make the following
simultaneous shifts, without changing $A_a$ and $\omega_a$:
\begin{align}\label{shift}
		A_I \; &\to \; A_I - \sum_{i=1}^{s} c_i^I\,A_i \, , \qquad 
		\ \omega_i \; \to\; \omega_i + \sum_{I=1}^{b_{2n-3}} c_i^I\,\omega_I \,,  \\
		\alpha_a^i \; &\to\;  \alpha_a^i + \sum_{I=1}^{b_{2n-3}} c_i^I\,q_I^a \, , \qquad 
		m_I^a\;  \to\; m_I^a - \sum_{i=1}^{s}  c_i^I \, v_{ai} \:,
\end{align}
where $c_i^I$ are arbitrary constants. This is the same freedom discussed in section \ref{sec53} and section 
\ref{bhsec9}. 
With these decompositions, we can then rewrite  \eqref{Cexpansion} as
\begin{equation}
	C_{2n-2} =  \left[\frac{2\pi}{V} \,\vol(X_{2n-1}) + \sum_{i=1}^{s} A_i \wedge \omega_i + \sum_{I=1}^{b_{2n-3}}   A_I \wedge \omega_I + \cdots\right] \frac{\nu_n L^{2n-2} }{(5-n){2\pi}}  N\, .
\end{equation}

Up until this point, all we have done is expand upon the Kaluza-Klein analysis discussed in \cite{Benvenuti:2006xg}. However, consider now 
putting the lower-dimensional theory on a spindle $\Sigma$, fibering the   internal manifold  $X_{2n-1}$ over it and building an $(2n+1)$-dimensional internal space $X_{2n-1}\, \hookrightarrow\, Y_{2n+1} \, \rightarrow\, \Sigma\,$. 
The fibration is achieved by turning on the  flavour magnetic charges
\begin{equation}
	\frac{1}{2\pi} \int_\Sigma \diff A_i = \frac{p_i}{m_+ m_-} \:,
\end{equation}
as explained in section \ref{sec:volform}. Next consider integrating $ \diff C_{2n-2}$ over supersymmetric $(2n-1)$-submanifolds $\Sigma_a$, that are fibrations $S_a\, \hookrightarrow\, \Sigma_a \, \rightarrow\, \Sigma\,$, where $S_a$ is defined to be the (toric) 
submanifold
such that $\int_{S_a} \omega_b = \delta_{ab}$ (see \eqref{Sigmaa} in the main text). Such integrals give the fluxes $\Nas_a$
\begin{equation}\label{fluxes_appendix}
	\Nas_a = \frac{5-n}{\nu_n L^{2n-2} } \int_{\Sigma_a} \diff C_{2n-2} \:.
\end{equation}
These fluxes depend only on the homology classes of $\Sigma_a$ in $Y_{2n+1}$ which, as explained in section \ref{sec53}, may be decomposed in terms of the homology classes of the cycles $\Sigma_I$ obtained by fibering $\mathcal{C}_I$ over the spindle plus that of the fibre $X_{2n-1}$ itself as
\begin{equation}
	[\Sigma_a] = \frac{q_0^a }{m_+ m_-}[X_{2n-1}] + \sum_{I=1}^{b_{2n-3}} q_I^a [\Sigma_I] =\frac{1 }{m_+ m_-}\sum_{i=1}^s p_i\,\alpha_a^i\, [X_{2n-1}] + \sum_{I=1}^{b_{2n-3}} q_I^a [\Sigma_I]\ .
\end{equation}
In particular this is the homology content of the last equation in \eqref{toricfluxrelations}, where $[X_{2n-1}]$ is the class of the fibre in $H_{2n-1}(Y_{2n+1},\Z)$. However, note that given the homology class $[\mathcal{C}_I] \in H_{2n-3}(X_{2n-1},\Z)$ in the fibre, the homology class $[\Sigma_I] \in H_{2n-1}(Y_{2n+1},\Z)$ in the total space is not uniquely identified but it depends on the specific representative we use for $[\mathcal{C}_I]$, because these will be twisted in different ways by the fibration. Indeed, in order for $[\Sigma_a]$ to be invariant under the shift of $\alpha_a^i$ in \eqref{shift}, correspondingly there must be a shift in $[\Sigma_I]$
\begin{equation}\label{hom_shift}
	[\Sigma_I] \, \to \, [\Sigma_I] - \frac{1}{m_+ m_-} \, \sum_{i=1}^{s} p_i\, c_i^I\,[X_{2n-1}] \:,
\end{equation}
which is exactly what parametrizes the ambiguity in choosing a representative for $[\mathcal{C}_I]$.

We now have everything we need to compute the fluxes \eqref{fluxes_appendix}:
\begin{align}\label{splitNa}
		\Nas_a &= \frac{5-n}{\nu_n L^{2n-2} } \left[\frac{1 }{m_+ m_-}\sum_{i=1}^s  p_i\,\alpha_a^i \int_{X_{2n-1}} \diff C_{2n-2} + \sum_{I=1}^{b_{2n-3}} q_I^a \int_{\Sigma_I} \diff C_{2n-2}\right]\nn\\
		&=  \frac{N}{m_+ m_- V}\sum_{i=1}^s  p_i\,\alpha_a^i \int_{X_{2n-1}} \vol(X_{2n-1}) + \sum_{I=1}^{b_{2n-3}} q_I^a \,\mathcal{N}_I\nn \\
		&= \frac{N}{m_+ m_-}\sum_{i=1}^s p_i\,\alpha_a^i + \sum_{I=1}^{b_{2n-3}} q_I^a\,\mathcal{N}_I \:,
\end{align}
where, in agreement with \eqref{fluxquant3}, \eqref{fluxquant4}, we introduced
\begin{align}\label{NIdef}
	\mathcal{N}_I \equiv \frac{5-n}{\nu_n L^{2n-2} } \int_{\Sigma_I} \diff C_{2n-2} &  = \frac{N}{2\pi} \left[\sum_{i=1}^{s} \int_\Sigma \diff A_i \int_{\mathcal{C}_I} \omega_i + \sum_{J=1}^{b_{2n-3}}  \int_{\Sigma} \diff A_J \int_{\mathcal{C}_I} \omega_J \right]\nn \\
	& = \frac{N}{m_+m_-}\sum_{i=1}^s p_i \int_{\mathcal{C}_I}\omega_i  + N\int_\Sigma \frac{\diff A_I}{2\pi}\, .
\end{align}
In particular this gives a formula for the ``baryonic fluxes'' $\mathcal{N}_I$ in terms of the Kaluza-Klein reduction. Notice that the 
second term $N\int_\Sigma {\diff A_I}/{2\pi}$ is what one would naturally call the ``baryonic flux'', but that in general also the first term contributes, coming 
from the flavour twisting $p_i$. 

Equation \eqref{splitNa} 
shows how to 
split the fluxes $\Nas_a$ into a ``flavour'' part and a ``baryonic'' part. It is clear that, while the fluxes $\Nas_a$ are unambiguous and dictate how the fields  are twisted over the spindle in the dual field theory\footnote{In the toric case there is a general prescription to identify these $\Nas_a$ with the twisting of the dual gauge theory, since this 
has a known relation to the GLSM. There is no such known general prescription in the non-toric setting.}, on the other hand
applying the shift invariance \eqref{shift} to \eqref{splitNa} results in an ambiguity in the flavour/baryonic splitting
\begin{align}
		\mathcal{N}_I \; &\to \; \mathcal{N}_I - \frac{N}{m_+ m_-}\sum_{i=1}^{s} c_i^I\,p_i \,,\nn\\
		\alpha_a^i \; &\to \;  \alpha_a^i + \sum_{I=1}^{b_{2n-3}} c_i^I\,q_I^a \:.
\end{align}
The transformations above encode the freedom to choose $s\times b_{2n-3}$ constants $c_i^I$ and, given that there are  $ b_{2n-3}$ fluxes $\mathcal{N}_I$, we can always choose them in such a way to fix the $\mathcal{N}_I$ to whatever we want ({\it e.g.} to zero) and we will actually still have some freedom left, parametrized by $(s-1)\, b_{2n-3}$ constants. The shift in the $\mathcal{N}_I$ can also be understood from \eqref{NIdef} by noticing that $\diff C_{2n-2}$ is invariant under the transformations \eqref{shift} but the cycle $\Sigma_I$ is not, and transforms as \eqref{hom_shift}.\footnote{One cannot see the shift in $\mathcal{N}_I$ directly from the final line of \eqref{NIdef}, as the information on how the cycle $\mathcal{C}_I$ has been fibred over $\Sigma$ has been lost in this expression.}


\providecommand{\href}[2]{#2}\begingroup\raggedright\endgroup

\end{document}